\RequirePackage{fix-cm}

\documentclass[twocolumn,final,natbib]{svjour3}          

\smartqed  
\usepackage{graphicx}
\usepackage{stfloats}

\graphicspath{{Figures/}}
\usepackage{float}
\usepackage{enumitem}

\usepackage[colorlinks,pdfpagelabels,bookmarksopen,bookmarksnumbered,citecolor=blue,urlcolor=blue,linkcolor=blue]{hyperref}
\setcitestyle{numbers,square}

\usepackage[utf8x]{inputenc} 

\usepackage{mathtools}
\usepackage{amsmath}
\usepackage{amssymb}
\usepackage{amsbsy}
\usepackage{mathrsfs}
\usepackage{bm}
\usepackage{relsize}

\usepackage[flushmargin]{footmisc} 

\usepackage{empheq}

\usepackage{color}
\usepackage{tcolorbox}
\allowdisplaybreaks
\tcbuselibrary{skins,breakable}
\newtcolorbox[auto counter, number freestyle={\noexpand\Roman{\tcbcounter}}]{mybox_alg}[2][]{%
    enhanced,
    breakable,
    fonttitle=\bfseries,
    leftupper=1mm,
    leftlower=1mm,
    colback=white!50!white,
    colbacktitle=white!50!white,
    coltitle=black,
    boxsep=1mm,
    lefttitle=3mm,
    float=h,
    title=Box~\thetcbcounter: #2,
    #1
}

\usepackage[noline,commentsnumbered,ruled]{algorithm2e}
\SetAlFnt{\scriptsize}

\usepackage[title]{appendix}

\usepackage[misc]{ifsym}

\newcommand{\Kappalarge}{{\Large{\pmb \kappa}}}
\newcommand{\hS}{{r}}
\newcommand{\shS}{{r}}
\newcommand{\bx}{{\bf x}}
\newcommand{\bbx}{{(\bf x)}}
\newcommand{\bxhat}{{\hat{\bf x}}}
\newcommand{\bbxhat}{{(\hat{\bf x})}}
\newcommand{\bbxchi}{{({\bf x},\chi)}}

\newcommand{\bchi}{{(\chi)}}
\newcommand{\domega}{{ \, d\Omega }}
\newcommand{\dgamma}{{ \, d\Gamma }}
\newcommand{\mcolon}{{ \, , }}
\newcommand{\mdot}{{ \, . }}
\newcommand{\Uconductivity}{{the nominal heat conduction energy density}}
\newcommand{\Usource}{{the nominal heat source energy density}}
\newcommand{\Uflux}{{the nominal heat flux energy density}}

\def \ifempty#1{\def\temp{#1} \ifx\temp\empty }
\newcommand{\gradT}[2][_{\chi}]{{
	\ifempty{#2}
		\bm{\nabla}\theta#1
	\else
		\bm{\nabla}\theta#1^{(#2)}
	\fi
}}

\journalname{ }
\date{ }
\def\makeheadbox{{%
\hbox to0pt{\vbox{\baselineskip=10dd\hrule\hbox
to\hsize{\vrule\kern3pt\vbox{\kern3pt
\hbox{This is a post-peer-review, pre-copyedit version of an article published in \bfseries{Computational Mechanics}.}
\hbox{The final authenticated version is available online at: \href{https://doi.org/10.1007/s00466-020-01850-0}{10.1007/s00466-020-01850-0}.}
\kern3pt}\hfil\kern3pt\vrule}\hrule}%
\hss}}}

\begin{document}

\title{Topology optimization of thermal problems in a nonsmooth variational setting: closed-form optimality criteria}

\author{Daniel$\ $Yago$^{1,2}$         \and
       Juan$\ $Cante$^{1,2}$           \and
       Oriol$\ $Lloberas-Valls$^{2,3}$  \and
       Javier$\ $Oliver$^{2,3}$
}

\authorrunning{Daniel Yago         \and
       		   Juan Cante           \and
       		   Oriol Lloberas-Valls  \and
       		   Javier Oliver} 

\institute{\begin{itemize}
           \item[\Letter] J. Oliver \at
                      	  \email{oliver@cimne.upc.edu} \\
           \item[1] Escola Superior d'Enginyeries Industrial, Aeroespacial i Audiovisual de Terrassa (ESEIAAT)\\
                         Technical University of Catalonia (UPC/Barcelona Tech), Campus Terrassa UPC, c/ Colom 11, 08222 Terrassa, Spain
           \item[2]      Centre Internacional de M\`{e}todes Num\`{e}rics en Enginyeria (CIMNE)\\ 
                         Campus Nord UPC, M\`{o}dul C-1 101, c/ Jordi Girona 1-3, 08034 Barcelona, Spain
           \item[3]     E.T.S d'Enginyers de Camins, Canals i Ports de Barcelona (ETSECCPB)\\
                         Technical University of Catalonia (UPC/Barcelona Tech), Campus Nord UPC, M\`{o}dul C-1, c/ Jordi Girona 1-3, 08034 Barcelona, Spain                   
           \end{itemize}	  	             	                  
}


\maketitle

\begin{abstract}
This paper extends the nonsmooth Relaxed Variational Approach (RVA) to topology optimization, proposed by the authors in a preceding work, to the solution of thermal optimization problems. First, the RVA topology optimization method is briefly discussed and, then, it is applied to a set of representative problems in which the thermal compliance, the deviation of the heat flux from a given field and the average temperature are minimized. For each optimization problem, the relaxed topological deri\-va\-tive (RTD) and the corresponding adjoint equations are presented. This set of expressions are then discretized in the context of the finite element method (FEM) and used in the optimization algorithm to update the characteristic function. 

Finally, some representative (3D) thermal topology optimization examples are presented to asses the performance of the proposed method and the Relaxed Variational Approach solutions are compared with the ones obtained with the \textit{level set method} in terms of the cost function, the topology design and the computational cost.


\end{abstract}

\section{Introduction}
\label{Sec_intro}

\subsection{Motivation and background}

During the last decades, a variety of topology optimization methods have been proposed in the literature. With no aim of being exhaustive, we could classify them into (i) homogenization methods, (ii) density based optimization (SIMP) methods, (iii) level set approaches, and (iv) evolutionary methods, among others. For further information the reader is addressed to reviews in \cite{Eschenauer2001,Rozvany2008,Sigmund2013,Dijk2013}.
Albeit these techniques were initially focused on structural problems, along time several of them have been extended to other problems, thus including thermal problems and a number of different applications in this field, e.g.:
\begin{enumerate}[label=(\alph*),noitemsep]
	\item \textit{Thermal compliance minimization}: focused on maximizing thermal diffusion in steady-state problems. \citet{Bendsoe2004} implemented the SIMP method for thermal optimization problem as an extension of structural optimization. This same problem was also addressed with ESO-based methods by \citet{Li1999}. Subsequently, \citet{Ha2005} suggested a level set method for the minimization of the thermal compliance via a Hamilton-Jacobi equation. Later, \citet{Zhuang2007} implemented the aforementioned problem using a topological derivative method. Alternatively, \citet{Gersborg-Hansen2006}, for the Finite Volume Method (FVM) together with a SIMP method, \citet{Gao2008}, for the ESO method, and \citet{Giusti2009}, for the topological derivative method, have developed the corresponding algorithms to include design-dependent effects of heat sources.\footnote{The magnitude of the heat source changes according to the material of the point.} Furthermore, \citet{Iga2009} and \citet{Yamada2011} included the heat convection effects in the design for maximizing thermal diffusivity using a homogenization design method and the modified phase-field method reported in \cite{Yamada2010}, respectively.
		
	\item \textit{Maximum/average temperature minimization}: looking for designs that reduce the temperature of thermal devices, while increasing their durability. With this goal in mind, researchers have proposed different objective functions to minimize either the average temperature or the maximum temperature in the design domain. \citet{Zhang2008} reported that the $p-norm$ of the temperature field in the design domain, approximates reasonably well the maximum temperature for a large enough $p$. \citet{Marck2012} proposed the minimization of the average temperature and its variance, via a SIMP method, by creating the Pareto front of the multi-objective thermal problem, thus leading to a reduction in the achieved temperature while avoiding temperature peaks. On the other side, \citet{Burger2013} minimized the average internal temperature in the whole design domain, by dissipating the generated heat through the introduction of distributed heat sources within the design domain. For the transient case, the minimization of the maximum temperature throughout the entire operating period was analyzed by \citet{Wu2019} via the SIMP method.
	 
	\item \textit{Multiple heat actions optimization}: which can be regarded as multi-objective problems where the cost function corresponds to the weighted sum of individual cost functions for each of the heat actions. In this context \citet{Li1999,Li2000} optimized some printed circuit boards (PCB) with the ESO method subjected to multiple heat source, by considering a functional proportional to the heat flux. Years later, \citet{Zhuang2007} proposed the optimization of some thermally conductive structures via a level set method by optimizing the weighted average of the quadratic temperature gradient.
	
	\item \textit{Multi-material thermal optimization}: thermal topology optimization has been also carried out taking into account three or more different materials. \citet{Zhuang2010} proposed a multi-material topology optimization for the heat conduction problem via a level set method. Later, \citet{Zhuang2015} used the SIMP method to optimize transient heat conduction problems.
	
	\item \textit{Heat flux manipulation optimization problems}: a precursor work on the field is the one by \citet{Narayana2012}, where multilayered optimized designs for thermal problems were presented. Later, \citet{Dede2013} proposed a homogenization-based method which optimizes the orientation of a micro-structure by modifying the effective conductivity tensor at each point. Following this line, \citet{Peralta2017} suggested a homogenization-based optimization, where the error in guiding the heat flux in given path is minimized, and successfully accomplished the optimization of a thermal concentrator. Finally, \citet{Fachinotti2018} extended the idea to \emph{black-and-white designs} via a SIMP optimization. 
	
\end{enumerate}

This work focuses on applying the Relaxed Variational Approach (RVA) to topology optimization, proposed by the authors in a previous work \citep{Oliver2019}, to thermal problems. The distinctive feature of RVA is that it keeps the original nonsmooth character of the characteristic function, the design variable, describing the material topology ($\chi:\Omega\rightarrow \{0,1\}$) but, in spite of this, a variational analysis can be conducted and, then, \emph{closed-form solutions} of the problem (equivalent to the Euler equations in smooth variational problems) can be readily obtained. 
The approach relies on the use of a specific topological sensitivity, \emph{the Relaxed Topological Derivative} (RTD), as an efficient and simple approximation to the geometrical (or exact) topological derivative (TD), which is consistently derived in the considered relaxed optimization setting.\footnote{based on a bi-material (soft/hard) approximation, or ersatz approach.}
Then, a robust and efficient \emph{Cutting\emph{\&}Bisection algorithm} is proposed for solving the obtained algebraic, non-linear, solutions in a sequential pseudo-time framework.

The goal here is, thus, to explore the possible extension of the benefits of the RVA, reported in \citep{Oliver2019} for structural problems, to the realm of thermal problems, typically: 
\begin{itemize}[noitemsep]
\item Avoid \emph{checkerboard patterns} and \emph{mesh-dependency} in the optimized solution.
\item Display \emph{black-and-white solutions}, instead of \emph{blurry black-gray-and-white} solutions, for the material distribution, without resorting to \emph{a posteriori filtering techniques}.
\item Achieve precise local optima, in a reduced number of iterations of the non-linear solution algorithm, thus leading to relevant diminutions of the associated computational cost. 
\item Involve \emph{general and easy-to-derive} sensitivities of the cost function in the resulting optimization algorithm.
\item Allow the control of the \emph{minimum width} of the material filaments in the optimized layout, thus incorporating \emph{manufacturing constraints} in the designs and precluding classical \emph{element/cell-size-dependence} in the obtained solutions, thus removing the well-known ill-posedness of the problem.
\end{itemize}

For this purpose three representative thermal optimization problems are explored in this work a) maximization of thermal diffusion, without boundary dependent properties, in steady-state thermal scenarios, b) thermal cloaking based on minimization of the deviation of the heat flux with respect to a target one and c) thermal cloaking based on minimizing the average temperature on a surface around the cloaked object. 

The remaining of this paper is structured as follows: in Section \ref{sec_review}, the considered Relaxed Variational Approach (RVA) to topology optimization is summarized in order to, both, supply to the reader the indispensable information and providing the work with the necessary completeness. Then, in Section \ref{sec_formulation_state}, a detailed specification of the RVA for thermal optimization problems is presented. Subsequently, a general optimization algorithm is described in Section \ref{sec_opt_alg}. The resulting formulation is then assessed, by its application to a set of thermal problems, first in terms of their formulation, in Section \ref{sec_topology_optimization_problems} and, then, in terms of their numerical application to specific 3D problems in Section \ref{sec_representative_numerical_simulations}. Finally, Section \ref{sec_conclusion} concludes with some final remarks.

\section{Relaxed Variational Approach (RVA) to topology optimization: a summary}\label{sec_review}

\subsection{Topology domain representation} \label{sec_relaxed_toplogical_optimization}

\begin{figure}
	\centering
	\includegraphics[width=6cm, height=4cm]{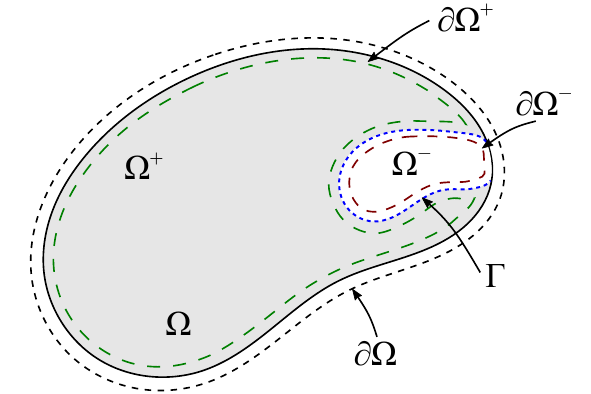}
	\caption{Representation of the analysis domain, $\Omega$, comprising two disjoint sub-domains $\Omega^+$ and $\Omega^-$. The external boundary of $\Omega$, $\partial\Omega$, is represented by a black dashed line, while the sub-domains boundaries, $\partial\Omega^+$ and $\partial\Omega^-$, are, respectively, depicted by long green and short red dashed lines. Finally, the common sub-domains border, $\Gamma$, is represented by a blue dotted line.}
	\label{fig_problem_setup3}
\end{figure}

Let the \emph{analysis domain}\footnote{Albeit the name \emph{design domain} is commonly used in topology optimization for $\Omega$, in this work distinction is made of the \emph{analysis domain}, the whole domain considered in the analysis, and the \emph{design domain}, the subset of $\Omega$  where the topology is going to be optimized (therefore changed from an initial layout). The reason is that, in some of the considered problems, a certain part of $\Omega$ is endowed with a fixed, predetermined, topology thus not being properly part of the design domain.}, $\Omega$, denote a fixed smooth open domain of $\mathbb{R}^n$ ($n=2$ or $3$), whose boundary $\partial\Omega$ is also smooth, composed in turn by two smooth open subdomains, $\Omega^+,\Omega^-\subset\Omega$, with $\overline{\Omega}^+\cup\overline{\Omega}^-=\overline{\Omega}$ and $\Omega^+\cap\Omega^-=\emptyset$.\footnote{$\overline{(\cdot)}$ denotes the closure of the open domain $(\cdot)$.} The first subdomain, $\Omega^+$, stands for the \textit{hard material} domain, made of a hard \emph{(high-conductive)} material ($\mathfrak{M}^+$), while subdomain, $\Omega^-$, denoted as the \textit{soft material} domain, is occupied by a soft (\emph{low-conductive}) material ($\mathfrak{M}^-$). These two subdomains are surrounded by their respective boundaries, $\partial\Omega^+$ and $\partial\Omega^-$, with $\partial\Omega^+\cap\partial\Omega^-=\Gamma$ (see Figure \ref{fig_problem_setup3}).

The standard nonsmooth \textit{characteristic function}, $\chi\bbx : \Omega \rightarrow \{0,1\}$, defining the topology of the analysis domain,\footnote{The characteristic function, $\chi$, is considered as the design variable in the topology optimization problem.} is then defined as 
\begin{equation} \label{eq_characteristic_function}
	\left\{
	\begin{split}
		&\Omega^{+}\coloneqq\{\mathbf{x}\in \Omega \;/\; \chi\bbx=1\} \\
		&\Omega^{-}\coloneqq\{\mathbf{x}\in \Omega \;/\; \chi\bbx=0\} 
	\end{split}
	\right. \mdot
\end{equation}

\begin{figure}
	\centering
	\includegraphics[width=8.4cm]{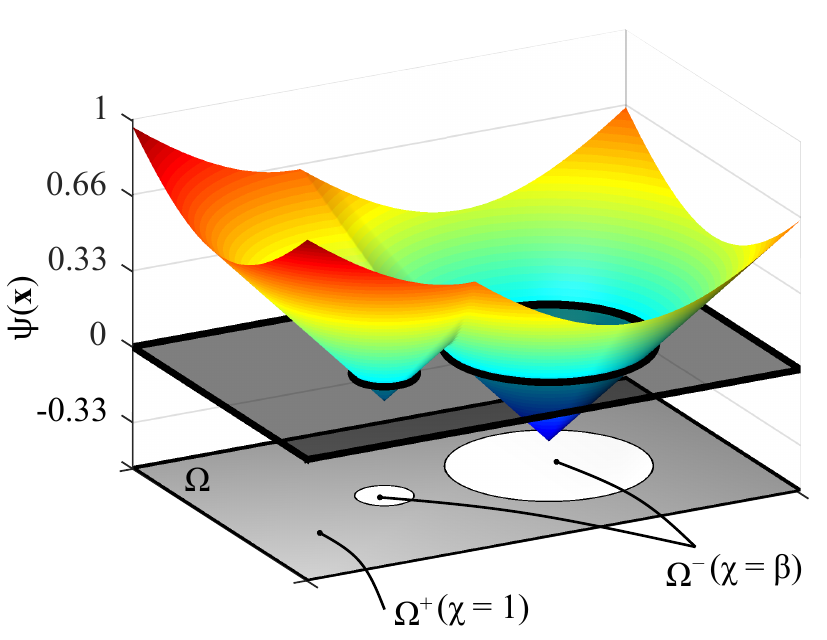}
	\caption{Topology representation in terms of the discrimination function, $\psi\bbx$.}
	\label{fig_problem_setup2}
\end{figure}

Alternatively, the topology can be implicitly defined through a smooth function (termed \textit{discrimination function} in \citet{Oliver2019}) $\psi\bbx:\Omega\rightarrow\mathbb{R}$, $\psi \in H^1({\Omega})$, defined as
\begin{equation} \label{eq_level_set}
	\left\{
	\begin{split}
		& \psi\bbx>0 \Longleftrightarrow \mathbf{x} \in \Omega^+\\
		& \psi\bbx<0 \Longleftrightarrow \mathbf{x} \in \Omega^-
	\end{split}
	\right. \mdot
\end{equation}
Then, the two aforementioned subdomains are implicitly defined through $\psi\bbx$ (see Figure \ref{fig_problem_setup2}) as
\begin{equation} \label{eq_domain_splitting}
	\left\{
	\begin{split}
		&\Omega^{+}\coloneqq\{\mathbf{x}\in \Omega \;/\; \psi\bbx>0\} \\
		&\Omega^{-}\coloneqq\{\mathbf{x}\in \Omega \;/\; \psi\bbx<0\} 
	\end{split}
	\right. \mcolon
\end{equation}
and the \textit{characteristic function}, $\chi_{\psi}\bbx : \Omega \rightarrow \{0,1\}$, defining the topology of the analysis domain, can be then expressed as
\begin{equation} \label{eq_heaviside_level_set}
	\chi_{\psi}\bbx={\cal H}(\psi\bbx) \mcolon
\end{equation}
where ${\cal H}(\cdot)$ stands for the Heaviside function evaluated at {$(\cdot)$}.\footnote{Henceforth, the subindex $\psi$ of the characteristic function, $\chi_{\psi}$, will be omitted.}

According to equations (\ref{eq_domain_splitting}) and (\ref{eq_heaviside_level_set}), the bi-valued \textit{characteristic function}, $\chi\bbx$, takes the value $1$ when the discrimination function is positive ($\psi\bbx>0$), i.e. when $\mathbf x\in\Omega^+$, and the value $0$ when $\psi\bbx<0$, i.e. when $\mathbf x\in\Omega^-$. This \emph{bi-valued} (\emph{black-and-white}) (black=1, white=0) character of $\chi$, is a fundamental feature of the RVA, and it is always held along the mathematical derivations keeping the nonsmooth character of  the design variable. However, the image-set $\{1,0\}$ is modified to $\{1,\beta\}$, by introducing the, here termed, \textit{relaxed Heaviside function} 
\begin{equation} \label{eq_shifted_Haviside_function}
	{\cal H}_{\beta}({x})=
	\left\{
	\begin{split}
		&1&for&\quad {x} > 0 \\
		&\beta&for&\quad {x} < 0 
	\end{split}\quad x\in{\mathbb R}\;;\quad\beta<<1
	\right.\mdot
\end{equation}
\begin{remark} \label{rmk_properties_interp}
\emph{In \emph{single-material topology optimization}, the value $\chi\bbx$ is commonly used to define the material property value ${\mathbb E}$, at point ${\bf x}$, in terms of the \emph{reference material property value $E$}, through ${\mathbb E}\bbx=\chi\bbx^m E$; with $m > 1$. Then, $\chi=1$ in $\Omega^+$ naturally defines a \emph{solid material} with properties ${\mathbb E}=\chi^m E=E$, whereas the value $\chi=0$ in $\Omega^-$, \emph{made of no-material (voids)}, defines null material properties ${\mathbb E\bbx}=\chi\bbx^m E=0$ in that domain.
In the present relaxed variational approach, instead, the shift of the low limit of $\chi$ to $\beta$ ($0<\beta<<1$), in equation (\ref{eq_shifted_Haviside_function}), relaxes that setting to a \emph{bi-material approach}, with $\Omega$ containing \emph{two different solid materials}: 1) a \emph{hard material}, in $\Omega^+$, with regular solid properties ${\mathbb E}=\chi^m E=E$, and 2) a \emph{soft material}, in $\Omega^-$, with very low material properties ${\mathbb E}=\chi^m E=\beta^m E$, which are scaled to values close to zero by the factor $\chi=\beta<<1$.\footnote{Thus, the single-material and the bi-material formulations converge asymptotically as $\beta\rightarrow 0$.} This qualifies the RVA as a \emph{relaxed} or \emph{ersatz/bi-material} approach. This fact will be retrieved later on in this work (see, for instance, equations (\ref{eq_conductivity_interp}) and (\ref{eq_heat_sources_interp})).}
\end{remark}

The topology optimization goal is, then, to minimize a functional or cost function $\mathcal{J}(\chi)$ subjected to one or more constraints and governed by the state equation, i.e.
\begin{equation} \label{eq_minimization_restricted}
	\left[\begin{split}
		&\underset{\chi\in{\mathscr{U}}_{ad}} {\operatorname{min}}\quad{\cal J}\left( \chi\right)\equiv\int_{\Omega}{j(\chi,\mathbf{x})}\domega &\quad(a) \\
		&\text{subject to:} \\
		&\hspace{1.2cm}{\cal C}(\chi)\equiv\int_{\Omega}{c(\chi,\mathbf{x})\domega}=0 &\quad(b) \\
		&\text{governed by:} \\
		&\hspace{1.2cm}{\textstyle state \; equation} &\quad(c) 
	\end{split}\right.
\end{equation}
where ${\mathscr{U}}_{ad}$ stands for the set of admissible solutions for $\chi$. Furthermore, ${\cal C}\bchi$ represents the constraint functional, which, in all the examples in this paper, will be the volume constraint,\footnote{The present \emph{Cutting}\&\emph{Bisection} algorithm is only intended for single constrained topology optimization problems. Furthermore, along this paper, only equality, pseudo-time evolving volume constraints are considered.} and the state equation will correspond to the energy balance in the domain $\Omega$, which will be described later in this paper (see equation (\ref{eq_problem_thermal_definition2})). Functionals (\ref{eq_minimization_restricted})-(a-b) are assumed to pertain to the following family 
\begin{equation} \label{eq_minimization_family}
 \mathcal{F}\bchi:L^2(\Omega) \rightarrow {\mathbb{R}} \; ;\quad	\mathcal{F}\bchi\equiv\int_{\Omega}{f(\chi,\mathbf{x})\domega} \mcolon
\end{equation}
the kernel $f(\cdot,\cdot)$ being sufficiently smooth, for differentiation purposes.
\subsection{Relaxed Topological Derivative (RTD)} \label{sec_relaxed_toplogical_derivative}

The RVA defines the \emph{Relaxed Topological Derivative} (RTD), as the sensitivity of the functional in equation (\ref{eq_minimization_family}). The RTD is derived as the change of the functional in terms of $\chi(\hat{\bf x})$, as the material at point $\bxhat$ is exchanged, per unit of the measure of a perturbed domain around $\bxhat$.
It can be computed in terms of the classical Fréchet derivative, $\frac{\partial (\cdot)}{\partial\chi}\bbxhat$, of the integral kernel, i.e.
\begin{equation} \label{eq_RTD_general}
	\dfrac{\delta {\cal F}(\chi)}{\delta \chi}\bbxhat=
	\left[\dfrac{\partial f({\chi},{\bf x})}{\partial {\chi}}\right]_{{\bf x}=\hat{\bf x}}\Delta \chi\bbxhat \mcolon
\end{equation}
where ${\Delta \chi}\bbxhat$ is termed the \emph{exchange function} and stands for the signed variation of $\chi\bbxhat$, due to that material exchange, i.e. 
\begin{equation} \label{eq_alpha}
	{\Delta\chi}({\bf x})=
	\left\{
	\begin{split}
		-(&1-\beta)<0  \;\;\;\;\textit{for} \;\;{\bf x}\in\Omega^{+}  \\ 
		 (&1-\beta)>0 \;\;\;\;\textit{for} \;\;{\bf x}\in\Omega^{-}
	\end{split} 
	\right. \mdot
\end{equation}
Details on the derivations can be found in \citet{Oliver2019}.

\subsection{Closed-form algebraic solutions} \label{sec_closed_form_box}
After some algebraic operations, the optimality condition for the \textit{constrained topology optimization problem} can be written as 
\begin{align} \label{eq_optimal_condition}
		&\dfrac{\delta {\cal L}(\chi,\lambda)}{\delta\chi}\bbx=
		\dfrac{\delta {{\cal J}}(\chi)}{\delta\chi}\bbx+\lambda\dfrac{\delta {\cal C}(\chi)}{\delta\chi}\bbx= \notag\\
		&=\left(\dfrac{\partial j \left(\chi,\mathbf x \right)}{\partial \chi}{\Delta\chi}\bbx
			+\lambda\,\text{sgn}(\Delta\chi\bbx)\right)>{0}\;\; \forall\mathbf{x}\in\Omega \mcolon
\end{align}
where $\lambda$ stands for a Lagrange multiplier enforcing restriction ${\cal C}(\chi)=0$, and ${\cal L}$ stands for the \emph{Lagrangian function} of the optimization problem (see \citet{Oliver2019} for additional information). Then, a \emph{closed-form} solution for the topology in equation (\ref{eq_heaviside_level_set}) can be computed as
\begin{equation} \label{eq_solution_psi_xi}
	\left\{
	\begin{split}
		&\psi\bbx\coloneqq\xi(\chi,{\bf x})-\lambda \quad \\
		&\chi \bbx={\cal H}_\beta(\psi\bbx)
	\end{split}
	\right.
	in \; \Omega \mcolon
\end{equation}
where $\xi(\chi,{\bf x})$ is termed the \emph{pseudo-energy}\footnote{The pseudo-energy, $\xi\bbxchi$, has normally dimensions of energy.} and it shall be specifically derived for each considered problem. Equations (\ref{eq_solution_psi_xi}) constitute a \emph{closed-form-algebraic} (non-linear fixed-point equation) solution of the problem, which are solved, for $\chi\bbx$ and $\lambda$, via the \emph{Cutting\emph{\&}Bisection algorithm} proposed in \cite{Oliver2019}. The resulting global algorithm is sketched in Box \ref{box_closed_form}, where the constraint equation is expressed in terms of the pseudo-time  $t\in[0,T]$, in the context of a time advancing strategy. Notice that the parameter $T$ stands for the pseudo-time corresponding to the final volume of the proposed topology optimization (pseudo-time dependent) procedure and must be set by the user.
\begin{remark}
\emph{The discrimination function $\psi\bbx$ in equation (\ref{eq_solution_psi_xi}) is subsequently smoothed through a \emph{Laplacian smoothing}, whose parameter $\epsilon$ determines the \emph{minimum filament width} of the resulting topology, thus removing the possible mesh dependency of the results and the ill-posedness of the problem. The reader is addressed to reference \cite{Oliver2019} for further details.}
\end{remark}

\vspace{3cm}

\begin{mybox_alg}[label=box_closed_form,nofloat]{Topology optimization: closed-form solution method}
	{\begingroup
	{\allowdisplaybreaks
	\abovedisplayskip=-6pt
	\belowdisplayskip=-6pt
	\begin{flalign} \label{eq_box_closedform}
		&\textit{Problem\footnotemark:}  \notag\\
		&\hspace{0.25cm}\left\{
		\begin{aligned}
			&\chi^{*} =
			\underset{\chi\in{\mathscr{U}}_{ad}}
			{\operatorname{argmin}}\  {\cal J}^{(h_e)}{(\chi)}\\
			&s.t.\quad{\cal C}(\chi)\equiv t-\dfrac{|\Omega^-(\chi_\psi)|}{|\Omega|}=0;\;\; t\in[0,T]\\
			&\hspace{0.8cm}{\textstyle state \; equation}
		\end{aligned}
		\right.   &(a)\notag\\
		&\textit{Lagrangian:}\quad\notag\\
		&\hspace{0.25cm}{\cal L}(\chi,\lambda)={\cal J}^{(h_e)}(\chi)+\lambda {\cal C}(\chi)&\ (b)\notag\\
		&\textit{Optimality criterion:}   &\notag\\
		&\hspace{0.5cm}\left\{
		\begin{aligned}
			&\dfrac{\delta {\cal L}(\chi,\lambda)}{\delta\chi}\bbx=-\left(\xi\bbxchi - \lambda \right)\\
			&{\cal C}(\chi)=0
		\end{aligned}
		\right. &(c)\notag\\
		&\textit{Shifting and normalization\footnotemark:}\notag\\
		&\hspace{0.25cm}\left\{
				\begin{aligned}
					&\hat{\xi}\bbx=\dfrac{\xi\bbx-\Delta_{shift}}{\Delta_{norm}}  &\quad\forall{\bf x}\in\Omega^{+} \\
					&\hat{\xi}\bbx=\dfrac{\xi\bbx}{\Delta_{norm}} &\quad\forall{\bf x}\in\Omega^{-} \\
				\end{aligned}
				\right. &(d)\notag\\
		&\textit{Closed-form solution:}\notag\\
		&\hspace{0.25cm}\left\{\begin{aligned}
					&\psi_{\chi}({\bf x},\lambda)\coloneqq \hat{\xi}\bbxchi -\lambda\\
					&\chi({\bf x},\lambda)={\cal H}_{\beta}\left[ \psi_{\chi}({\bf x},\lambda) \right] \\
					&{\cal C}(\chi({\bf x},\lambda))=0
				\end{aligned}\right. &(e)\notag\\
		&\textit{Topology:}\notag\\
		&\hspace{0.25cm}\left\{
		\begin{aligned}
			&\Omega^{+}(\chi)\coloneqq\{\mathbf{x}\in \Omega \;/\; \psi_\chi\mathbf{(x,\lambda)}>0 \}\\
			&\Omega^{-}(\chi)\coloneqq\{\mathbf{x}\in \Omega \;/\; \psi_\chi\mathbf{(x,\lambda)}<0  \}\\
			&\Gamma(\chi)\ \ \coloneqq\{\mathbf{x}\in \Omega \;/\;  \psi_\chi\mathbf{(x,\lambda)}=0 \} 
		\end{aligned}
		\right.
	\end{flalign}}
	\endgroup}
\end{mybox_alg}
\addtocounter{footnote}{-2}
\stepcounter{footnote}\footnotetext{From now on, superscript $(\cdot)^{(h_e)}$ refers to results obtained from approximations via finite element calculations of typical mesh-size $h_e$.}
\stepcounter{footnote}\footnotetext{Shifting and normalization operations in terms of $\Delta_{shift}$ and $\Delta_{norm}$ (standing, respectively, for the minimum value and the range of $\xi$ at $t=0$) are introduced for the purposes of providing \emph{algorithmic time consistency} to the problem at $t=0$. It can be proven that those operations do not alter the problem solution.}

\section{Formulation of the state problem} \label{sec_formulation_state}

\begin{figure*}[htb]
	\centering
	\includegraphics[width=14cm, height=4.5cm]{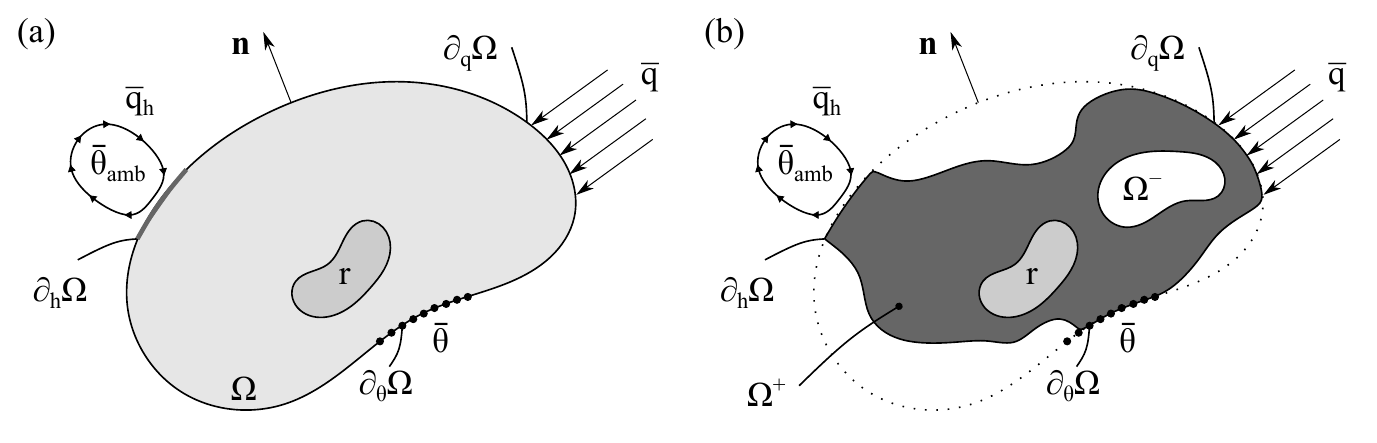}
	\caption{Thermal problem sketch: (a) fixed analysis domain $\Omega$ with boundary conditions (in which the temperature $\overline{\theta}\bbx$, the normal heat flux $\overline{q}\bbx$ or the convective heat flux $\overline{q_h}\bbx$ can be prescribed at $\partial_{\theta}\Omega$, $\partial_q\Omega$ and $\partial_h\Omega$, respectively) and (b) Hard and soft material domains, $\Omega^+$ and $\Omega^-$, respectively, with the same boundary conditions.}
	\label{fig_thermal_problem2}
\end{figure*}

In the context of the  relaxed (bi-material) approach referred to in Remark \ref{rmk_properties_interp}, both the unknowns (temperatures) and data of the optimization problem (material properties) depend on the topology layout, that is, on the characteristic function, $\chi$. Then, let $\Omega$ be the analysis domain, whose boundary $\partial\Omega$ is made of three mutually disjoint subsets, $\partial\Omega=\partial_{\theta}\Omega\cup\partial_{q}\Omega\cup\partial_{h}\Omega$, as depicted in Figure \ref{fig_thermal_problem2}, with $\partial_{\theta}\Omega$ of nonzero Lebesgue measure. Boundaries $\partial_{\theta}\Omega$, $\partial_q\Omega$ and $\partial_h\Omega$ are, respectively, those subsets of  $\partial\Omega$, where temperature, $\overline{\theta}\bbx$, heat fluxes, $\overline{q}\bbx={\bm q}\bbx\cdot\bf{n}$ and convective heat fluxes, $h \left(\theta\bbx-{\theta_{amb}}\bbx\right)={\bm q}\bbx\cdot\bf{n}$, are prescribed.

The steady-state thermal problem, for the temperature distribution $\theta\bbxchi$, states the heat energy balance in the analysis domain, $\Omega$, and it can be formulated as
\begin{equation} \label{eq_problem_thermal_definition2}
	\left[\begin{split}
		&{\text{Find }} \theta\bbxchi {\text{, such that}}\\
		&\hspace{0.00cm}\left \{
		\begin{split}
			& -\bm{\nabla}\cdot\bm{q}\bbxchi+\hS\bbxchi=0 \ &in \; \Omega\\
			&\bm{q}\bbxchi\cdot\mathbf{n}={\overline{q}}\bbx \ &on \; \partial_{q}\Omega\\
			&\theta\bbxchi=\overline{\theta}\bbx \ &on \; \partial_{\theta}\Omega\\
			&\bm{q}\bbxchi\cdot\mathbf{n} = h\ \left(\theta\bbxchi-\theta_{amb}\bbx\right) \ &on \; \partial_{h}\Omega
		\end{split}
		\right. \mcolon \hspace{-0.2cm}
	\end{split} \right.\hspace{-0.2cm}
\end{equation}
where $\bm{q}\bbxchi$ stands for the heat flux, $\hS\bbxchi$ is the heat source function and ${\overline{q}}\bbx$ stands for the prescribed heat flux on the boundaries of $\Omega$. Additionally, $h$ denotes the heat transfer coefficient, ${\theta_{amb}}\bbx$ corresponds to the ambient temperature imposed at $\partial_{h}\Omega$ and $\mathbf{n}$ defines the unit outwards normal.

The conductive material is governed by the Fourier's law, i.e. $\bm{q}\bbxchi = - \Kappalarge\bbxchi\cdot\gradT{}\bbx$, where $\Kappalarge$ stands for the symmetric second order thermal conductivity tensor and $\gradT{}\bbx$ is the thermal gradient tensor.\footnote{$\Kappalarge=\kappa\mathbf{I}$ for isotropic conductive materials.} Both, the conductivity, $\Kappalarge\bbxchi$, and the heat source, $\hS\bbxchi$, are postulated, in terms of the characteristic function, $\chi$, (see Remark \ref{rmk_properties_interp}) as follows:
\begin{empheq}[left=\empheqlbrace]{align}
	& \Kappalarge_\chi\bbx = \chi_{\kappa} ^{m_{\kappa}} \bbx \Kappalarge\bbx\; ; \quad m_{\kappa} > 1 \label{eq_conductivity_interp}\\
	& \hS_{\chi}\bbx = \chi_{_{\shS}} ^{m_{\shS}}\bbx\hS\bbx\; ; \quad \;\; m_{\shS} \geq 1\label{eq_heat_sources_interp}
\end{empheq} 
with 
\begin{empheq}[left=\empheqlbrace]{align}
	& \chi_{\kappa}\bbx={\cal H}_{\beta_\kappa}(\chi)\coloneqq\left\{
		\begin{aligned}
			&1\quad           &\text{if }\mathbf{x}\in\Omega^{+} \\
			&\beta_{\kappa}   &\text{if }\mathbf{x}\in\Omega^{-}
		\end{aligned} \right. \\
	& \chi_{\shS}\bbx={\cal H}_{\beta_\shS}(\chi)\coloneqq\left\{
		\begin{aligned}
			&1\quad           &\text{if }\mathbf{x}\in\Omega^{+} \\
			&\beta_{\shS}        &\text{if }\mathbf{x}\in\Omega^{-}
		\end{aligned} \right. \mcolon \label{eq_chi_heat_source}
\end{empheq}
where $\chi_{\kappa}$ and $\chi_{\shS}$ stand for the \emph{relaxed characteristic functions} for the thermal conductivity, $\Kappalarge$, and the heat source, $\hS$, respectively. Associated to the \emph{relaxation factor}, $\beta$, of every property, we define  the contrast factor, $\alpha$, through $\beta_{(\cdot)}=\alpha_{(\cdot)} ^{1/{m_{(\cdot)}}}\Longrightarrow\alpha_{(\cdot)}=\beta_{(\cdot)}^{m_{(\cdot)}}$. Different values of $m_{(\cdot)}$ may be required for the topology optimization procedure, depending on the material interpolation.

Alternatively, the thermal problem stated in equation (\ref{eq_problem_thermal_definition2}) can be written in variational form as
\begin{empheq}[left=\empheqlbrack,right=\hspace{-0.1cm}]{align}
	&\text{Find the temperature field  ${\pmb \theta}_\chi\in{\cal{U}}(\Omega)$ such that} \notag \\
	& \hspace{0.25cm} a(w,\theta_\chi) = l(w) \quad \forall w\in {\cal V}(\Omega) \label{eq_weak_problem2}\\
    &\text{where} \notag \\
	&\hspace{0.25cm}a(w,\theta_\chi) = \int_{\Omega}{\bm{\nabla} w\bbx\cdot\Kappalarge_\chi\bbx\cdot\bm{\nabla}\theta_\chi\bbx}\domega + \notag \\
	&\hspace{1.5cm} +\int_{\partial_{h}\Omega}{h\ w\bbx\theta_\chi\bbx}\dgamma \mcolon \label{eq_lhs_thermal_problem}\\
	&\hspace{0.25cm}l(w) = - \int_{\partial_{q}\Omega}{w\bbx{\overline{q}}\bbx}\dgamma + \notag \\
	&\hspace{1.25cm}+ \int_{\partial_{h}\Omega}{h\ w\bbx\theta_{amb}\bbx}\dgamma + \notag \\
	&\hspace{1.25cm}+ \int_{\Omega}{w\bbx \hS_\chi\bbx}\domega \mcolon \label{eq_rhs_thermal_problem}
\end{empheq}
where the set of admissible temperature fields is ${\cal{U}}(\Omega) \coloneqq \left\{\theta\bbx \; / \;  \theta\in H^1(\Omega) , \; \theta = \overline{\theta} \; on \; \partial_{\theta}\Omega \right\}$, and the space of admissible virtual temperature fields is given by ${\cal V}(\Omega)  \coloneqq \left\{w\bbx      \; / \; w      \in H^1(\Omega) , \; w = 0                 \; on \; \partial_{\theta}\Omega  \right\}$. Equations (\ref{eq_weak_problem2}) to (\ref{eq_rhs_thermal_problem}) are discretized via the Finite Element Method as shown in Appendix \ref{sec_finite_element_discretization}.

\section{Optimization algorithm} \label{sec_opt_alg}

\begin{figure*}
	\centering
	\includegraphics[width=16.5cm, height=4.5cm]{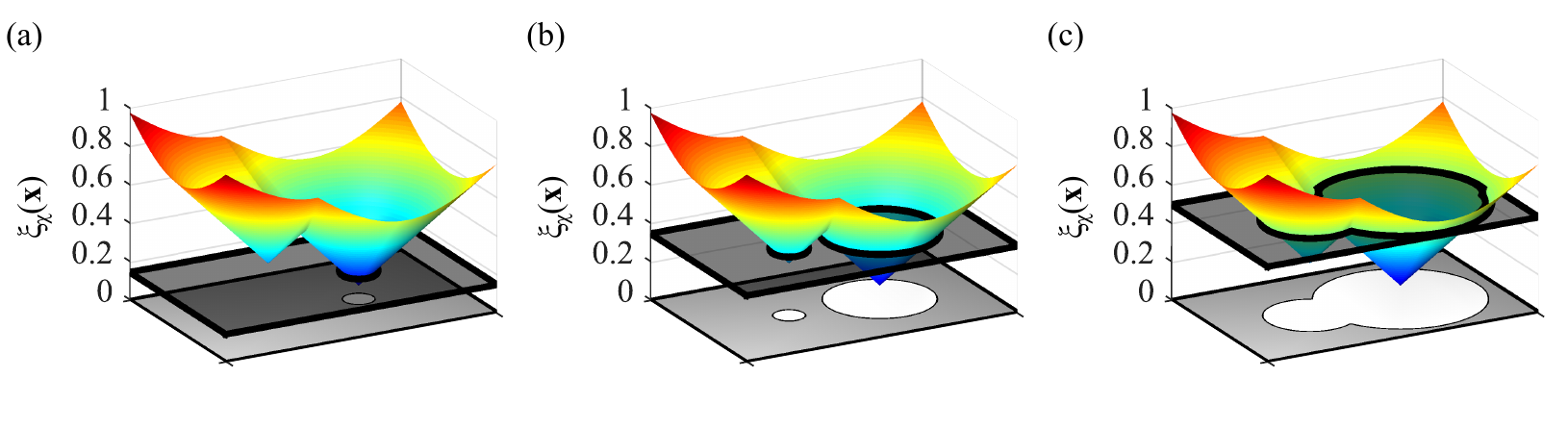}
	\caption{Cutting and bisection iterative algorithm. Visual representation for different $\lambda$: (a) cutting plane at $\lambda_1=0.15$, (b) Cutting plane at $\lambda_2=0.35$ and (c) Cutting plane at $\lambda_3=0.50$. As it can be observed, the ratio of soft domain, $\frac{|\Omega^-|}{|\Omega|}$, increases with the Lagrange multiplier. Therefore, $|\Omega^-(\lambda_1)|<|\Omega^-(\lambda_2)|<|\Omega^-(\lambda_3)|$.}
	\label{fig_bisection}
\end{figure*}
	
The algorithm to obtain the optimal characteristic function distribution, $\chi \bbx$,\footnote{{The solution $\chi$, resulting from the optimization process, must lie in the subset of admissible solutions, ${\mathscr{U}}_{ad}$, corresponding  to the tackled single-material (state) thermal problem (i.e. for $\beta\rightarrow0$). Then, the subset is defined as ${\mathscr{U}}_{ad} = \{ \chi \; / \; \Omega^+(\chi)\subset\Omega, \; 
		\partial_{\theta}\Omega\cap\partial\Omega^+\bchi\neq\emptyset, \;
		\partial_q\Omega\subset\partial\Omega^+\bchi, \;
		\partial_h\Omega\subset\partial\Omega^+\bchi \}$.}} is based on the \emph{Cutting\emph{\&}Bisection technique}, shown in Algorithm \ref{alg_bisection_method}, in the context of the \emph{pseudo-time-advancing strategy}. The strategy, described in \citet{Oliver2019}, is sketched in Algorithm \ref{alg_opt_code}. The number of time-steps of this methodology is related to the robustness and computational cost of the problem: the more time-steps, the more robust the solution is, although the computational cost of the optimization is higher. Then, it is up to the user to impose a feasible time evolution based on his/her own experience.
		
	{\setlength{\intextsep}{0pt}
				\begin{algorithm}
					\caption{Optimization algorithm} \label{alg_opt_code}
					\KwData{Given the mesh, state equation, boundary conditions and objective function}
					\KwResult{Find $\chi_n$ for $\mathcal{T}\coloneqq\{t_0,t_1,\ldots,t_n,\ldots,T\}$}
					\Begin{
					 Initialization of the design variables\;
					 \For{$n\leftarrow 1$ \KwTo $n_{steps}$}{
						 	Initialization of step n\;
						 	$i\leftarrow0$\;
							\While{Topology and Lagrange multiplier tolerances are not satisfied}{
								Solve the equilibrium equation using FEM\;
								Compute the relaxed topological sensitivity (RTD) using the adjoint method\;
								Modify the sensitivity (\emph{Shifting and normalization})\; Regularize the sensitivity by a Laplacian smoothing\;
								Compute the Lagrangian multiplier using a \emph{bisection algorithm} (algorithm \ref{alg_bisection_method})\;
								Update the discrimination function\;
								Update the characteristic function\;
								$i\leftarrow i+1$\;
							}
							$\chi_n\leftarrow$ current characteristic function\;
					 }
				}
				\end{algorithm}
				}
		
	For practical purposes, the \emph{Laplacian regularization} is applied to the \emph{pseudo-energy density}, $\xi$ (sensitivity), instead of the \emph{discrimination function}, $\psi=\xi-\lambda$ , since the regularization does not affect the (constant) Lagrange multiplier $\lambda$. In this way, it is required only once for each iteration of the algorithm \ref{alg_opt_code} (outer loop), instead of at every iteration of the \emph{Cutting}\&\emph{Bisection} algorithm \ref{alg_bisection_method} (inner loop). This minor modification translates into a significant reduction in the computational cost of the \emph{bisection algorithm}.
				
	In addition, the procedure to compute the Lagrange multiplier, imposing the constraint equation of (\ref{eq_box_closedform})-(a), is illustrated in Figure \ref{fig_bisection}. A modified Marching Cubes method, detailed in \citet{Oliver2019}, is used to numerically compute the 0-level iso-surface of the \emph{discrimination function}, $\psi$. Through this technique, the element hard-phase volume can be obtained, along with the constraint value, $\cal{C}$.

			{\setlength{\intextsep}{0pt}
			\begin{algorithm}
				\SetAlgoSkip{}
				\caption{\emph{Cutting}\&\emph{Bisection} iterative algorithm} \label{alg_bisection_method}
				\KwData{Given the mesh, the regularized energy density $\xi_\tau\bbxchi$ and the pseudo-time $t_{n}$}
				\KwResult{Find $\lambda_{n}$ such that the constraint equation is fulfilled}
				\Begin{
			 $j\leftarrow0$\;
			 \While{Volume constraint is not satisfied}{
				  		Update the Lagrangian multiplier\;
				  		Compute the corresponding discrimination function\;
				  		Compute the corresponding characteristic function\;
				  		Compute the corresponding volume constraint\;
				  		$j\leftarrow j+1$\;
					 }
				}
				\end{algorithm} }

\section{Topology optimization problems}
\label{sec_topology_optimization_problems}

\subsection{Thermal compliance problem}
\label{sec_structural_compliance_problems}
Let us now consider the \textit{maximal thermal diffusivity (minimal thermal compliance)} topology optimization problem. This goal can be achieved by minimizing the negative of the total potential energy, i.e.:
\begin{equation} \label{eq_cost_function_compliance}
	\left[
	\begin{split}
		&\underset{\chi\in{\mathscr{U}}_{ad}} {\operatorname{min}}{\cal J}(\theta_{\chi}(\mathbf{x},t))\equiv
						-\left(\frac{1}{2}a_{\chi}(\theta_{\chi},\theta_{\chi}) -l(\theta_{\chi})\right)\equiv \\
		&\hspace{2.6cm}\equiv\frac{1}{2}l(\theta_{\chi}(\mathbf{x},t))
						&(a) \\
		&\text{subject to:}  \\   
		&\hspace{0.75cm}{\mathcal C}(\chi,t)\coloneqq t-\dfrac{\vert\Omega^-\vert(\chi)}{\vert\Omega\vert}=0\;;\quad t\in[0,1]   &(b) \\
		&\text{governed by:} \\
		&\hspace{0.75cm} a(w,\theta_\chi) = l(w) \quad \forall w\in {\cal V}(\Omega) \, , \, \forall \theta_\chi\in {\cal U}(\Omega) &(c)
	\end{split} \right. \mdot
\end{equation}

This problem belongs to the class of problems considered in equation (\ref{eq_minimization_restricted}) with
\begin{equation} \label{eq_compliance_cost_function}
	\begin{split}
		{\cal J}(\theta_\chi)&\equiv \frac{1}{2} l(\theta_{\chi})=\\
			&=\frac{1}{2} \Big(\int_{\Omega} \hS\theta_{\chi}\domega-\int_{\partial_{q}\Omega}	{\overline{{q}}}\theta_{\chi}\dgamma- \\
			&\hspace{3cm}-\int_{\partial_{h}\Omega}{h\theta_{amb}}\theta_{\chi}\dgamma\Big)=\\
			&=\frac{1}{2} a_{\chi}(\theta_{\chi},\theta_\chi)\equiv\\
			&\equiv \frac{1}{2} \Big(\int_{\Omega} \gradT{} \cdot{\Kappalarge}_{\chi}\cdot\gradT{}\domega-\\
			&\hspace{3.25cm}-\int_{\partial_{h}\Omega}{h\theta_{\chi}}\theta_{\chi}\dgamma \Big)=\\
			&=\int_\Omega {\cal U_\chi}\domega - \int_{\partial_{h}\Omega}{h\theta_{\chi}}\theta_{\chi}\dgamma
	\end{split}
\end{equation}
where equations (\ref{eq_lhs_thermal_problem}) and (\ref{eq_rhs_thermal_problem}) have been considered for $w\equiv\theta_\chi$, and ${\cal U}_\chi$ can be identified as the \emph{actual thermal energy density } (${\cal U}_{\chi}=\frac{1}{2}\gradT{}\cdot{\Kappalarge}_{\chi}\cdot\gradT{}$). Comparing equations (\ref{eq_compliance_cost_function}) and (\ref{eq_minimization_restricted}), we can identify
\begin{equation} \label{eq_identification1}
	j(\chi,{\bf x})\equiv\frac{1}{2}\gradT{} \cdot{\Kappalarge}_{\chi}\bbx\cdot\gradT{}={\cal U}_\chi\bbx \mdot
\end{equation}

The corresponding finite element discretization counterpart of the problem in equation (\ref{eq_cost_function_compliance}) reads
\begin{equation}\label{eq_discrete_form}
	\left[
	\begin{split}
		&\underset{\chi\in{\mathscr{U}}_{ad}}{\operatorname{min}}\ {\cal J}^{(h_e)}(\theta_{\chi}(t))\equiv \frac{1}{2}\mathbf{f}^{T}\hat{\pmb{\theta}}_{\chi}(t)\quad&(a)  \\
		&\text{subject to:} \\
		&\hspace{0.75cm}{\mathcal C}(\chi,t)\coloneqq t-\dfrac{\vert\Omega^-\vert(\chi)}{\vert\Omega\vert}=0\;;\quad t\in[0,1]   &(b) \\
		&\text{governed by:}  \\
		&\hspace{0.75cm}{\mathbb K}_{\chi}\hat{\pmb{\theta}}_{\chi}=\mathbf{f} &(c)
	\end{split} \right. \mcolon
\end{equation}
where $h_e$ stands for the typical size of the finite element mesh, and $\mathbf{f}^{T}\hat{\pmb{\theta}}_{\chi}(t)$ denotes the thermal compliance. Bear in mind that the discretization of the state equation for the thermal problem (\ref{eq_equilibrium}) has been also considered in the previous minimization problem.

\subsubsection{Topological sensitivity of the cost function} \label{sec_sensitivity_compliance}

The \textit{adjoint method} \cite{Lions2011} for sensitivity analysis is used in this paper to compute the relaxed topological derivative (RTD) of the cost-function, ${\cal J}^{(h_e)}(\theta_{\chi})$, in equation (\ref{eq_discrete_form})-(a), without explicitly computing the sensitivity of the nodal temperature field ($\partial {\theta}_{\chi} \slash \partial \chi$). 

Let $\overline{\cal J}^{(h_e)}(\chi)$ be the extended cost function of ${\cal J}^{(h_e)}(\chi)$ defined as
\begin{equation} \label{eq_rephrased}
	\overline{\cal J}^{(h_e)}(\chi)=\frac{1}{2} \mathbf{f}^{T}\hat{\pmb{\theta}}_{\chi}-
            \hat{\mathbf{w}}^T \left( {\mathbb K}_{\chi}\hat{\pmb{\theta}}_{\chi}-\mathbf{f}\right) \mcolon
\end{equation}
where $\hat{\mathbf{w}}$ stands for the solution of the adjoint state problem. Then, the sensitivity of the cost function results, after using the RTD, in the following
\begin{equation}\label{eq_diferentiation}
	\begin{split}
		&\dfrac{\delta\overline{\cal J}^{(h_e)}(\chi)}{\delta\chi}\bbxhat=	\left(\frac{1}{2}\mathbf{f}^{T}-\hat{\mathbf{w}}^T{\mathbb K}_{\chi}\right)\dfrac{\delta\hat{\pmb{\theta}}_{\chi}}{\delta{\chi}}\bbxhat+\\
		&\hspace{1cm}+\left(\frac{1}{2}\dfrac{\delta\mathbf{f}^T _{\chi}}{\delta{\chi}}\bbxhat\hat{\pmb{\theta}}_{\chi}-\hat{\mathbf{w}}^T\dfrac{\delta {\mathbb K}_{\chi}}{\delta\chi}\bbxhat\hat{\pmb{\theta}}_{\chi} + \hat{\mathbf{w}}^T\dfrac{\delta\mathbf{f}_{\chi}}{\delta{\chi}}\bbxhat\right) \mdot
	\end{split}
\end{equation}
After some algebraic manipulation, accounting for the adjoint state equation, one arrives to
\begin{equation} \label{eq_top_derivative_compliance}
	\dfrac{\delta\overline{\cal J}^{(h_e)}(\chi)}{\delta\chi}\bbxhat = \left[\dfrac{\delta\mathbf{f}^T _{\chi}}{\delta{\chi}}\bbx\hat{\pmb{\theta}}_{\chi}- \hat{\pmb{\theta}}_{\chi}^T\dfrac{\delta {\mathbb K}_{\chi}}{\delta\chi}\bbx\hat{\pmb{\theta}}_{\chi}\right]_{\mathbf{x}=\hat{\mathbf{x}}} \mdot
\end{equation}

Finally, equation (\ref{eq_top_derivative_compliance}) is discretized using the FEM expressions of equations (\ref{eq_rephrased0})-(\ref{eq_final_derivative0}), as detailed in Appendix \ref{App_thermal_complaince}, as
\begin{equation} \label{eq_compliance_topological_derivative}
	\begin{split}
		\dfrac{\delta{\overline{\cal J}^{(h_e)}(\theta_{\chi})}}{\delta\chi}\bbxhat&=m_{\shS}\left({\chi}_{\shS}\bbxhat\right) ^{m_{\shS}-1}{\overline{\cal U}_{ \shS}}\bbxhat\Delta\chi_{_{\shS}}\bbxhat-\\
		&-2m_{\kappa} \left(\chi_{\kappa}\bbxhat\right) ^{m_{\kappa}-1}{\overline{\cal U}}\bbxhat{{\Delta\chi}_{\kappa}({\hat{\bf x}}) } \mcolon
	\end{split}
\end{equation}
where $\overline{\cal U}\bbxhat$ is \textit{\Uconductivity} and ${\overline{\cal U}_{ \shS}}\bbxhat$ is \textit{\Usource}, which are respectively written as 
\begin{equation} \label{eq_compliance_energies}
	\left\{
	\begin{split}
		&\overline{\cal U}\bbxhat=\dfrac{1}{2}\left( \gradT{} \cdot{\Kappalarge}\cdot \gradT{}\right)\bbxhat\quad&(a) \\
		&{\overline{\cal U}_{ \shS}}\bbxhat = \left(\hS\theta_{\chi}\right)\bbxhat \quad&(b) 
	\end{split} \right. \mdot
\end{equation}

\subsubsection{Closed-form solution} \label{sec_closedform_compliance}
In Box \ref{box_thermal_diffusivity}, the \emph{pseudo-energy density}, $\xi \bbxchi$, to be considered for the closed-form solution in Box \ref{box_closed_form}, is presented.

\begin{mybox_alg}[label=box_thermal_diffusivity,nofloat,text width=0.44\textwidth]{Topology optimization of thermal compliance problems}
	{\begingroup
	{\allowdisplaybreaks
	\abovedisplayskip=-6pt
	\belowdisplayskip=0pt
	\begin{flalign} \label{eq_compliance}
		&\textit{Problem:}  \notag\\
		&\hspace{0.25cm}\left[
		\begin{aligned}
			&\chi^{*} =
			\underset{\chi\in{\mathscr{U}}_{ad}}
			{\operatorname{argmin}}\  {\cal J}^{(h_e)}{(\chi)}\coloneqq {\bf f}^T{\hat{\pmb{\theta}}}_{\chi}\\
			&s.t.\quad{\mathcal C}(\chi,t)\coloneqq t-\dfrac{\vert\Omega^-\vert(\chi)}{\vert\Omega\vert}=0\;;\quad t\in[0,1] \\
			&\hspace{0.8cm}{\mathbb K}_{\chi}\hat{\pmb{\theta}}_{\chi}=\mathbf{f}
		\end{aligned}
		\right.   &(a)\notag\\
		&\textit{Energy density:}\quad\notag\\
		&\hspace{0.5cm}\xi\bbxchi = \gamma_1 \bbxchi \, {\overline{\cal U}}\bbx
								   -\gamma_2 \bbxchi \, {\overline{\cal U}_{ \shS}}\bbx &(b) \notag\\
		&\hspace{0.25cm}\text{where}\notag\\
		&\hspace{0.5cm}\left\{
		\begin{aligned}
			&\overline{\cal U}\bbxhat=\dfrac{1}{2}\left(\gradT{} \cdot{\Kappalarge}\cdot \gradT{}\right)\bbxhat\ge 0; \\
			&{\overline{\cal U}_{ \shS}}\bbxhat = \left(\hS\theta_{\chi}\right)\bbxhat \\
			&\gamma_1 = 2m_{\kappa} \left(\chi_{\kappa}\bbx\right) ^{m_{\kappa}-1} {{\Delta\chi}_{\kappa}\bbx } \\
			&\gamma_2 = m_{\shS}\left({\chi}_{\shS}\bbx\right) ^{m_{\shS}-1} \Delta\chi_{_{\shS}}\bbx 
		\end{aligned} \right. 
	\end{flalign}}
	\endgroup}
\end{mybox_alg}

\subsection{Thermal cloaking in terms of heat flux}
\label{sec_thermal_cloaking}

\begin{figure*}[pt]
	\centering
	\includegraphics[width=16.5cm, height=6cm]{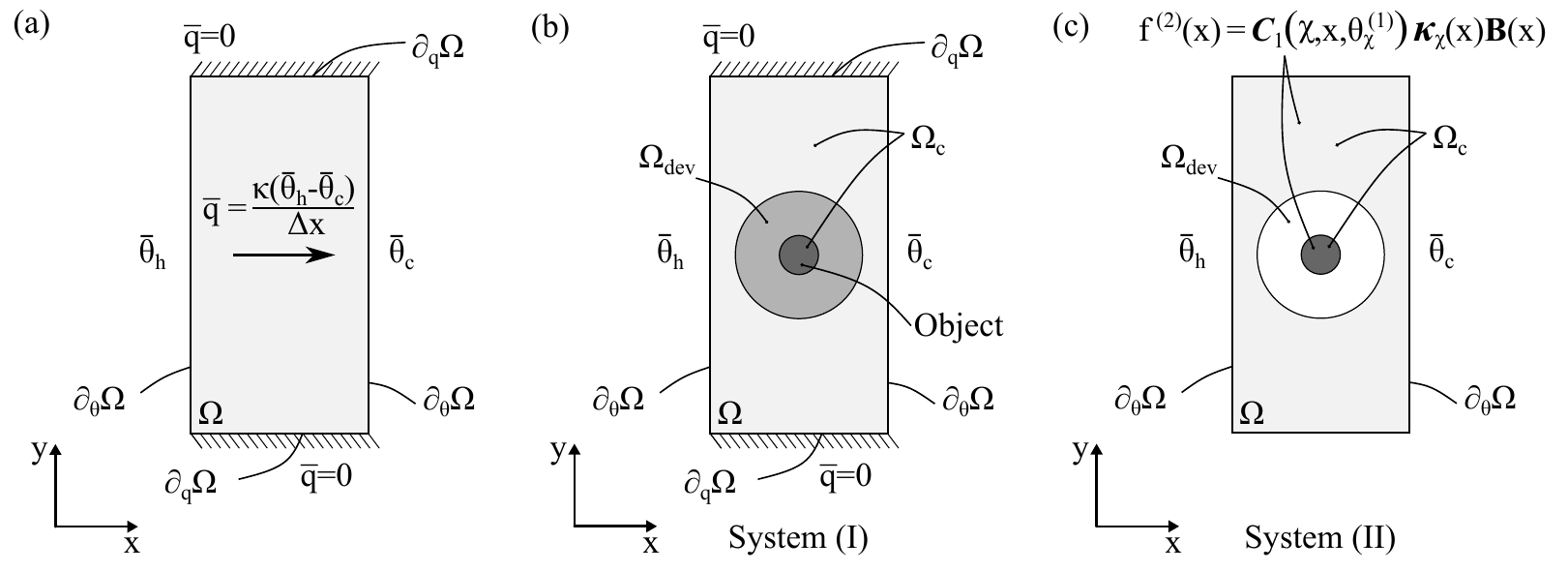}
	\caption{Thermal cloaking problem: (a) homogeneous problem setting where a constant uniform heat flux over all the domain $\Omega$ is observed, (b) topology optimization domain with boundary conditions of system (I), and (c) topology optimization domain with boundary conditions of system (II). The objective is to minimize the perturbation of an object placed in the center of the domain $\Omega$. For that reason, it is surrounded by a cloaking device, in dark gray, which must be optimized.}
	\label{fig_flux_cloaking}
\end{figure*}

We now consider an object whose thermal properties may differ from the properties of the surrounding material $\Omega$. Then, the main objective is to thermally cloak the object, colored in black (see Figure \ref{fig_flux_cloaking}), from being detected by an external thermal detecting device, measuring the deviation between the constant heat flux, theoretically observed on the 3D homogeneous domain $\Omega$, and the actual flux in the non-homogeneous domain containing the cloaked object. Under the assumption that there is no body that alters the flux, the heat flux entering across the left face of $\Omega$ should be constant and equal to that exiting across the right face. In addition, the unperturbed domain presents a known homogeneous heat flux field. Thus, the goal of this topology optimization problem is to find the optimal topology of the surrounding cloaking device, $\Omega_{dev}$, displayed in dark gray, that mitigates the perturbation of the object in the heat flux field so as to resemble the original homogeneous heat-flux.

The problem setting is illustrated in Figure \ref{fig_flux_cloaking}, in which the constant given heat flux is prescribed via the equivalent Dirichlet conditions on both vertical sides, i.e. the temperature is prescribed to a high value, $\overline{\theta}_h$, and a low value, $\overline{\theta}_c$, at the left and right sides, respectively (see Figure \ref{fig_flux_cloaking}-(a)). Adiabatic conditions are assumed on the other two boundaries. Figure \ref{fig_flux_cloaking}-(b) depicts the setting and boundary conditions when the object to be hidden is placed inside the analysis domain, $\Omega$.
The corresponding topology optimization problem is written as the minimization of the deviation (measured through a L2-norm) between the constant heat flux and the actual heat flux in domain $\Omega_{c}\equiv \Omega\setminus\Omega_{dev}$, which reads as
\begingroup
\allowdisplaybreaks
\begin{align} \label{eq_flux_cloaking}
	\left[
	\begin{aligned}
		&\min_{\chi\in{\mathscr{U}}_{ad}}{\cal J}({\theta}_\chi({\bf x},t))
		=\left\| {\bf q}_{\chi}({\bf x},{\theta}_\chi)-{\overline{\bf q}}\bbx \right\| _{L_2 (\Omega_{c})}= \\
		&\hspace{2cm}=\left(\int_{\Omega_{c}}\left|{\bf q}_{\chi}({\bf x},{\theta}_\chi)-{\overline{\bf q}}\bbx\right|^2 \;d\Omega\right)^\frac{1}{2} \;&(a)\\
		&\text{subject to}:\quad\\
		&\hspace{0.75cm}{\mathcal C}(\chi,t)\coloneqq t-\dfrac{\vert\Omega^-\vert(\chi)}{\vert\Omega\vert}=0\;;\quad t\in[0,1] \;&(b) \\
		&\text{governed by:} \\
		&\hspace{0.75cm} a(w,\theta_\chi) = l(w) \quad \forall w\in {\cal V}(\Omega) \, , \, \forall \theta_\chi\in {\cal U}(\Omega) &(c)
	\end{aligned} \right.
\end{align}
\endgroup
where, in equation (\ref{eq_flux_cloaking})-(a), ${\bf q}_{\chi}({\bf x},{\theta}_\chi)$ stands for the heat flux vector, which depends on the topology, whereas ${\overline{\bf q}}\bbx$ corresponds to the prescribed (original) heat flux at the same point.

This problem belongs to the class of problems with the functional considered in equation (\ref{eq_minimization_family}), which can be generalized as
\begin{equation} \label{eq_minimization_family2}
	\mathcal{F}_\chi\equiv\left(\int_{\Omega}{f(\chi,\mathbf{x})\domega}\right)^p
\end{equation}
where $p>0$ stands for an exponential factor. Then, the relaxed topological derivative (RTD) proposed in equation (\ref{eq_RTD_general}) can be rewritten as
\begin{equation} \label{eq_RTD_general2}
	\dfrac{\delta {\cal F}(\chi)}{\delta \chi}\bbxhat=p \ {\cal F}(\chi)^{p-1}
	\left[\dfrac{\partial f({\chi},{\bf x})}{\partial {\chi}}\right]_{{\bf x}=\hat{\bf x}}\Delta \chi\bbxhat \mdot
\end{equation}
Therefore, the functional (\ref{eq_flux_cloaking})-(a) is related to (\ref{eq_minimization_family2}) by
\begin{equation} \label{eq_cost_cloaking}
	\begin{split}
		{\cal J}({\theta}_\chi)&\equiv\left(\int_{\Omega_{c}}\left|{\bf q}_{\chi}({\bf x},{\theta}_\chi)-{\overline{\bf q}}\bbx\right|^2 \;d\Omega\right)^\frac{1}{2}= \\
		&=\left(\int_{\Omega}{1_{\Omega_c}\bbx}\left| -{\Kappalarge}_{\chi}\bbx\cdot{\bm{\nabla}\theta}_{\chi}\bbx - {\overline{\bf q}}\bbx\right|^2 \;d\Omega\right)^\frac{1}{2}
	\end{split}	
\end{equation}
with $p=1/2$. Comparing equations (\ref{eq_cost_cloaking}), (\ref{eq_minimization_family2}) and (\ref{eq_minimization_restricted}) we can readily identify
\begin{equation} \label{eq_identification2}
	j(\chi,{\bf x})\equiv{1_{\Omega_c}\bbx}\left|{\bf q}_{\chi}({\bf x},{\theta}_\chi)-{\overline{\bf q}}\bbx\right|^2 \quad \forall{\bf x}\in\Omega \mcolon
\end{equation}
with ${1_{\Omega_c}\bbx}:\Omega\rightarrow\{0,1\}$ being the indicator function of the subdomain $\Omega_c\subset\Omega$, which is equal to 1 for any point contained in $\Omega_c$, and 0 for any point outside the subdomain $\Omega_c$. 

Let us now discretize the cost function, ${\cal J}(\theta_{\chi}(t))$, using the FEM expressions defined in Appendix \ref{sec_finite_element_discretization}, which yields to
\begin{equation}\label{eq_discrete_form2}
	\begin{split}
		&\underset{\chi\in{\mathscr{U}}_{ad}}{\operatorname{min}}\ {\cal J}^{(h_e)}(\theta_{\chi}(t))\equiv \\
		&\hspace{0.25cm}\equiv\left(\int_{\Omega}\resizebox{0.6\hsize}{!}{%
		${1_{\Omega_c}\bbx}\left| -{\Kappalarge}_{\chi}\bbx \mathbf{B}\bbx\hat{\pmb{\theta}}_\chi^{(1)} - {\overline{\bf q}}\bbx\right|^2 \;d\Omega$}\right)^\frac{1}{2} \mcolon 
	\end{split}
\end{equation}
where the constraint equation and the state equation are identical to those shown in equation (\ref{eq_flux_cloaking})-(b-c).

\subsubsection{Topological sensitivity of the cost function} \label{sec_top_sensitivity_flux}

Mimicking the procedure described in Section \ref{sec_sensitivity_compliance}, we include the discretized version of the state equation (\ref{eq_flux_cloaking})-(c) into the discretized cost function (\ref{eq_discrete_form2}), in order to express the extended cost function, $\overline{{\cal J}}^{(h_e)}(\chi)$, as 
\begin{equation}
	\begin{split}
		\overline{{\cal J}}^{(h_e)}(\chi)=&\left(\int_{\Omega}\resizebox{0.55\hsize}{!}{%
		${1_{\Omega_c}\bbx}\left| -{\Kappalarge}_{\chi}\bbx \mathbf{B}\bbx\hat{\pmb{\theta}}_\chi^{(1)} - {\overline{\bf q}}\bbx\right|^2 \;d\Omega$}\right)^\frac{1}{2} - \\
		 &- \hat{\bf w}^T \left( {\mathbb K}_{\chi}\hat{\pmb{\theta}}_{\chi}^{(1)}-{\mathbf{f}}^{(1)}\right) \mcolon
	\end{split}
\end{equation}
where $\hat{\mathbf{w}}$ is the solution of the adjoint state problem. Once the extended cost function is defined, we proceed to derive it using the Relaxed Topological Derivative as
\begin{equation}\label{eq_diferentiation_1}
	\begin{split}
		\dfrac{\delta\overline{\cal J}^{(h_e)}(\chi)}{\delta\chi}\bbxhat= &
			-\left(\hat{\mathbf{w}}^T{\mathbb K}_{\chi}+\bm{C}_1{\Kappalarge}_{\chi}\nabla\right)\dfrac{\delta{\hat{\pmb{\theta}}}_{\chi}^{(1)}}{\delta{\chi}}\bbxhat- \\
			& -	\bm{C}_1\dfrac{\delta {\Kappalarge}_{\chi}(\chi)}{\delta\chi}\bbxhat{\bm{\nabla}\theta}^{(1)}_{\chi}\bbxhat -\\
			& -\hat{\mathbf{w}}^T\dfrac{\delta {\mathbb K}_{\chi}}{\delta\chi}\bbxhat\hat{\pmb{\theta}}_{\chi}^{(1)}
			+\hat{\mathbf{w}}^T\dfrac{\delta\mathbf{f}_{\chi}^{(1)}}{\delta{\chi}}\bbxhat
	\end{split}
\end{equation}
where $\bm{C}_1\left(\chi,\bxhat,{\theta}_\chi^{(1)}\right)$ is 
\begin{equation}
	\bm{C}_1\left(\chi,\bxhat,{\theta}_\chi^{(1)}\right)=\dfrac{{1_{\Omega_c}\bbxhat} \ \left({\bf q}_{\chi}\left(\bxhat,{\theta}^{(1)}_\chi\right)-{\overline{\bf q}}\bbxhat\right)}{{\cal J}^{(h_e)}(\chi)} \mdot
\end{equation}

We must now solve the \emph{adjoint state problem} of equation (\ref{eq_diferentiation_1}) for $\hat{\mathbf{w}}=\hat{\pmb{\theta}}_{\chi}^{(2)}$. Thus, in contrast to the first optimization problem, that has been shown in Section \ref{sec_structural_compliance_problems}, the \textit{original thermal system (I)} has to be supplemented with an \textit{auxiliary thermal system (II)} (see Figure \ref{fig_flux_cloaking}). Both systems are governed by the thermal problem (equation (\ref{eq_equilibrium})) with the same stiffness matrix ${\mathbb K}_{\chi}$ but different actions and solutions $\hat{\pmb{{\theta}}}_{\chi}^{(1)}$ and $\hat{\pmb{{\theta}}}_{\chi}^{(2)}$, respectively, defined as
\begin{equation} 
	\left\{
	\begin{split}
		&{\mathbb K}_{\chi}\ \hat{\pmb{{\theta}}}_{\chi}^{(1)}={\bf f}^{(1)}\quad& &(\text{system  I})\\
		&{\mathbb K}_{\chi}\ \hat{\pmb{{\theta}}}_{\chi}^{(2)}={\bf f}^{(2)}\quad& &(\text{system II})\\
	\end{split}
	\right.
\end{equation}
where 
\begin{equation} \label{eq_flux_cloaking_system2}
	\begin{split}
		{\bf f}^{(2)} &= -\int_{\Omega}\mathbf{N}^T\left({\bf{x}}\right)\dfrac{\delta {\cal J}^{(h_e)}({\theta}_\chi^{(1)})}{\delta{\theta}_{\chi}}({\bf{x}})\domega = \\ 
			& = -\int_{\Omega}\mathbf{N}^T({\bf{x}})\bm{C}_1\left(\chi,{\bf x},{\theta}_\chi^{(1)}\right){\Kappalarge}_{\chi}\bbx{\bf B}\bbx \domega \mdot
	\end{split}
\end{equation}

By simplifying the first term of equation (\ref{eq_diferentiation_1}), and after some algebraic manipulations, detailed in Appendix \ref{App_flux_cloaking}, the relaxed topological sensitivity of the cost function can be expressed as a sum of \emph{energy densities}, i.e.
\begin{equation} \label{eq_sensitivity_cloaking_flux}
	\begin{split}
		\dfrac{\delta\overline{\cal J}^{(h_e)}(\chi)}{\delta\chi}\bbxhat = &
			+2\gamma_1 \bbxchi{\overline{\cal U}}_{1-2}\bbxhat -\\
			&-\gamma_2 \bbxchi {\overline{\cal U}_{ \shS}}\bbxhat
			+\gamma_1 \bbxchi{\overline{\cal U}}_{\bf q}\bbxhat \mcolon
	\end{split}
\end{equation}
where $\overline{\cal U}_{1-2}\bbxhat$, ${\overline{\cal U}_{ \shS}}\bbxhat$ and ${\overline{\cal U}}_{\bf q}\bbxhat$ are, respectively, \textit{\Uconductivity}, \textit{\Usource} and \textit{\Uflux}, which are given by
\begin{equation} \label{eq_energies_cloaking}
	\left\{\begin{split}
		&\overline{\cal U}_{1-2}\bbxhat=\dfrac{1}{2}\left(\gradT{1}\cdot{\Kappalarge}\cdot\gradT{2}\right)\bbxhat\quad&(a) \\
		&{\overline{\cal U}_{ \shS}}\bbxhat = \left(\hS\theta_{\chi}^{(2)}\right)\bbxhat \quad&(b) \\
		&{\overline{\cal U}}_{\bf q}\bbxhat = \left(\bm{C}_1{\Kappalarge}\cdot\gradT{1}\right)\bbxhat \quad&(c) 
	\end{split} \right.
\end{equation}
and
\begin{equation}
	\left\{\begin{split}
		&\gamma_1 \bbxchi = (1-\beta_{\kappa})m_{\kappa} \left(\chi_{\kappa}\bbx\right) ^{m_{\kappa}-1} \\
		&\gamma_2 \bbxchi = (1-\beta_{\shS})m_{\shS}\left({\chi}_{\shS}\bbx\right) ^{m_{\shS}-1}
	\end{split} \right. \mdot
\end{equation}

\subsubsection{Closed-form solution}
The \emph{problem-dependent energy density}, $\xi\bbxchi$, of the original functional ${\cal J}^{(h_e)}$ (equation (\ref{eq_discrete_form2})) is illustrated in Box \ref{box_flux_cloaking}, analogously to Box \ref{box_thermal_diffusivity}.

\begin{mybox_alg}[label=box_flux_cloaking,nofloat,text width=0.45 \textwidth]{Topology optimization of heat flux cloaking}
	\noindent{
	\begingroup
	\allowdisplaybreaks
	\abovedisplayskip=-6pt
	\belowdisplayskip=0pt	
	\begin{flalign} \label{eq_compliance_flux_cloaking}
		&\textit{Problem:}  \notag\\
		&\hspace{0.25cm}\left[
		\begin{aligned}
			&\chi^*=\underset{\chi\in{\mathscr{U}}_{ad}}
						{\operatorname{argmin}} \ {\cal J}^{(h_e)}({\chi})= \\
			&\hspace{0.4cm}=\left(\int_{\Omega}\resizebox{0.6\hsize}{!}{%
						${1_{\Omega_c}\bbx}\left| -{\Kappalarge}_{\chi}\bbx \mathbf{B}\bbx\hat{\pmb{\theta}}_\chi^{(1)} - {\overline{\bf q}}\bbx\right|^2 \;d\Omega$}\right)^\frac{1}{2} \\
			&s.t.\quad{\mathcal C}(\chi,t)\coloneqq t-\dfrac{\vert\Omega^-\vert(\chi)}{\vert\Omega\vert}=0\;;\quad t\in[0,1] \\
			&\hspace{0.8cm}{\mathbb K}_{\chi}\hat{\pmb{\theta}}_{\chi}^{(i)}=\mathbf{f}^{(i)} \ ; \quad  i=\{1,2\}
		\end{aligned}
		\right.   &\ (a)\notag\\
		&\textit{Energy density:} \notag\\
		&\hspace{0.25cm}\xi\bbxchi = \gamma_1\bbxchi \ \left(2{\overline{\cal U}}_{1-2}\bbx + {\overline{\cal U}_{\mathbf{q}}}\bbx \right)- \notag\\ 
		&\hspace{1.6cm} -\gamma_2 \bbxchi \ {\overline{\cal U}_{ \shS}}\bbx &\ (b)\notag\\
		&\hspace{0.25cm}\text{where} \notag\\[-5pt]
		&\hspace{0.5cm}\left\{\begin{aligned}
				&\overline{\cal U}_{1-2}\bbxhat=\dfrac{1}{2}\left(\gradT{1}\cdot{\Kappalarge}\cdot\gradT{2}\right)\bbxhat; \\
				&{\overline{\cal U}_{\shS}}\bbxhat = \left(\hS\theta_{\chi}^{(2)}\right)\bbxhat; \\
				&{\overline{\cal U}_{\mathbf{q}}}\bbxhat = \resizebox{0.65\hsize}{!}{%
							$ \left(\frac{{1_{\Omega_c}\bbx}\left({\bf q}_{\chi}\left({\theta}_\chi^{(1)}\right)-{\overline{\bf q}}\right){\Kappalarge}\cdot\gradT{1}}{{\cal J}^{(h_e)}\left(\chi,{\theta}_\chi^{(1)}\right)}\right)\bbxhat$ } \\
				&\gamma_1 \bbxchi = (1-\beta_{\kappa})m_{\kappa} \left(\chi_{\kappa}\bbx\right) ^{m_{\kappa}-1} \\
				&\gamma_2 \bbxchi = (1-\beta_{\shS})m_{\shS}\left({\chi}_{\shS}\bbx\right) ^{m_{\shS}-1} 
		\end{aligned}\right. 
	\end{flalign}
	\endgroup
	}
\end{mybox_alg}

\subsection{Thermal cloaking in terms of temperature average and variance} \label{sec_temp_var_min}

Let us now consider a hot object whose temperature is higher than the environment temperature, $\theta_{amb}$. The goal is to cloak the object for an external thermal detecting device, located at some distance from it (like a thermal camera). The cloaked object might be then easily detected if the temperature along a virtual plane, between the object and the observer, changes significantly with respect to the ambient temperature. Thus, the goal is to find the optimal layout of a surrounding cloaking device, which minimizes the perturbation of the temperature on this plane.

\begin{figure*}
	\centering
	\includegraphics[width=13cm, height=6cm]{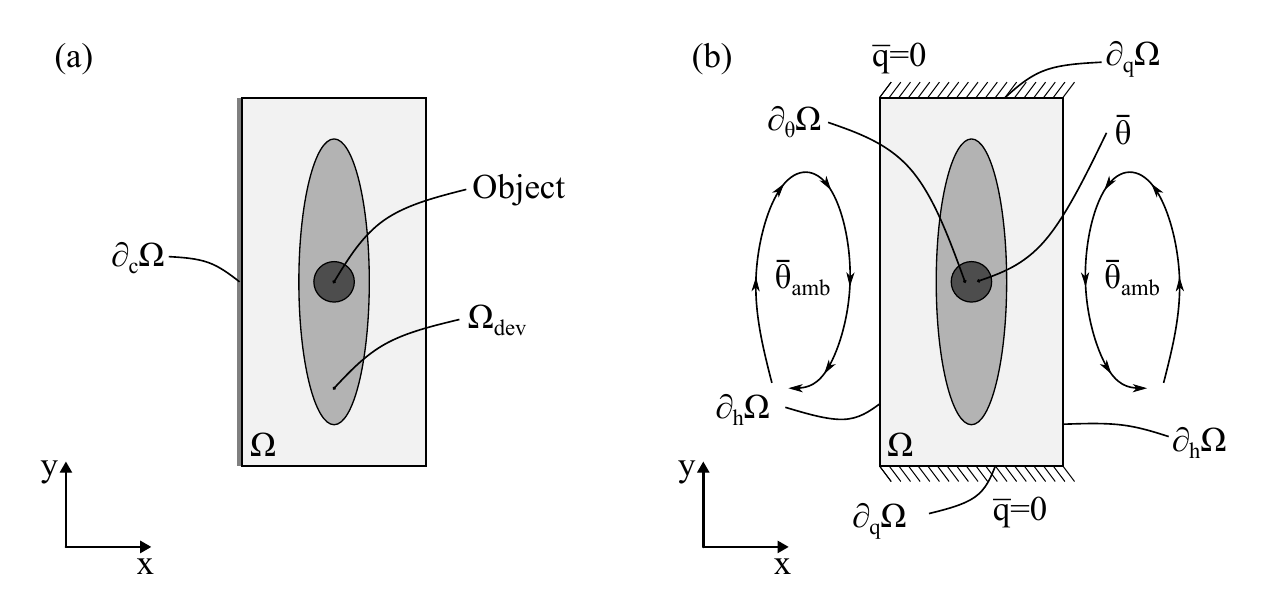}
	\caption{Average and variance temperature minimization: (a) representation of the subdomains surrounding the object to be cloaked (the cloaking device, $\Omega_{dev}$, is displayed in dark gray, while the left edge, where the average value and the variance of the temperature are minimized, is denoted by $\partial_c\Omega$) and (b) problem setting with boundary conditions. The domain, $\Omega$, corresponds to the control volume in which optimization will be carried out, which includes the object prescribed at a high temperature, $\overline{\theta}$. The left and right sides are subject to convective boundary conditions, while adiabatic conditions are assumed on top and bottom sides of the domain.}
	\label{fig_temp_cloaking}
\end{figure*}

The setting of the problem is sketched in Figure \ref{fig_temp_cloaking}, in which $\Omega$ represents the region of concern, the small black region, placed at the center, represents the object to be cloaked, and the surrounding ellipsoid, colored in gray, corresponds to the cloaking device, $\Omega_{dev}$. In addition, the vertical left edge, referred as the \textit{cloaking port}, $\partial_c\Omega$, illustrates the plane where the temperatures are measured by the observer. The temperature of the object is prescribed at a high temperature $\overline{\theta}>{\theta}_{amb}$ on its surface, $\partial_{\theta}\Omega$, and natural convective boundary conditions are applied on the left and right edges, $\partial_h\Omega$. On the other two faces, adiabatic conditions are considered.

The optimal topology will be achieved with a \emph{multi-objective optimization} via two cost functionals. The first functional addresses the minimization of the average temperature on the cloaking port, $\partial_c\Omega$, while the second is responsible of minimizing the variance of the temperature on the same face, ensuring an homogeneous temperature on the left edge.
The topological optimization problem, evaluated via a \emph{weighted sum} of the functionals, is expressed as
\begin{equation} \label{eq_cloaking_minimi_obj}
	\left[ 
	\begin{split}
		&\min_{\chi\in{\mathscr{U}}_{ad}}{\cal J}(\theta_\chi)=
		\omega\,{\cal J}_{\text{av}}(\theta_\chi) + (1-\omega)\,{\cal J}_{\text{vr}}(\theta_\chi) &(a)\\
		&\text{subject to}:\\
		&\hspace{0.75cm}{\mathcal C}(\chi,t)\coloneqq t-\dfrac{\vert\Omega^-\vert(\chi)}{\vert\Omega\vert}=0;\; t\in[0,1] &(b)\\
		&\text{governed by}:\\
		&\hspace{0.75cm} a(w,\theta_\chi) = l(w) \quad \forall w\in {\cal V}(\Omega) \, , \, \forall \theta_\chi\in {\cal U}(\Omega) &(c)
	\end{split}\right.
\end{equation}
where ${\cal J}_{\text{av}}(\theta_\chi)$ corresponds to the objective function of the \emph{average temperature minimization}, while ${\cal J}_{\text{vr}}(\theta_\chi)$ corresponds to the objective function of the \emph{temperature variance minimization}. The coefficient $\omega$ represents the weight between these two objective functions. Therefore, we are simultaneously optimizing, for a given weighting coefficient $\omega$, both functionals and achieving an \emph{optimal trade-off} from these objective functions. If this weight is changed, a different optimal solution will be obtained. Thus, given a set of weight values, the optimal solutions of each optimization problem define the classical Pareto front \cite{ATHAN1996}.

According to \citet{Marler2004}, a convenient transformation of the original objective functions is through its ranges. This \textit{normalization} is given as follows
\begin{equation} \label{eq_normalization_multiobj}
	{\widetilde{\mathcal J}}_i(\chi)=\dfrac{{{\mathcal J}}_i(\chi)-{{\mathcal J}}_i^\circ}{{{\mathcal J}}_i^{max}-{{\mathcal J}}_i^\circ} \quad \text{for } i=\{\text{av,vr}\}
\end{equation}
where ${\widetilde{\mathcal J}}_i(\chi)$ represents the transformed objective function, ${{\mathcal J}}_i^\circ$ denotes the utopia point\footnote{The utopia point ${{\mathcal J}}_i^\circ$ defined as ${{\mathcal J}}_i^\circ = \min_\chi {{\mathcal J}}_i(\chi) \quad \forall \chi \in {\mathscr U}_{ad}$ is an unattainable optimal point and it may be prohibitively expensive to compute. In these cases, an approximation is used.} and ${{\mathcal J}}_i^{max}$ corresponds to the maximum objective function value.\footnote{The maximum objective function value corresponds either to the maximum value that minimizes the other objective functions, ${{\mathcal J}}_i^{max}=\max_j {{\mathcal J}}_i(\chi_j^*) \quad j\neq i$, or the absolute maximum of ${{\mathcal J}}_i(\chi)$.} This \textit{normalization} yields non-dimensional objective functions values between zero and one. We have chosen to normalize the functionals with respect to the minimum value when minimizing only each objective functional ${{\mathcal J}}_i(\chi)$ (Utopia point) and the maximum value obtained from the minimization of the other functional ${{\mathcal J}}_i(\chi_j^*)$. Therefore, two extra optimization problems must be done for $\omega=1$ and $\omega=0$. From the first problem, ${\mathcal J}_{av}^\circ$ and ${{\mathcal J}}_{vr}^{max}$ are obtained, and from the second, ${{\mathcal J}}_{av}^{max}$ and ${\mathcal J}_{vr}^\circ$.

According to this scalarization approach, the transformed optimization problem is written as follows
\begin{equation} \label{eq_cloaking_minimi_obj2}
	\left[
	\begin{split}
		&\min_{\chi\in{\mathscr{U}}_{ad}}\widetilde{\cal J}(\theta_\chi)=
		 \omega\,\dfrac{{\cal J}_{\text{av}}(\theta_\chi)-{\mathcal J}_{av}^\circ}{{{\mathcal J}}_{av}^{max}-{\mathcal J}_{av}^\circ} + \\
		 &\hspace{2.25cm}+(1-\omega)\,\dfrac{{\cal J}_{\text{vr}}(\theta_\chi)-{\mathcal J}_{vr}^\circ}{{{\mathcal J}}_{vr}^{max}-{\mathcal J}_{vr}^\circ} &(a)\\
		&\text{subject to}:\\
		&\hspace{0.75cm}{\mathcal C}(\chi,t)\coloneqq t-\dfrac{\vert\Omega^-\vert(\chi)}{\vert\Omega\vert}=0;\; t\in[0,1] &(b)\\
		&\text{governed by}:\\
		&\hspace{0.75cm} a(w,\theta_\chi) = l(w) \quad \forall w\in {\cal V}(\Omega) \, , \, \forall \theta_\chi\in {\cal U}(\Omega) &(c)
	\end{split} \right.
\end{equation}

Thanks to the use of a multi-objective scheme, the topological sensitivity of both terms may be computed independently, as it will be shown below.

\subsubsection{Average temperature minimization} \label{sec_temperature_min}

\begin{figure*}[pb]
	\centering
	\includegraphics[width=14cm, height=7.3cm]{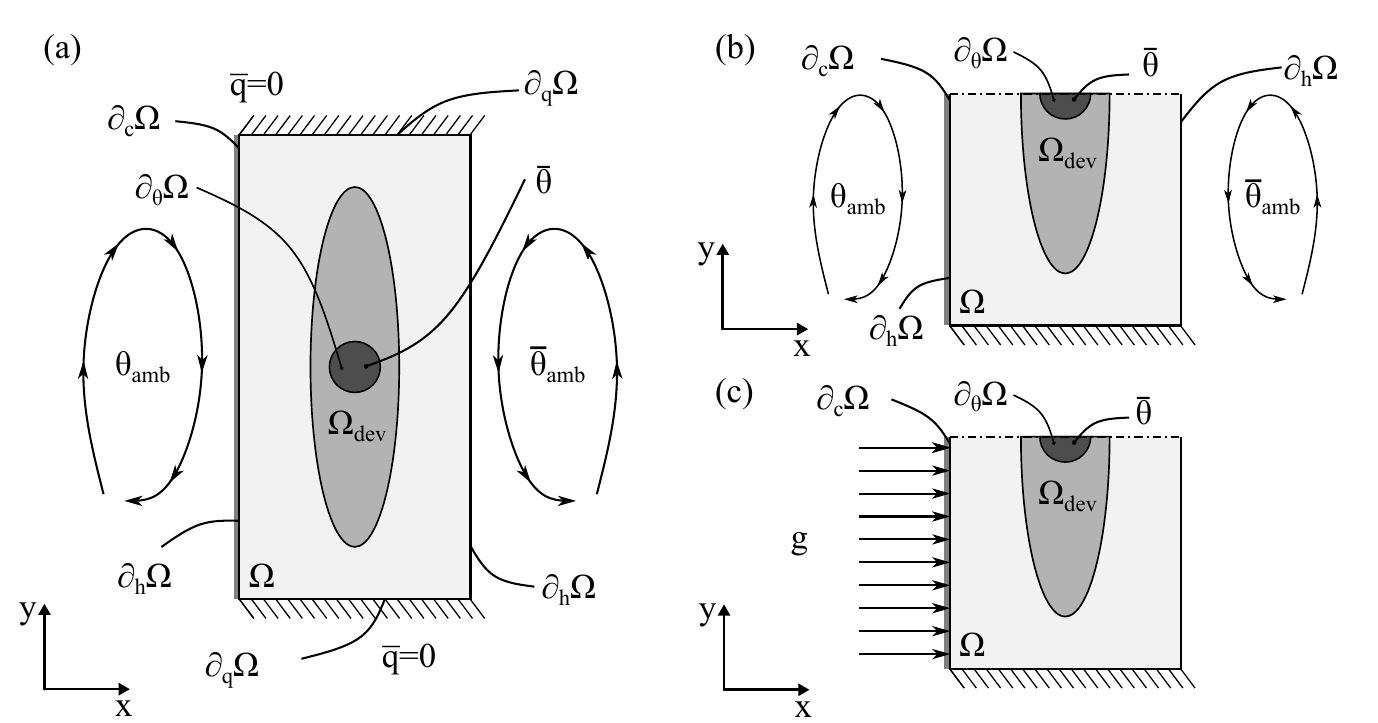}
	\caption{Average temperature minimization: (a) problem setting, (b) system (I) (half-domain), and (c) system (II), where $g\coloneqq f^{(2)}$ (half-domain). The optimal design of the cloaking device, in gray, must achieve a reduction in the average temperature of the left surface, $\partial_c\Omega$.}
	\label{fig_temp_min_cloaking}
\end{figure*}

Let us now focus on the first objective function which deals with the minimization of the average temperature over the cloaking port, $\partial_c\Omega$, by designing the cloaking device (drawn in gray in Figure \ref{fig_temp_min_cloaking}). The corresponding optimization problem, subjected to the same constraint equation and ruled by the thermal state equation of equation (\ref{eq_cloaking_minimi_obj}), is given as
\begin{equation} \label{eq_temperature_min1}
	\begin{split}
		\min_{\chi\in{\mathscr{U}}_{ad}}{\cal J}_{\text{av}}(\theta_\chi)&= C_2 \ \int_{\partial_c\Omega} \theta_\chi\bbx\dgamma = \\ 
		&= C_2 \int_{\partial\Omega} {1}_{\partial_c\Omega}\bbx\,\theta_\chi\bbx\dgamma \mcolon 
	\end{split}\\
\end{equation}
where the integrated temperature is normalized with the corresponding Lebesgue measure, $C_2=\left(\int_{\partial_c\Omega} \dgamma\right)^{-1}$, and ${1}_{\partial_c\Omega}\bbx$ stands for the the indicator function on the subset $\partial_c\Omega$, to enforce the minimization over the whole boundary.

Discretizing the topology optimization problem (\ref{eq_temperature_min1}) via the finite element method, we finally obtain
\begin{equation} \label{eq_temperature_min}
	\begin{split}
		\min_{\chi\in{\mathscr{U}}_{ad}}{\cal J}^{(h_e)}_{\text{av}}(\theta_\chi)&= C_2 \int_{\partial\Omega} {1}_{\partial_c\Omega}\bbx\, {\bf N}\bbx\hat{\pmb{\theta}}_\chi^{(1)} \dgamma = \\
		&= C_2 {\bf 1}_{\partial_c\Omega}^T\hat{\pmb{\theta}}_\chi^{(1)} \mcolon 
	\end{split} 
\end{equation}
whose extended functional is then derived according to Section \ref{sec_relaxed_toplogical_derivative} in order to compute the topological sensitivity of the cost function. Following the same steps as in Section \ref{sec_sensitivity_compliance}, and applying the \emph{adjoint method} with  $\hat{\mathbf{w}}=-C_2\hat{\pmb{\theta}}_{\chi}^{(2)}$ to avoid computing the temperature derivative with respect to the design variable, one finds that problem (\ref{eq_temperature_min}) also requires the resolution of an \emph{auxiliary state equation} (system (II)) in addition to the \emph{original state equation} (system (I)), which read as
\begin{equation} \label{eq_temp_min_systems}
	\left\{
	\begin{split}
		&{\mathbb K}_{\chi}\ \hat{\pmb{{\theta}}}_{\chi}^{(1)}={\bf f}^{(1)}\quad& &(\text{system  I})\\
		&{\mathbb K}_{\chi}\ \hat{\pmb{{\theta}}}_{\chi}^{(2)}={\bf f}^{(2)}\quad& &(\text{system II})\\
	\end{split}
	\right.
\end{equation}
where 
\begin{equation} \label{eq_temp_minimization_system2}
	{\bf f}^{(2)} = - {\bf 1}_{\partial_c\Omega} = -\int_{\partial\Omega}\mathbf{N}^T({\bf{x}}){1}_{\partial_c\Omega}\bbx\dgamma \mdot
\end{equation}

Introducing the solution of the two state equations, $\hat{\pmb{{\theta}}}_{\chi}^{(1)}$ and $\hat{\pmb{{\theta}}}_{\chi}^{(2)}$, into the corresponding relaxed topological derivative of the cost function, and after some algebraic manipulations, detailed in Appendix \ref{App_temp_min}, one obtains the expression of the \emph{pseudo-energy density}, expressed as
\begin{equation} \label{eq_xi_average_temp}
	\xi_{\text{av}}\bbxchi = \gamma_1\bbxchi \ {\overline{\cal U}}_{1-2}\bbx +\gamma_2\bbxchi \ {\overline{\cal U}_{\shS-2}}\bbx \mcolon
\end{equation}
where $\overline{\cal U}_{1-2}\bbxhat$ and ${\overline{\cal U}_{\shS-2}}\bbxhat$ correspond respectively to \textit{\Uconductivity} and \textit{\Usource}, and $\gamma_1\bbxchi$ and $\gamma_2\bbxchi$ are respectively the coefficient of these \emph{energy densities}, which depend on the \emph{characteristic function} and the properties of the material. In summary
\begin{equation} \label{eq_energy_average_temp}
	\left\{
	\begin{split}
		&\overline{\cal U}_{1-2}\bbxhat=\dfrac{1}{2}\left(\gradT{2}\cdot{\Kappalarge}\cdot\gradT{1}\right)\bbxhat &(a) \\
		&{\overline{\cal U}_{\shS-2}}\bbxhat = \left(\hS\theta_{\chi}^{(2)}\right)\bbxhat &(b) \\
		&\gamma_1 \bbxchi = -2 C_2 (1-\beta_{\kappa}) m_{\kappa} \left(\chi_{\kappa}\bbx\right) ^{m_{\kappa}-1}  &(c) \\
		&\gamma_2 \bbxchi = C_2 (1-\beta_{\shS})  m_{\shS}\left({\chi}_{\shS}\bbx\right) ^{m_{\shS}-1} &(d)
	\end{split} \right. \mdot
\end{equation}

\subsubsection{Temperature variance minimization} \label{sec_temperature_var_min}

The second objective function deals with the minimization of the temperature variance over the cloaking port, $\partial_c\Omega$, so the main goal is to design a cloaking device that \emph{homogenizes the temperature on a desired surface}. This optimization problem is written as follows
\begin{equation} \label{eq_temperature_var_min}
	\begin{split}
		&\min_{\chi\in{\mathscr{U}}_{ad}}{\cal J}_{\text{vr}}(\theta_\chi)=C_3 \int_{\partial_c\Omega} \left(\theta_\chi\bbx-{\cal J}_{\text{av}}(\theta_\chi)\right)^2 \dgamma \\
		& \hspace{1cm}= C_3 \int_{\partial\Omega} {1}_{\partial_c\Omega}\bbx\,\left(\theta_\chi\bbx-{\cal J}_{\text{av}}(\theta_\chi)\right)^2\dgamma \mcolon 
	\end{split}
\end{equation}
where the coefficient $C_3$ is equal to the inverse of the measure of the surface, i.e. $C_3 = \left(\int_{\partial_c\Omega} \dgamma\right)^{-1}$, and, as commented before, the temperature variance is only minimized on a part of the boundary of the domain described by the indicator function of the surface $\partial_c\Omega$, ${1}_{\partial_c\Omega}\bbx$.

Applying the FEM discretization (\ref{eq_shape_temp}) to expression (\ref{eq_temperature_var_min}), we finally reach to
\begin{equation} \label{eq_temperature_var_min2}
	\begin{split}
		&\min_{\chi\in{\mathscr{U}}_{ad}}{\cal J}^{(h_e)}_{\text{vr}}(\theta_\chi) = \\
		&=
		C_3 \int_{\partial\Omega} \resizebox{0.6\hsize}{!}{%
						${1}_{\partial_c\Omega}\bbx\,\left({\bf N}\bbx\hat{\pmb{\theta}}_\chi^{(1)}-{\bf N}\bbx{\mathbb{I}}\,{\cal J}^{(h_e)}_{\text{av}}\left(\theta_\chi^{(1)}\right)\right)^2$} \dgamma= \\
		&= C_3 \resizebox{0.75\hsize}{!}{%
						$\left(\hat{\pmb{\theta}}_\chi^{(1)}-{\mathbb{I}}\,{\cal J}^{(h_e)}_{\text{av}}\left(\theta_\chi^{(1)}\right)\right)^T \mathbb{M}_{\partial_c\Omega}  \left(\hat{\pmb{\theta}}_\chi^{(1)}-{\mathbb{I}}\,{\cal J}^{(h_e)}_{\text{av}}\left(\theta_\chi^{(1)}\right)\right) $} 
	\end{split}
\end{equation}
with
\begin{equation}
	\mathbb{M}_{\partial_c\Omega} = \int_{\partial\Omega} {\bf N}^T\bbx {1}_{\partial_c\Omega} \bbx {\bf N}\bbx \dgamma \mcolon
\end{equation}
where $\mathbb I$ represents an all-ones vector with the same length as $\hat{\pmb{\theta}}_\chi^{(1)}$. Equation (\ref{eq_temperature_var_min2}) is subject to the volume constraint in equation (\ref{eq_cloaking_minimi_obj})-(b) and governed by the thermal state equation (\ref{eq_cloaking_minimi_obj})-(c). Now, mimicking the procedure followed for the first functional of equation (\ref{eq_cloaking_minimi_obj})-(a) in Section \ref{sec_temperature_min}, we proceed to compute the RTD of the expression (\ref{eq_temperature_var_min2}) via the \emph{adjoint method} with $\hat{\mathbf{w}}=-C_3\hat{\pmb{\theta}}_{\chi}^{(3)}$, and introducing the RTD of the average temperature ${\cal J}^{(h_e)}_{\text{av}}\left(\theta_\chi^{(1)}\right)$ with the corresponding \emph{adjoint state problem}, equation (\ref{eq_temp_min_systems})-(system (II)).

Finally, one can obtain three \emph{state equations}, being the first two equations mutual to both optimizations problems. Thus, the \textit{original thermal system (I)} is supplemented with two \textit{auxiliary thermal system}: (II) and (III) (where $g$ in Figure \ref{fig_temp_min_cloaking} corresponds to $f^{(2)}$ for the \emph{first auxiliary system}, while it is equal to $f^{(3)}$ for the \emph{second auxiliary system}), which are described by
\begin{equation}
	{\mathbb K}_{\chi}\hat{\pmb{\theta}}_{\chi}^{(i)}=\mathbf{f}^{(i)} \ ; \quad  i=\{1,2,3\}
\end{equation}
with
\begin{equation} \label{eq_temp_minimization_system3}
	\begin{split}
		{\bf f}^{(3)} &= - 2 \mathbb{M}_{\partial_c\Omega}^T {\mathcal T}_\chi\left(\theta_\chi^{(1)}\right) \\
		&=-2 \int_{\partial_c\Omega}\mathbf{N}^T {1}_{\partial_c\Omega}\bbx\,\left(\theta_\chi^{(1)}\bbx-{\cal J}^{(h_e)}_{\text{av}}\left(\theta_\chi^{(1)}\right)\right) \dgamma \mcolon
	\end{split}
\end{equation}
where ${\mathcal T}_\chi\left(\theta_\chi^{(1)}\right)$ corresponds to $ \hat{\pmb{\theta}}_\chi^{(1)}-{\mathbb{I}}\,{\cal J}^{(h_e)}_{\text{av}}\left(\theta_\chi^{(1)}\right)$. 

After replacing the solutions of both \emph{auxiliary systems}, $\hat{\pmb{{\theta}}}_{\chi}^{(2)}$ and $\hat{\pmb{{\theta}}}_{\chi}^{(3)}$, into the RTD of ${\cal J}^{(h_e)}_{\text{vr}}(\theta_\chi)$ and simplifying the consequent terms, the corresponding \emph{spatial energy density}, $\xi\bbxchi$, can be written as
\begin{equation} \label{eq_xi_variance_temp}
	\begin{split}
		\xi_{\text{vr}}\bbxchi =& \gamma_3 {\overline{\cal U}}_{1-2}\bbx +\gamma_4 {\overline{\cal U}_{ \shS-2}}\bbx + \\
			& +\gamma_5{\overline{\cal U}}_{1-3}\bbx +\gamma_6{\overline{\cal U}_{ \shS-3}}\bbx
	\end{split}
\end{equation}
where $\overline{\cal U}_{i-j}\bbxhat$ is \textit{\Uconductivity} for i-th and j-th temperature fields ($i,j=\{1,2,3\}$) and ${\overline{\cal U}_{\shS-k}}\bbxhat$ corresponds to \textit{\Usource} for the k-th temperature field ($k=\{1,2,3\}$), which are respectively written as 
\begin{equation} \label{eq_energy_cloaking_var_temp}
	\left\{
	\begin{split}
		&{\overline{\cal U}}_{1-2}\bbxhat = \dfrac{1}{2}\left(\gradT{2}\cdot{\Kappalarge}\cdot\gradT{1}\right)\bbxhat\\
		&{\overline{\cal U}_{ \shS-2}}\bbxhat = \left(\hS\theta_{\chi}^{(2)}\right)\bbxhat \\
		&{\overline{\cal U}}_{1-3}\bbxhat = \dfrac{1}{2}\left(\gradT{3}\cdot{\Kappalarge}\cdot\gradT{1}\right)\bbxhat\\
		&{\overline{\cal U}_{ \shS-3}}\bbxhat = \left(\hS\theta_{\chi}^{(3)}\right)\bbxhat
	\end{split} \right. \mcolon
\end{equation}
and $\gamma_i$ for $i=\{3,4,5,6\}$ are the corresponding coefficients, defined as
\begin{equation} 
	\left\{
	\begin{split}
		&\gamma_3 \bbxchi = 4C_3C_2(1-\beta_{\kappa})m_{\kappa} \left(\chi_{\kappa}\bbx\right) ^{m_{\kappa}-1} {\mathcal A} &(a) \\
		&\gamma_4 \bbxchi = -C_3C_2(1-\beta_{\shS})m_{\shS}\left({\chi}_{\shS}\bbx\right) ^{m_{\shS}-1} {\mathcal A} &(b) \\
		&\gamma_5 \bbxchi = -2C_3(1-\beta_{\kappa})m_{\kappa} \left(\chi_{\kappa}\bbx\right) ^{m_{\kappa}-1} &(c) \\
		&\gamma_6 \bbxchi =  C_3(1-\beta_{\shS})m_{\shS}\left({\chi}_{\shS}\bbx\right) ^{m_{\shS}-1} &(d)
	\end{split} \right. \mcolon
\end{equation}
where
\begin{equation}
	{\mathcal A} = \left({\mathcal T}_\chi\left(\theta_\chi^{(1)}\right)\right)^T \mathbb{M}_{\partial_c\Omega} {\mathbb{I}} \mdot
\end{equation}

For additional details, the reader is addressed to Appendix \ref{App_temp_var_min} where intermediate steps are presented.

\subsubsection{Temperature multi-objective minimization}

\paragraph{Topological sensitivity of the cost function}

\begin{mybox_alg}[label=box_temp_var_min,float*=!b,text width=0.94\textwidth]{Topology optimization for average and variance temperature minimization}
	{\begingroup
	{\allowdisplaybreaks
	\abovedisplayskip=-6pt
	\belowdisplayskip=0pt	
	\begin{flalign} \label{eq_compliance_temp_cloaking_multi}
			&\textit{Problem:}  \notag\\
			&\hspace{0.5cm}\left[
			\begin{aligned}
				&\chi^*=\underset{\chi\in{\mathscr{U}}_{ad}}{\operatorname{argmin}} \ \widetilde{\cal J}^{(h_e)}(\chi)=\omega\widetilde{\cal J}^{(h_e)}_{\text{av}}(\chi)+(1-\omega)\widetilde{\cal J}^{(h_e)}_{\text{vr}}(\chi) \\
				&s.t.\quad{\mathcal C}(\chi,t)\coloneqq t-\dfrac{\vert\Omega^-\vert(\chi)}{\vert\Omega\vert}=0\;;\quad t\in[0,1]\\
				&\hspace{0.8cm}{\mathbb K}_{\chi}\hat{\pmb{\theta}}_{\chi}^{(i)}=\mathbf{f}^{(i)} \ ; \quad  i=\{1,2,3\}
			\end{aligned}
			\right.   &\ (a)\notag\\
			&\textit{Energy density:}\quad\notag\\
			&\hspace{0.5cm}\begin{aligned}
								\xi\bbxchi = & \omega C_4 \left[ \gamma_1 \ {\overline{\cal U}}_{1-2}\bbx +\gamma_2 \ {\overline{\cal U}_{\shS-2}}\bbx \right] + \\
								& \hspace{3cm} +(1-\omega) C_5 \left[\gamma_3 \ {\overline{\cal U}}_{1-2}\bbx  + \gamma_4 \ {\overline{\cal U}_{\shS-2}}\bbx + \gamma_5 \ {\overline{\cal U}}_{1-3}\bbx +\gamma_6 \ {\overline{\cal U}_{\shS-3}}\bbx\right]
							\end{aligned} &\ (b)\notag\\
			&\hspace{0.5cm}\text{where} \notag\\[-5pt]
			&\hspace{1cm}\left\{\begin{aligned}
								&{\overline{\cal U}}_{1-2}\bbxhat = \dfrac{1}{2}\left(\gradT{2}\cdot{\Kappalarge}\cdot\gradT{1}\right)\bbxhat ;
								&&\; {\overline{\cal U}}_{1-3}\bbxhat = \dfrac{1}{2}\left(\gradT{3}\cdot{\Kappalarge}\cdot\gradT{1}\right)\bbxhat\\
								&{\overline{\cal U}_{\shS-2}}\bbxhat = \left(\hS\theta_{\chi}^{(2)}\right)\bbxhat ;
								&&\; {\overline{\cal U}_{\shS-3}}\bbxhat = \left(\hS\theta_{\chi}^{(3)}\right)\bbxhat \\
								& \gamma_1 \bbxchi = -2 C_2 (1-\beta_{\kappa}) m_{\kappa} \left(\chi_{\kappa}\bbx\right) ^{m_{\kappa}-1} ; 
								&& \gamma_2 \bbxchi = +C_2 (1-\beta_{\shS})  m_{\shS}\left({\chi}_{\shS}\bbx\right) ^{m_{\shS}-1}\\
								& \gamma_3 \bbxchi = +4C_3C_2(1-\beta_{\kappa})m_{\kappa} \left(\chi_{\kappa}\bbx\right) ^{m_{\kappa}-1}{\mathcal A}
								&& \gamma_4 \bbxchi = -C_3C_2(1-\beta_{\shS})m_{\shS}\left({\chi}_{\shS}\bbx\right) ^{m_{\shS}-1}{\mathcal A}\\
								& \gamma_5 \bbxchi = -2C_3(1-\beta_{\kappa})m_{\kappa} \left(\chi_{\kappa}\bbx\right) ^{m_{\kappa}-1}
								&& \gamma_6 \bbxchi = +C_3(1-\beta_{\shS})m_{\shS}\left({\chi}_{\shS}\bbx\right) ^{m_{\shS}-1} \\
								&{\mathcal A} = \left({\mathcal T}_\chi\left(\theta_\chi^{(1)}\right)\right)^T \mathbb{M}_{\partial_c\Omega} {\mathbb{I}}; && 
							\end{aligned} \right.
	\end{flalign}}%
	\endgroup}%
\end{mybox_alg}

Taking into account the expressions obtained in Sections \ref{sec_temperature_min} and \ref{sec_temperature_var_min}, we can define the \emph{energy distribution} of the original problem (equation (\ref{eq_cloaking_minimi_obj})) as a linear combination of equations (\ref{eq_xi_average_temp}) and (\ref{eq_xi_variance_temp}), yielding to
\begin{equation} \label{eq_linear_comb}
	\xi\bbxchi=\omega\,\xi_{\text{av}}\bbxchi +(1-\omega)\,\xi_{\text{vr}}\bbxchi \mcolon
\end{equation}
where the parameter $\omega$ adjusts the weight of each objective function (or sensitivity). As previously mentioned, the sensitivity corresponds to the \emph{weighted sum} of the sensitivities of the two problems.  

Since each term of the original multi-objective problem (\ref{eq_cloaking_minimi_obj}) has been normalized with its range (equation (\ref{eq_normalization_multiobj})), the sensitivity of the scalarized multi-objective problem (\ref{eq_cloaking_minimi_obj2})-(a) includes some extra terms with respect to equation (\ref{eq_linear_comb}) to account for it, i.e. the sensitivity is expressed as
\begin{equation}
	\widetilde{\xi}\bbxchi = \omega\,C_4\,\xi_{\text{av}}\bbxchi + (1-\omega)\,C_5\,\xi_{\text{vr}}\bbxchi \mcolon
\end{equation} 
where
\begin{equation}
	\left\{
	\begin{split}
		C_4 = \dfrac{1}{{{\mathcal J}}_{av}^{max}-{\mathcal J}_{av}^\circ} \\
		C_5 = \dfrac{1}{{{\mathcal J}}_{vr}^{max}-{\mathcal J}_{vr}^\circ}
	\end{split} \mdot \right.
\end{equation}

As explained before, each topology optimization problem requires \textit{auxiliary thermal systems}. We must solve two and three thermal systems for the average temperature minimization and the temperature variance minimization, respectively. However, \textit{the auxiliary thermal system} of the first minimization problem (\ref{eq_temperature_min}) is included into the second minimization problem (\ref{eq_temperature_var_min2}). Therefore, only the following 3 thermal systems must be solved,
\begin{equation}
	\left\{
	\begin{split}
		&{\mathbb K}_{\chi}\ \hat{\pmb{{\theta}}}_{\chi}^{(1)}={\bf f}^{(1)} &(a)\\
		&{\mathbb K}_{\chi}\ \hat{\pmb{{\theta}}}_{\chi}^{(2)}={\bf f}^{(2)}=-{\bf 1}_{\partial_c\Omega} &(b)\\
		&{\mathbb K}_{\chi}\ \hat{\pmb{{\theta}}}_{\chi}^{(3)}={\bf f}^{(3)}=- 2 \mathbb{M}_{\partial_c\Omega}^T \left(\hat{\pmb{\theta}}_\chi^{(1)}-{\mathbb{I}}\,{\cal J}^{(h_e)}_{\text{av}}\left(\theta_\chi^{(1)}\right)\right) &(c)
	\end{split}
	\right.
\end{equation}

\paragraph{Closed-form solution} 

The \emph{energy distribution}, $\xi\bbxchi$, of this topology optimization problem is stated in Box \ref{box_temp_var_min}. This function combines the \emph{energy distributions} presented in equations (\ref{eq_xi_average_temp}) and (\ref{eq_xi_variance_temp}).

\section{Representative numerical simulations}
\label{sec_representative_numerical_simulations}

\begin{figure*}[pb]
	\centering
	\includegraphics[width=15cm, height=8cm]{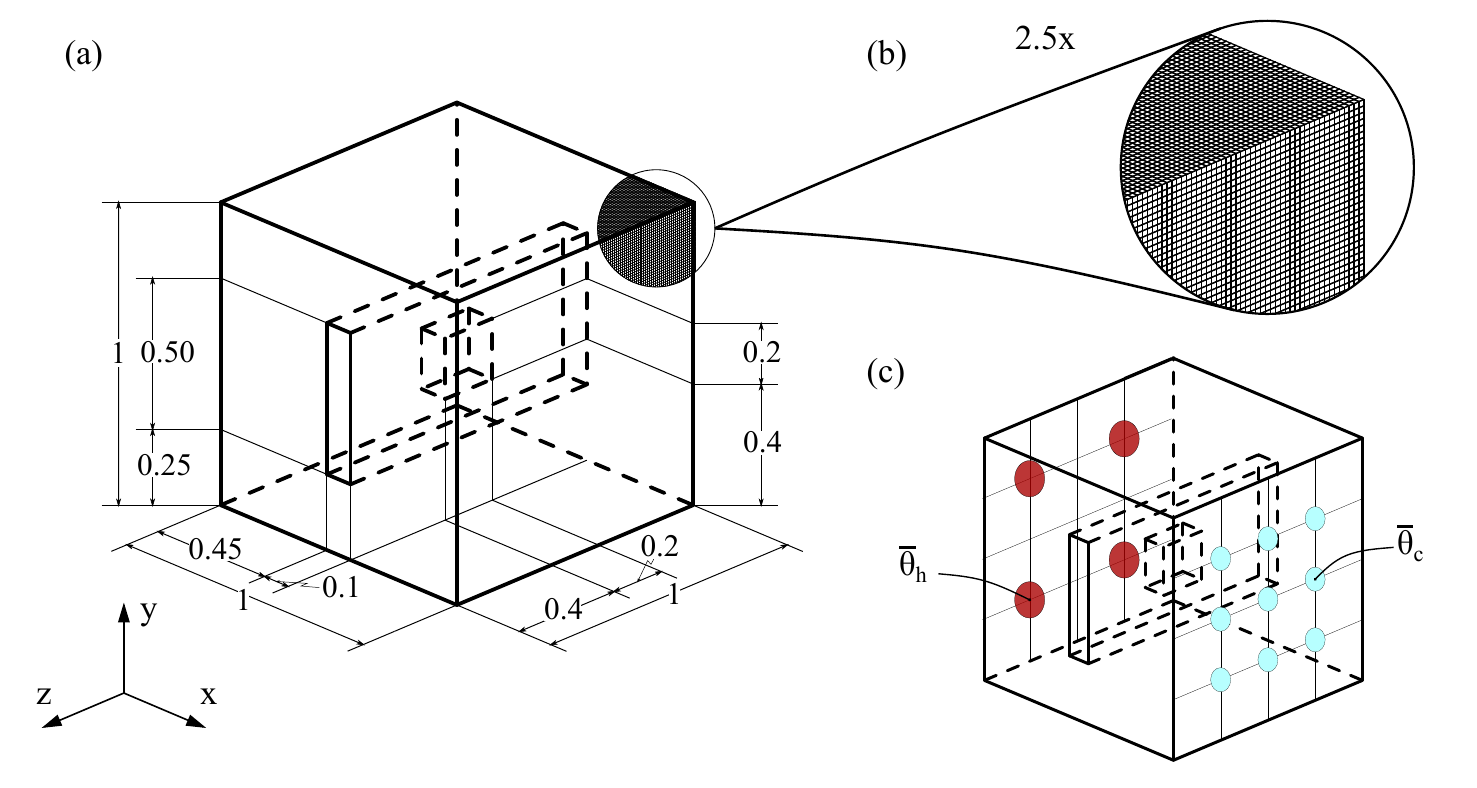}
	\caption{Thermal heat conductor: (a) Setup of the analysis domain, (b) Detailed mesh based on hexahedral finite elements and (c) Boundary conditions of the problem. The temperature is prescribed to $\overline{\theta}_h$ at the four circular regions on the left face (colored in red) while it is set to $\overline{\theta}_c$ at the nine circular regions on the right face (colored in blue). The other surfaces are assumed to be adiabatic.}
	\label{fig_domain_thermal_complaince}
\end{figure*}

In this section,  a number of 3D numerical examples to assess the performance of the proposed methodology are presented. 
Unless otherwise specified, all simulations are done using an isotropic thermal material with a normalized conductivity $\kappa=1 W/(K\,m)$ and a null heat source ($\hS=0 W/m^3$). When needed, the heat transfer coefficient is set to $h=1 W/(K\,m^2)$ and the ambient temperature is fixed to ${\theta}_{amb}=283.15 K$. The material contrast factor and the corresponding exponent are set to $\alpha=10^{-3}$ and $m=5$\footnote{The exponential parameters $m_i$ are set on the basis of the authors’ experience.}, respectively. The used relaxation factor is $\beta=2.51\cdot10^{-1}$. $Tol_{\chi}=10^{-1}$, $Tol_{\lambda}=10^{-1}$ and $Tol_{\cal C}=10^{-3}$ are the used tolerances. In all cases, eight-node hexahedral ($Q_1$) finite elements are used in the solution of the thermal state equation.

\subsection{Thermal compliance minimization. 3D thermal conductor.}
\label{sec_thermal_compliance_conductor}

This example refers to the minimization of the thermal compliance, as explained in Section \ref{sec_structural_compliance_problems}, in a thermal component, e.g. heat pipes for a CPU heat sink, in a cubic domain subject to specific Dirichlet conditions. The aim is to display the potential of the present methodology for obtaining the optimal topology for heat conduction in a complex analysis domain. 

The analysis domain, illustrated in Figure \ref{fig_domain_thermal_complaince}, is a cube, $1\text{x}1\text{x}1 \ m$, with a rectangular hole all the way across it, with dimensions $0.1$x$0.5$x$1$ m, located in the center and oriented in the z direction. A small prismatic volume, $0.1$x$0.2$x$0.2$ m, is set in the center of the domain as part of the initial domain. The radii of the left and right circular areas, highlighted in Figure \ref{fig_domain_thermal_complaince}-(c), are $R_h=0.075$ m and $R_c=0.05$ m, respectively. The domain is discretized with a structured mesh of $120\text{x}120\text{x}120$ hexahedral elements (mesh size $h_e=8.3\cdot10^{-3}$ m), which leads to 1.648.512 hexahedra (see Figure \ref{fig_domain_thermal_complaince}-(b)). 

\begin{figure*}[pt]
	\centering
	\includegraphics[width=17cm, height=7.5cm]{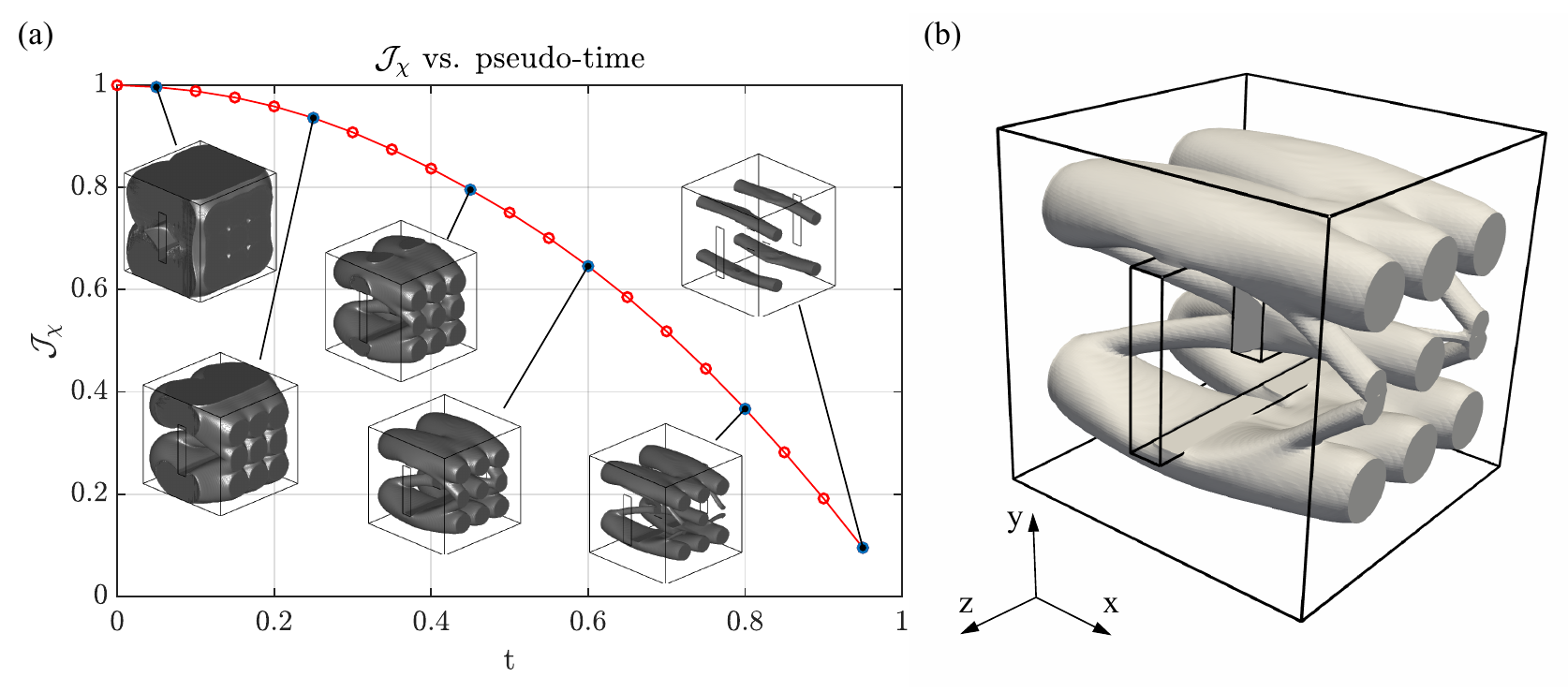}
	\caption{Thermal heat conductor. Thermal compliance minimization: (a) Cost function and topology evolution, (b) Topology for $t=\frac{|\Omega^-\bchi|}{|\Omega|}=0.75$.}
	\label{fig_thermal_compliance_evolution}
\end{figure*}

It is assumed that the four areas, colored in red and located on the left surface, with a prescribed temperature of $\overline{\theta}_h=293$ K are connected with four CPU's IHS. The other nine areas, at temperature $\overline{\theta}_c=278$ K, colored in blue, and located on the right face, are coupled to the cooling system (heat sink). Adiabatic boundary conditions are assumed on the other faces.

For the Laplacian smoothing (see Appendix \ref{sec_finite_element_discretization}),  a value of $\tau=1$ is used, resulting in a parameter $\varepsilon=8.3\cdot10^{-3}$ m. The time interval of interest $[0,0.95]$ is discretized in 19 equally spaced steps.

In Figure \ref{fig_thermal_compliance_evolution}-(a), the evolution of the cost-function, $\cal{J}_\chi$, and some representative optimal topologies are illustrated in terms of the pseudo-time, ($t=\frac{|\Omega^-\bchi|}{|\Omega|}$). As it could be expected, while the soft material increases, the cost function decreases. In Figure \ref{fig_thermal_compliance_evolution}-(b), an intermediate optimal design, when the hard material is the 25\% of the total analysis domain, is presented. 
The topologies in Figure \ref{fig_thermal_compliance_evolution}-(a) show how the hot regions are connected with the cold ones, minimizing the thermal compliance.
In the limit case of imposing very little conductive material (high values of $t$), the obtained optimal topology connects the hot and cold faces with only four (thin) heat pipes (see also Online Resource 1).

\begin{figure*}[pb]
	\centering
	\includegraphics[width=17cm, height=7.5cm]{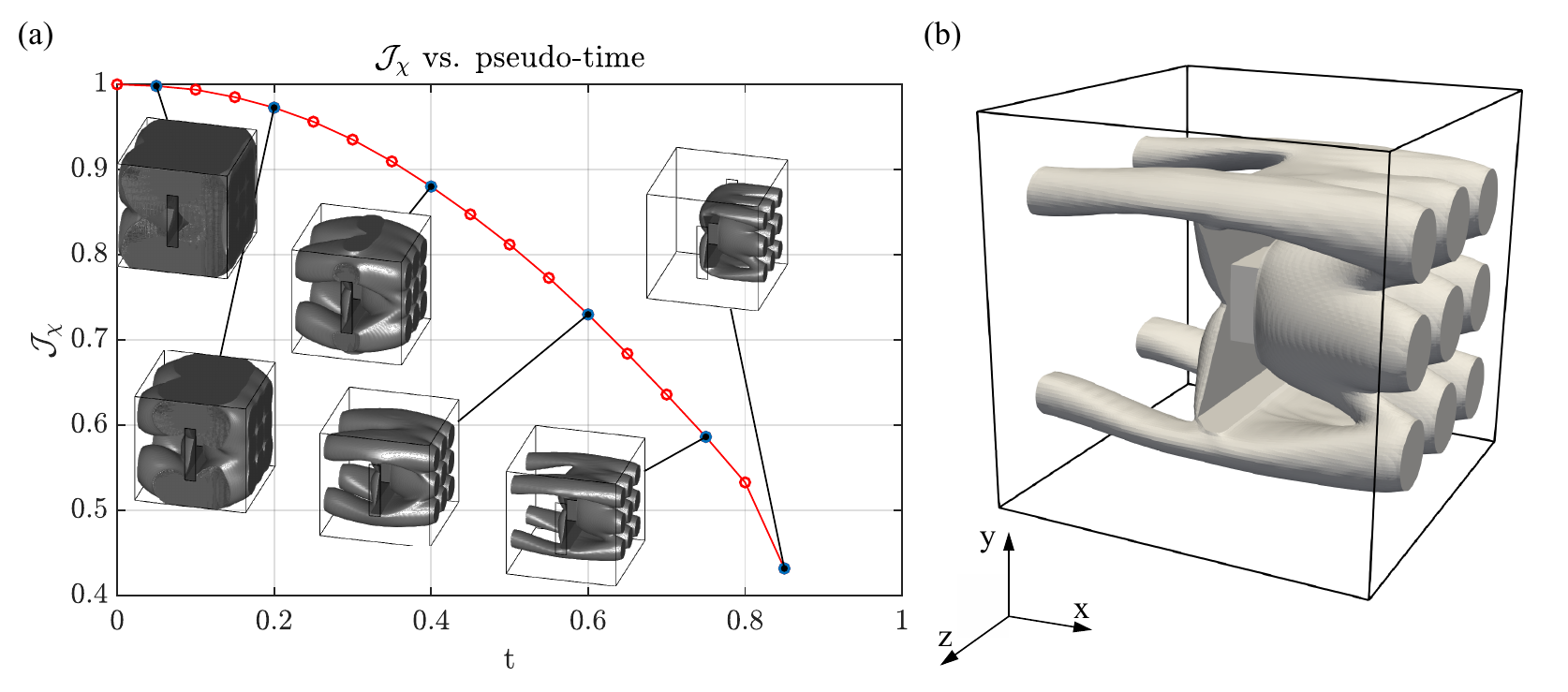}
	\caption{Thermal heat conductor. Thermal compliance minimization including heat source: (a) Cost function and topology evolution, (b) Topology for $t=\frac{|\Omega^-\bchi|}{|\Omega|}=0.8$.}
	\label{fig_thermal_compliance_evolution_hsource}
\end{figure*}

Let us now modify this numerical example in order to consider a not null heat source ($r\neq0$) inside the design domain, $\Omega$. Then, a heat source of $r=1 kW/m^3$ is considered in the small prismatic volume, located at the center of the domain (see Figure \ref{fig_domain_thermal_complaince}), which cannot be removed from the hard material domain. The contrast factor for the heat source is set to $\alpha=1e-3$, and the exponent is set to $m=1$. Both the boundary conditions and the mesh dicretization are kept unchanged with respect to the definition of the example. In addition, the same value of $\tau$ is used for the Laplacian smoothing. Nevertheless, the time interval of interest $[0,0.85]$, in this case, is discretized in 17 equally spaced steps.

Mimicking Figure \ref{fig_thermal_compliance_evolution}, Figure \ref{fig_thermal_compliance_evolution_hsource}-(a) illustrates the evolution of the cost-function throughout the topology optimization in terms of the pseudo-time, $t$, along with some optimal topologies. The optimal topology for $t=0.8$ is displayed in Figure \ref{fig_thermal_compliance_evolution_hsource}-(b). Due to the incorporation of the heat source in the central prismatic volume, a major change in the optimal topologies between the two presented situations is observed. In the last situation, the volume, in which the heat source is added, is also connected to the cold regions on the right side of the domain in order to dissipate as much heat as possible. In addition, the connection between hot and cold regions, observed in Figure \ref{fig_thermal_compliance_evolution} for high values of $t$, gets removed in favor of a better connection to the heat source.

\subsection{Thermal cloaking optimization}

\begin{figure*}[pb]
	\centering
	\includegraphics[width=12cm, height=8cm]{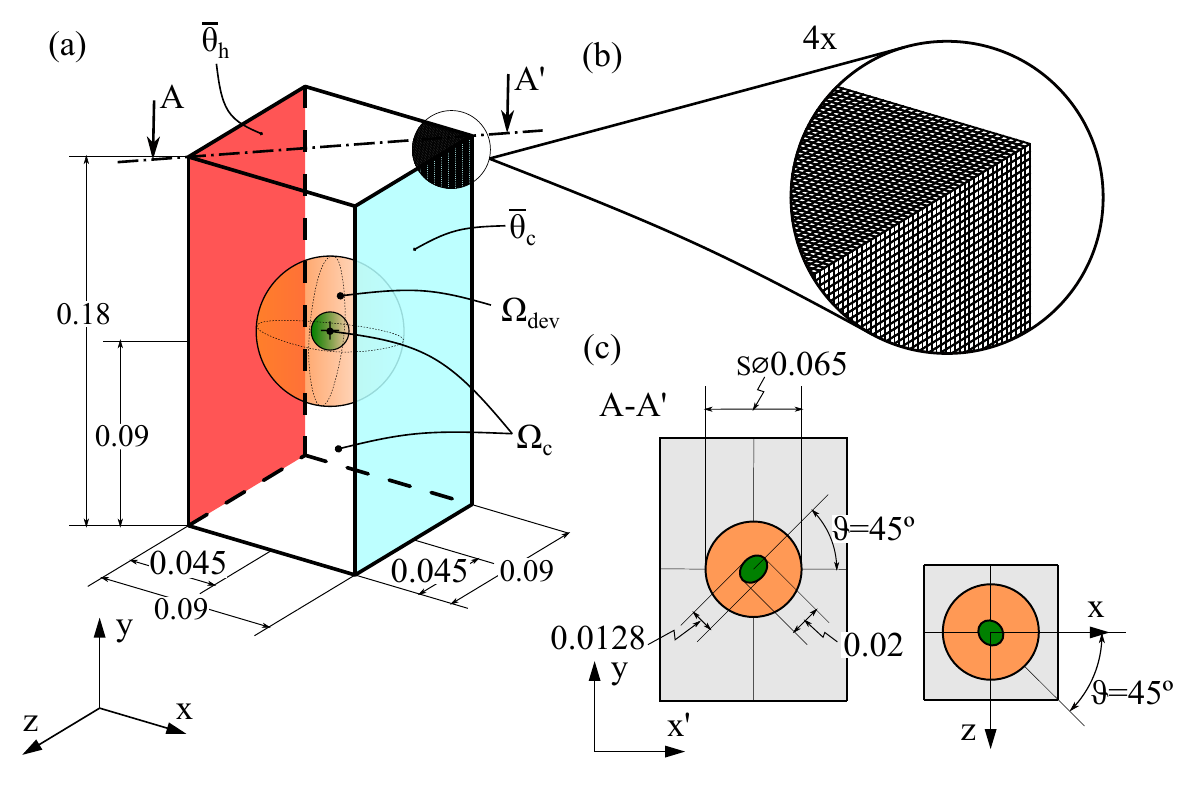}
	\caption{Heat flux cloaking device: (a) Analysis domain, with boundary conditions and dimensions, (b) Detailed mesh and (c) Dimensional details. The cloaked object in green, placed at the center of the domain, is surrounded by the cloaking device, $\Omega_{dev}$, in orange, whose design is optimized. The temperature on the left surface is set to $\overline{\theta}_h$, while the right one is set to $\overline{\theta}_c$.}
	\label{fig_domain_thermal_flux_cloaking}
\end{figure*}

\subsubsection{Thermal cloaking via heat flux manipulation. 3D heat flux cloaking device.}
\label{sec_thermal_cloaking_flux_device}

The optimization of a 3D thermal cloaking device, surrounding the object to be cloaked, is now addressed.
The goal is to design the optimal topology of the cloaking device by means of the manipulation of the heat flux around it, as detailed in Section \ref{sec_thermal_cloaking}.
This problem, inspired in the pioneering work by \citet{Fachinotti2018}, can be considered a 3D extension of this work, with the heat flux prescribed to a given constant value. 
For the solution of the problem, a square prismatic domain $\Omega$, with dimensions $0.09$x$0.18$x$0.09$ (in meters), is defined and discretized with a structured mesh of $100$x$200$x$100$ hexahedral elements (Figure \ref{fig_domain_thermal_flux_cloaking}).
The non-dimensional regularization parameter $\tau$ is equal to 0.1 and the pseudo-time interval $[0,0.08]$ is discretized in 8 steps.

\begin{figure*}[pt]
	\centering
	\includegraphics[width=16cm, height=7.5cm]{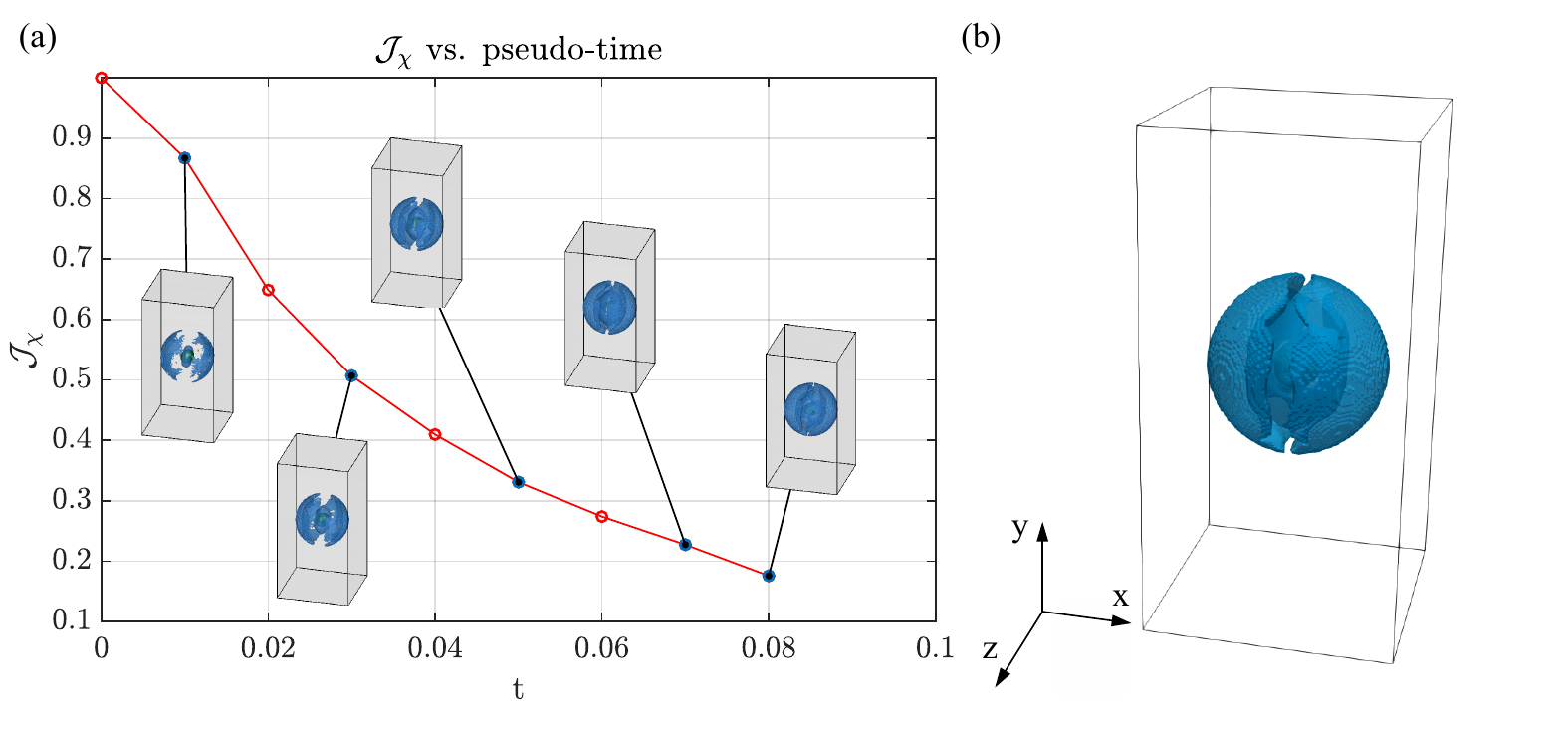}
	\caption{Heat flux cloaking device: (a) Cost function and topology evolution, and (b) Topology for $t=\frac{|\Omega^-|}{|\Omega|}=0.08$.}
	\label{fig_thermal_flux_cloaking_evolution}
\end{figure*}

\begin{figure*}[pb]
	\centering
	\includegraphics[width=16cm, height=8.5cm]{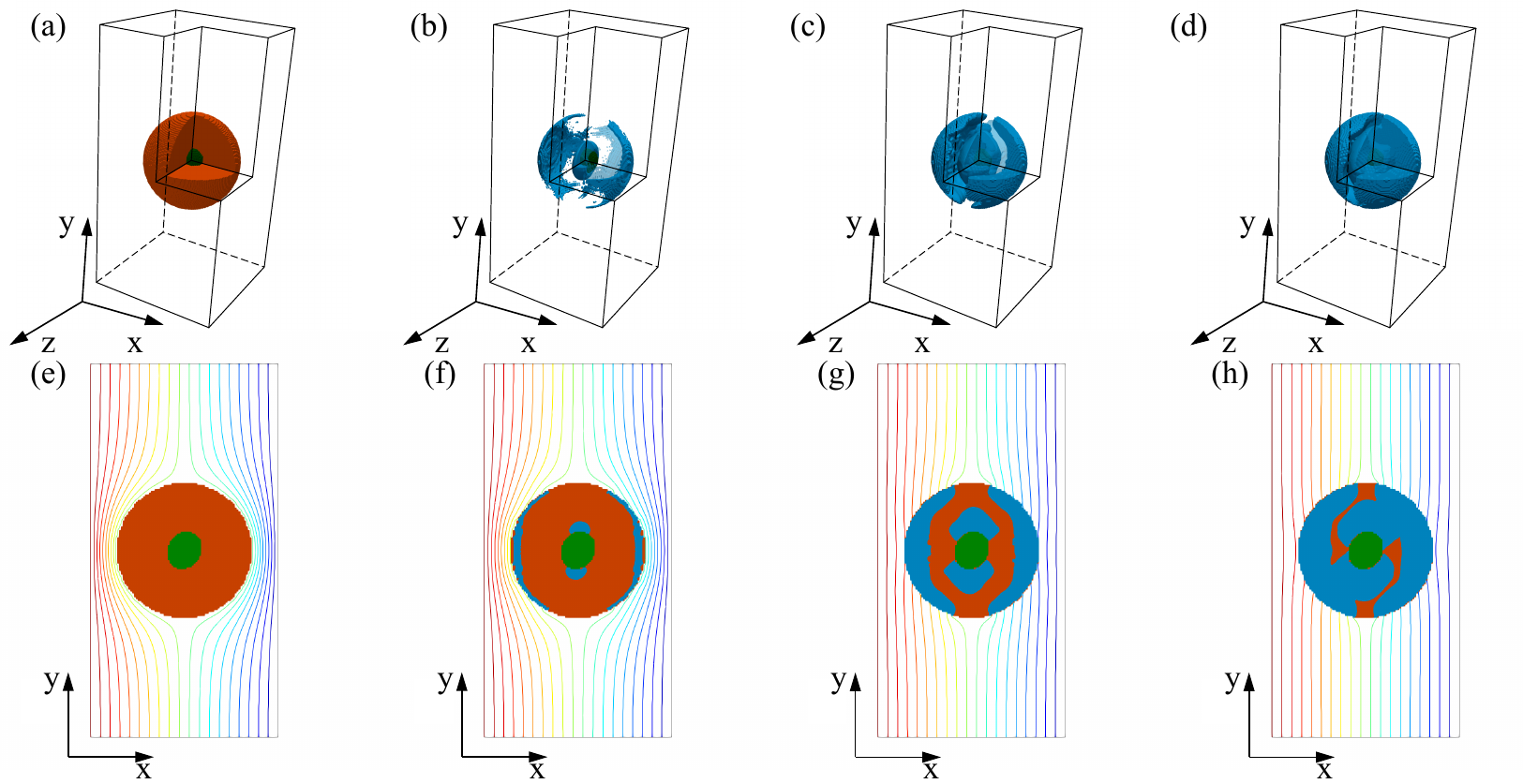}
	\caption{Heat flux cloaking device: figures (a)-(d): 3D view of intermediate topologies, in terms of the soft (low conductive) material counterpart, at steps 0, 1, 5 and 8, respectively. Figures (e)-(h): evolution of the isotherms and layout of the cloaking device at the middle x-y plane, for the same representative steps. (Color legend: blue$\rightarrow$soft material, orange$\rightarrow$hard material and green$\rightarrow$cloaked object).}
	\label{fig_thermal_flux_cloaking_slice}
\end{figure*}

Domain, $\Omega$, is partitioned in three distinct regions, as illustrated in Figure \ref{fig_domain_thermal_flux_cloaking}: 1) the cloaked object is an ellipsoid, colored in green, located at the center of the analysis domain (the principal axes of the ellipsoid are $d_1=0.02m$ and $d_2=d_3=0.0128m$, the main axis being oriented 45$^{\circ}$ with respect to the x and y axes); b) a sphere of diameter $d=0.065m$, shaded in orange, corresponding to the cloaking device to be designed (design domain, $\Omega_{dev}$), and c) the remaining part of the analysis domain, colored in gray in Figure \ref{fig_domain_thermal_flux_cloaking}-(c).  Regions 1 and 3 correspond to domain $\Omega_c\equiv\Omega\setminus\Omega_{dev}$, and the optimization goal is to keep the original homogeneous heat flux constant and unaffected by the cloaking device in these regions. 

The conductivity in $\Omega_c$ and $\Omega_{dev}$ is $\kappa=0.57$ $W/(mK)$ and $\kappa=403W/(mK)$, respectively. In order to obtain a conductivity of $\kappa=0.22W/(mK)$ in the soft phase of region 2, a contrast factor of $\alpha=5.459\cdot 10^{-4}$ is considered, equivalently, $m=5$ and   $\beta=0.886$ are also considered.

The temperatures on the left and right surfaces of the domain are prescribed to $\overline{\theta}_h=321.85K$ and $\overline{\theta}_c=283.15K$, respectively. The other surfaces are assumed to be adiabatic. Under these boundary conditions and assuming an homogeneous isotropic thermal material of $\kappa=0.57W/(mK)$ for the whole domain, the homogeneous temperature gradient in the x-direction results in a constant horizontally heat flux $\overline{\bm{q}}=[245.1,\ 0,\ 0]\ W/m^2$, which corresponds to the target heat flux in $\Omega_c$.

In Figure \ref{fig_thermal_flux_cloaking_evolution}-(a), the evolution of the cost function, including some representative optimal topologies, is presented. A detail of the optimal layout for $t=8\%$ is illustrated in Figure \ref{fig_thermal_flux_cloaking_evolution}-(b).
In Figures \ref{fig_thermal_flux_cloaking_slice}-(a-d), the topology design evolution of the cloaking device is plotted for different intermediate time steps \footnote{removing an octant of the total domain as well as the hard material for a better visualization of the topology.}(see also Online Resource 2). 
Figures \ref{fig_thermal_flux_cloaking_slice}-(e-h) represent the isotherms and the optimal topology layout of both material phases, obtained at the slice  parallel to the x-y plane, and centered along z-axis.
As it can be observed in the figure, isotherms tend to reach an homogeneous temperature gradient configuration\footnote{The isotherms for the homogeneous case are vertical, equally spaced, isolines from $\overline{\theta}_h$ to $\overline{\theta}_c$.} as $t$ increases (and, thus, more low-conductivity material is used in the cloaked domain). 
Also it can be observed that the optimal design of the cloaking device, and the way it works, are, by no means, obvious. The incoming horizontal heat flux is modified, by the combination of the low and high conductive materials in $\Omega_{dev}$, into two different structures: a low-conductive shell and a low-conductive toroid-like domain. The thickness of the shell structure increases along time, and strongly modifies the heat flux near the left and right faces of the cloaking device $\Omega_{dev}$, as it can be observed in Figures \ref{fig_thermal_flux_cloaking_slice}-(f) and \ref{fig_thermal_flux_cloaking_slice}-(g). The toroid surrounds the cloaked object and controls the heat flux inside it, see Figure \ref{fig_thermal_flux_cloaking_slice}-(b).

\subsubsection{Thermal cloaking via average and variance temperature minimization. 3D thermal cloaking device.}
\label{sec_thermal_cloaking_device}

\begin{figure*}[pb]
 	\centering
 	\includegraphics[width=16cm, height=8cm]{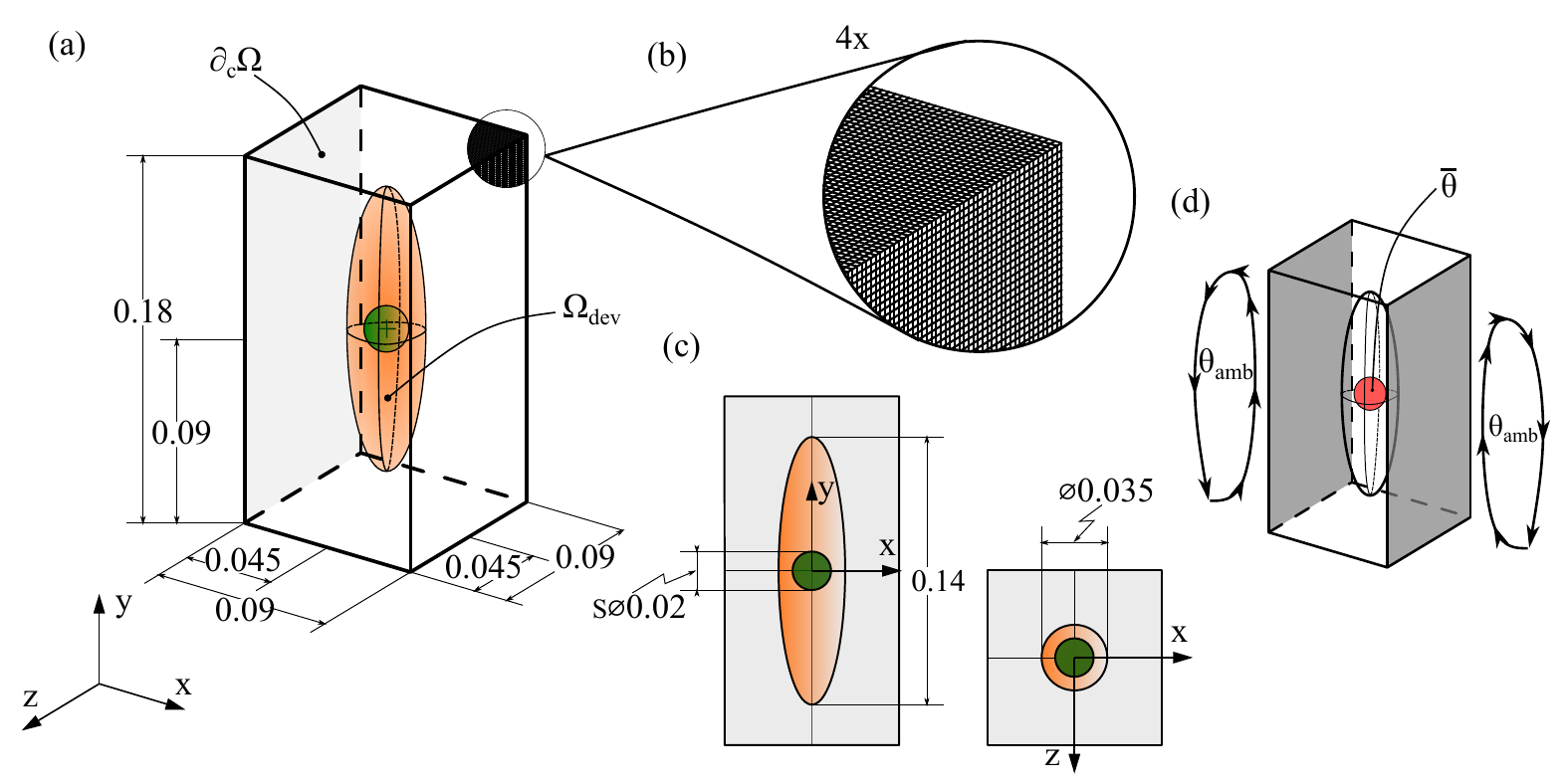}
 	\caption{Thermal cloaking device: (a) Analysis domain with its dimensions, (b) Detail of the mesh, (c) Details of dimensions and (d) Boundary conditions. The cloaked object, in green, prescribed to a high temperature $\overline{\theta}$ is surrounded by the cloaking device, in orange, which must distribute the heat to minimize the average and the variance of the temperature on the left face, $\partial_c\Omega$.}
 	\label{fig_domain_thermal_temperature_cloaking}
 \end{figure*}

Now, a thermal cloaking device is again designed but, this time, aiming at minimizing the average and variance temperature, on a virtual plane at the surface of the analysis domain, in which the values and distribution of temperature are measured by an external device (a thermal camera, for instance). 
The cloaking device, in $\Omega_{dev}$, should mitigate the distortion produced on the virtual plane by the (hot) cloaked object.
The setup of the problem is displayed in Figure \ref{fig_domain_thermal_temperature_cloaking}. The dimensions of the prismatic domain, $\Omega$, are the same than in the previous example, but a slightly finer finite element mesh is used ($150$x$300$x$150$ linear hexahedral elements). 
Taking advantage of the symmetries, only a quarter of the domain is discretized.

\begin{figure*}[pt]
	\centering
	\includegraphics[width=16cm, height=7.5cm]{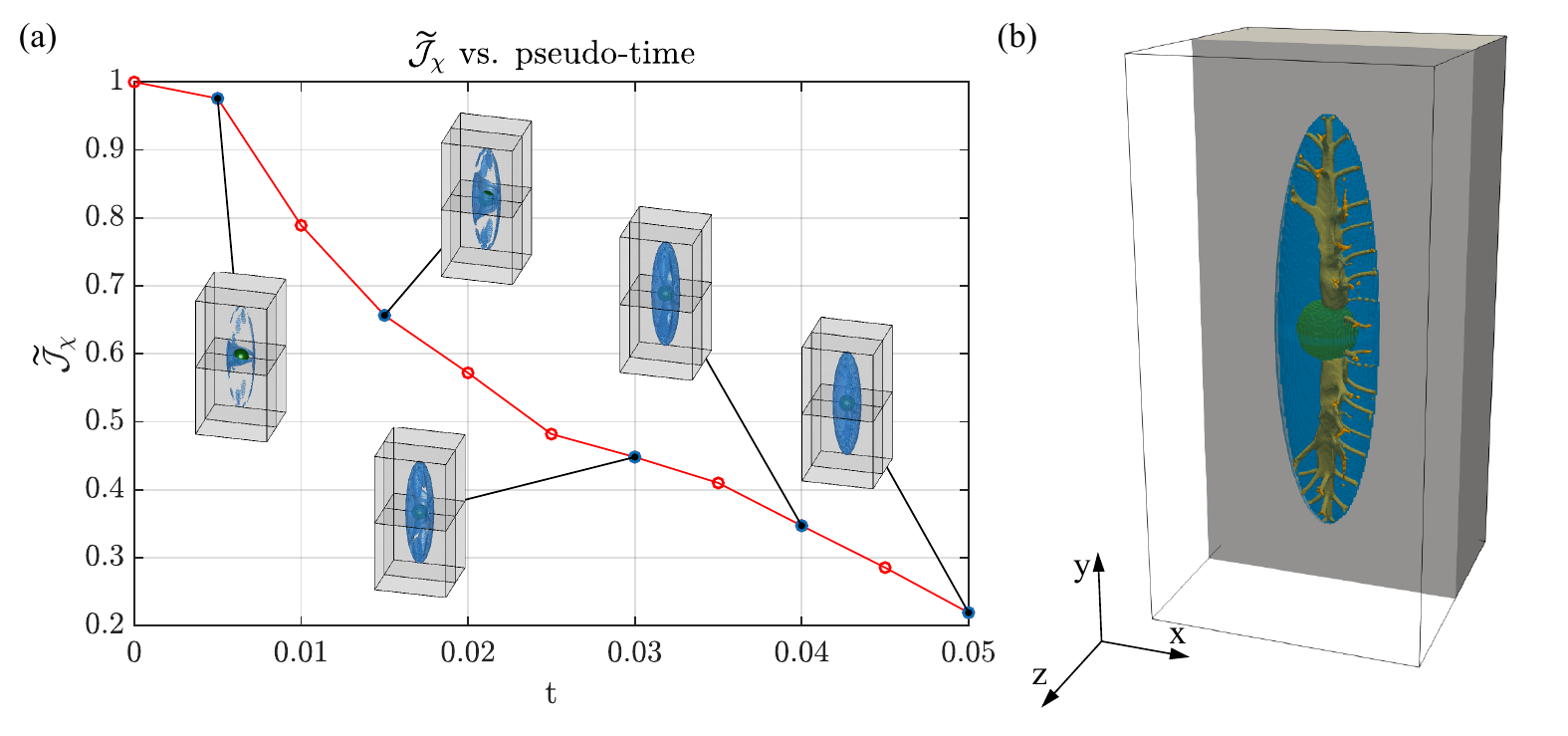}
	\caption{Thermal cloaking device: (a) Cost function and topology evolution, and (b) Topology for $t=\frac{|\Omega^-|}{|\Omega|}=0.05$.}
	\label{fig_domain_thermal_temperature_cloaking_evolution}
\end{figure*}

The domain is again partitioned in three different regions, see Figure \ref{fig_domain_thermal_temperature_cloaking}-(c). 
The innermost region is a sphere of radius $R=0.01m$ (the hot object to be cloaked, colored in green), which is completely surrounded by region 2, an ellipsoid shaded in orange (the cloaking device,  $\Omega_{dev}$), of dimensions $d_x=d_z=0.035m$ and $d_y=0.14m$ (see Figure \ref{fig_domain_thermal_temperature_cloaking}). The remaining volume of $\Omega$ defines region 3. The material properties of each region are the same as the ones described in Section \ref{sec_thermal_cloaking_flux_device}. The conductivity of regions 1 and 3 is set to $\kappa=0.57$ $W/(mK)$, while it is set to $\kappa=403W/(mK)$ for the hard material in $\Omega_{dev}$. The contrast factor in $\Omega_{dev}$ is $\alpha=5.459\cdot 10^{-4}$. The temperature of the cloaked object is set to $\overline{\theta}=313K$.
Left and right surfaces are subjected to a convective flux described by $h=1W/(Km^2)$ and ${\theta}_{amb}=283K$. 
The other surfaces are assumed to be adiabatic (see Figure \ref{fig_domain_thermal_temperature_cloaking}-(d)). 
The regularization parameter is $\tau=0.1$, and the time interval $[0,0.05]$ is split into 10 equally spaced pseudo-time steps.

Following the scheme detailed in Section \ref{sec_temp_var_min}, the optimization problem (\ref{eq_cloaking_minimi_obj}) has to be solved three times (for $\omega=0$,  $\omega=1$, and $\omega=0.5$, respectively). From the results of the first two optimizations, the values of ${\mathcal J}_{av}^\circ=308.6K$, ${{\mathcal J}}_{vr}^{max}=7.4\cdot10^{-2}K^2$, ${{\mathcal J}}_{av}^{max}=310.4K$ and ${\mathcal J}_{vr}^\circ=9\cdot10^{-3}K^2$, have been determined. 
In this specific case, the results of the second problem are not required, since the maximum average temperature is obtained in the first iteration and the utopia point of the variance can be approximated as ${\mathcal J}_{vr}^\circ=0K^2$. 
Finally, completing the objective function (\ref{eq_cloaking_minimi_obj2}) with the previous parameters, the third optimization problem is solved for $\omega=0.5$.

\begin{figure*}[pt]
	\centering
	\includegraphics[width=17.5cm, height=9cm]{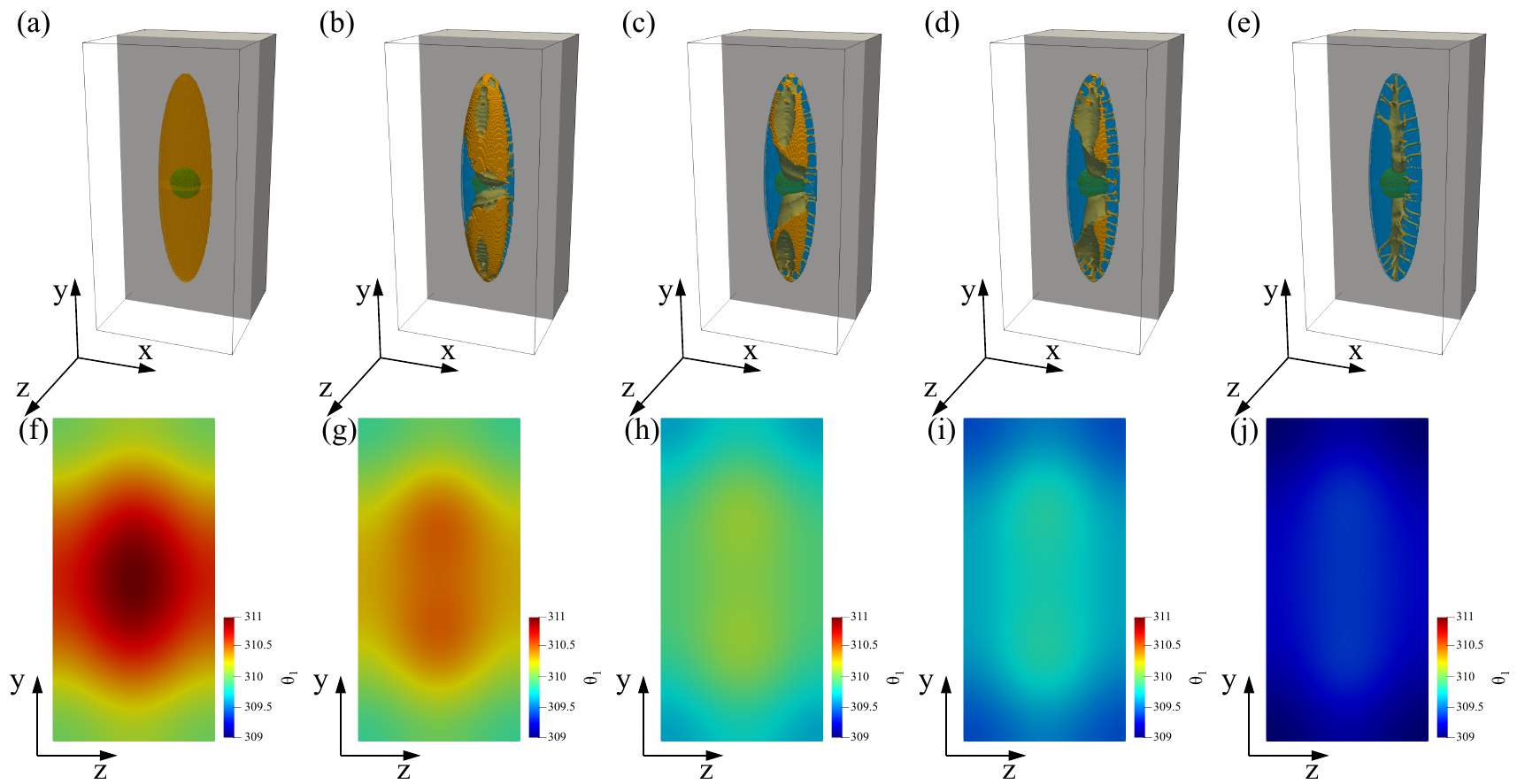}
	\caption{Thermal cloaking device: figures (a)-(e): 3D view of intermediate configurations, illustrated by the soft and hard materials of the cloaking device, for steps 0, 3, 6, 8 and 10. Figures (f)-(j): Evolution of the temperature field of the left y-z plane, for the same representative pseudo-time steps. (Color legend: blue$\rightarrow$soft material, orange$\rightarrow$hard material and green$\rightarrow$sphere).}
	\label{fig_domain_thermal_temperature_cloaking_slice}
\end{figure*}

The cost function evolution and intermediate topologies are displayed in Figures \ref{fig_domain_thermal_temperature_cloaking_evolution} and \ref{fig_domain_thermal_temperature_cloaking_slice}. In Figures \ref{fig_domain_thermal_temperature_cloaking_slice}-(a) to \ref{fig_domain_thermal_temperature_cloaking_slice}-(e), the design evolution of the cloaking device shows how the hard material (colored in orange), which initially completely fills the design domain, is progressively replaced by an insulating material (the low-conductive, soft, material colored in blue), see also Online Resource 3. 
The final optimal layout of the cloaking device, presented in Figure \ref{fig_domain_thermal_temperature_cloaking_evolution}-(b), where half of the domain has been removed for the sake of clarity, resembles a sort of "spine", linked with the rest of the domain at its right side while the links at the left side are scarce and limited to the top and bottom of the "spine". Therefore, the internal heat generated by the cloaked object is, on one hand, transmitted to the top and bottom regions of the left surface ($\partial_c\Omega$) and, on the other, to the complete right surface where the heat is dissipated by natural convection. The distribution of temperatures obtained on the left surface, see Figures \ref{fig_domain_thermal_temperature_cloaking_slice}-(f) to \ref{fig_domain_thermal_temperature_cloaking_slice}-(j), confirms that as the hard (high-conductive) material tends to vanish, the temperature resulting in an uniform temperature distribution approaching the ambient temperature, $\theta_{amb}$. This "a posteriori" analysis, explains the role of that, by no means obvious, resulting thermal cloaking analysis.

\subsection{Computational assessment. Variational closed-form solution vs. level set method}
\label{sec_computational_assessment}

This section, is devoted to analyze the computational performance of the \emph{nonsmooth relaxed variational approach} to topology optimization, based on the Relaxed Topological Derivative (RTD), used in this work for thermal problems, with respect to a \emph{level set method} driven by the same Relaxed Topological Derivative. To illustrate the comparison, the example described in Section \ref{sec_thermal_compliance_conductor} is analyzed with both methods. The comparisons are established in terms of the cost function values and the relative computational cost, which, in  turn, is evaluated in terms of the number of iterations that each method requires to converge with the same tolerances ($Tol_\chi=10^{-1}$ and $Tol_{\cal C}=10^{-3}$).\footnote{The comparison is done in terms of the number of iterations, instead of the computational time, as the computational cost per iteration is almost equivalent for the two approaches. Additionally, the number of iterations remains independent of the platform.}
For a fair comparison, the time interval $[0,0.9]$ and the number of steps, 18, are used for both methods.

The level set function, $\phi\bbx$, in the level set method, is updated through a time-evolving (Hamilton-Jacobi) equation \cite{Allaire2005a}, while the volume constraint is satisfied by means of a Lagrangian multiplier updating scheme\footnote{The \emph{Cutting\emph{\&}Bisection} algorithm in Section \ref{sec_opt_alg} is then replaced by the standard Augmented Lagrangian update, see equation (\ref{eq_iterative_process})-(c). At convergence, the volume constraint is fulfilled at he prescribed tolerance.} \cite{Simo1992}.
The time evolution process continues until both the topology, defined via the characteristic function, and volume tolerances are satisfied. 
Therefore, the level set function is iteratively updated as follows (see \cite{Oliver2019} for more details)
\begin{equation} \label{eq_iterative_process}
	\left\{
	\begin{split}
		&\phi^{(i+1)}({\bf x})=\phi^{(i)}({\bf x})-\dfrac{\Delta t}{\Delta \chi^{(i)}({\bf x})}\dfrac{\delta {\cal L}({\chi}^{(i)},\lambda^{(i)})}{\delta {\chi}^{(i)}}({\bf x}) &(a)\\
		&\chi^{(i+1)}={\cal H}_{\beta}\left(\phi^{(i+1)}({\bf x})\right) &(b)\\
		&\lambda^{(i+1)}=\lambda^{(i)}+\rho\ {\cal C}(\chi(\phi^{(i)})) &(c)
	\end{split}
	\right. \mcolon
\end{equation}
where $\dfrac{\delta {\cal L}({\chi}^{(i)},\lambda^{(i)})}{\delta {\chi}^{(i)}}({\bf x})$ corresponds to the relaxed topological derivative (RTD) of the Lagrangian and $\rho\in\mathbb{R}^+$ is a suitable penalty value.

We emphasize that the parameter $\Delta t$, in equation (\ref{eq_iterative_process})-(a), has a remarkable effect in the convergence rate of this method. 
For very small values, the method will require many iterations until convergence is achieved while, for large values of $\Delta t$, results oscillate or even diverge. This parameter has to be tuned for every problem to find the optimal (convergent and large enough) value of $\Delta t$.
After this, a value of $\Delta t=1\cdot10^{-1}$ has been established for the considered problem as the optimal one for the comparison purposes. The penalty is set to $\rho=5\cdot10^{-2}$.

\begin{figure*}[pt]
	\centering
	\includegraphics[width=13.5cm, height=15cm]{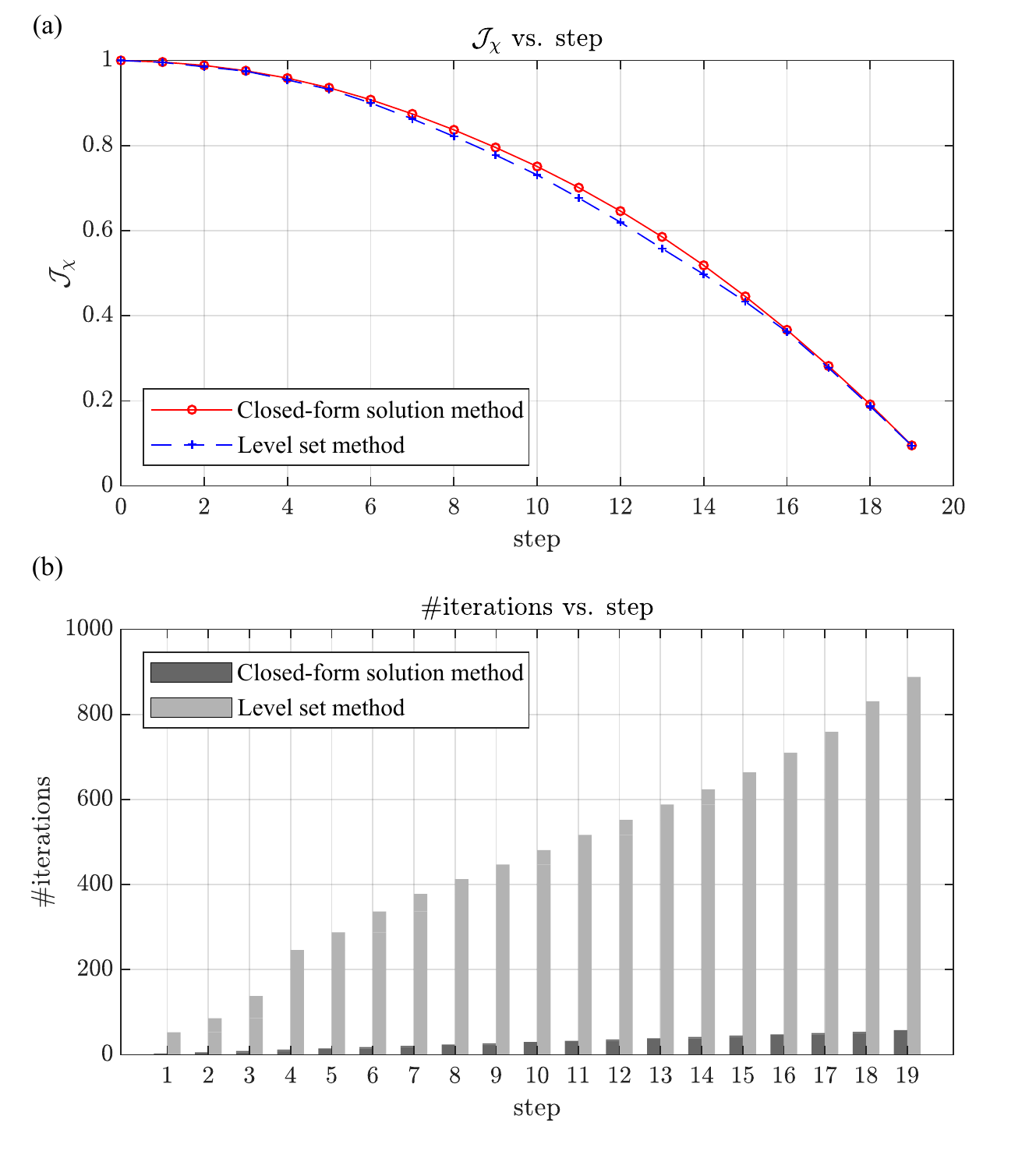}
	\caption{Thermal heat conductor. Non-smooth variational closed-form method vs level set method: (a) Cost-function evolution, and (b) Computational cost in terms of the number of iterations.}
	\label{fig_Bi_HJ_comparison}
\end{figure*}

The results of the comparison, as for the cost function is concerned, are depicted in Figure \ref{fig_Bi_HJ_comparison}. The cost function evolution, displayed in Figure \ref{fig_Bi_HJ_comparison}-(a), shows close results for both methods, although the result for some steps may be slightly different. However, significant improvements, in terms of the total computational cost, are obtained using the closed-form solutions of the proposed approach, with respect to level set method. This is represented in Figure \ref{fig_Bi_HJ_comparison}-(b), where the accumulative number of iterations is illustrated. From these results, it can be concluded that the nonsmooth variational approach, is more than an order of magnitude (up to 15 times) faster than the level set method, while obtaining similar results in terms of optimal topologies and cost function. Moreover, the computational cost (number of required iterations) seems to be uniform along the steps for the \emph{nonsmooth closed-form solution approach}.

\section{Concluding remarks} \label{sec_conclusion}

In this paper, the nonsmooth variational approach to relaxed topology optimization, proposed in \citet{Oliver2019} for structural problems, has been extended and applied to solve thermal topology optimization problems involving the analysis of 3D heat conducting components and thermal cloaking devices. From this work the following conclusions cab be displayed:
\begin{itemize}
	\item[$\bullet$] The RVA technique can be readily extended from structural problems to thermal ones. One, evident, reason for this is that, in spite that the physics, and technical applications in both sets of problems are very different, the mathematical settings in which they are inserted are similar. However, problems like thermal cloaking, tackled in this work, which have not a clear counterpart in structural analysis, have been successfully solved here.
	
	\item[$\bullet$] The \emph{Cutting}\&\emph{Bisection} technique used to solve the resulting, fixed point algebraic closed-form, equations has been tested here beyond the original structural scenario, in which they were  overall positive or negative. Here, the technique has proven to efficiently work both for constant-sign \emph{energy densities} (Section \ref{sec_structural_compliance_problems}) but, also, in sign-changing cases (Sections \ref{sec_thermal_cloaking} and \ref{sec_temp_var_min}). This dissipates one of the unknowns pending on this subject. The success of this algorithm strongly relies on the unique-valued character of the energy functions, $\xi$, as it happens in all considered problems of this work.
	
	\item[$\bullet$] As in the structural problems case, the obtainment of the  \emph{closed-form optimality criteria solutions} only requires the \emph{formulation of the cost function}, the corresponding \emph{energy density}, and a \emph{pseudo-time}  (vo\-lume-driven) \emph{advancing scheme}. The Relaxed Topological Derivative, as sensitivity for the optimization problem, can be systematically and simply derived via the classical \emph{adjoint method}, as proven in the presented applications.  
	
	\item[$\bullet$] The presented numerical examples confirm that the proposed approach provides \emph{smooth black-and-white topology designs}, also for thermal optimization problems. Mesh-size dependency and checkerboards effects are effectively removed by the the minimum material filament size control via the \emph{Laplacian smoothing technique}, so that post-process filtering algorithms are not necessary. 
	
	\item[$\bullet$] In Sections \ref{sec_thermal_cloaking_flux_device} and \ref{sec_thermal_cloaking_device} the approach proves amenable to achieve complex non-trivial topology layouts, far from being intuitive, and even impossible to obtain without suitable numerical computational methods.
	
	\item[$\bullet$] In alignment with what was reported in \cite{Oliver2019} for structural optimization, the computational cost of the considered method for thermal optimization problems turns out to be much smaller (more than 15 times for the test considered here) when compared with an, equivalent, level set method (Ha\-milton-Jacobi update scheme based on the same Relaxed Topology Derivative).
\end{itemize}
	
In summary, the considered topological optimization methodology, based on
\begin{itemize}
		\item[$1)$] Optimizing the distribution of the nonsmooth characteristic function in a variational setting, 
		\item[$2)$] Resorting the easy-to-derive Relaxed Topological Derivative as sensitivity, and 
		\item[$3)$] 
		Obtaining closed-form optimality criteria, to be numerically solved using a robust \emph{Cutting}\&\emph{Bisection} algorithm, in a pseudo-time advancing scheme.
\end{itemize}		
		 
When applied to complex thermal problems, the proposed methodology exhibits the same encouraging features than in structural problems. Its extension to other families of topology optimization problems is an ongoing research that will be presented in future works.

\begin{acknowledgements}
This research has received funding from the European Research Council (ERC) under the European Union’s Horizon 2020 research and innovation programme (Proof of Concept Grant agreement n 874481) through the project “Computational design and prototyping of acoustic metamaterials for target ambient noise reduction” (METACOUSTIC).
The authors also acknowledge financial support from the Spanish Ministry of Economy and Competitiveness, through the research grant DPI2017-85521-P for the project “Computational design of Acoustic and Mechanical Metamaterials” (METAMAT) and through the “Severo Ochoa Programme for Centres of Excellence in R\&D” (CEX2018-000797-S).
D. Yago acknowledges the support received from the Spanish Ministry of Education through the FPU program for PhD grants. 
\end{acknowledgements}

\begin{appendices}

	\numberwithin{equation}{section}
	
	\section{Finite element discretization} \label{sec_finite_element_discretization}
		
	The finite element method (FEM) is used to discretize and solve the state-equation (\ref{eq_weak_problem2}) and the required adjoint problems. The temperature field in $\Omega$ is approximated via $C_0$ shape functions as follows\footnote{Voigt's vector/matrix notation is used in what follows.}:
	\begin{equation} \label{eq_shape_temp}
		\mathbf{\theta}_{\chi}\bbx\equiv\mathbf{N}_{\theta}\bbx\hat{\pmb{\theta}}_\chi
	\end{equation}
	where $\mathbf{N}_{\theta}\bbx$ is the, temperature, shape-function matrix and $\hat{\pmb{\theta}}_\chi$ corresponds to the nodal temperature vector. Equivalently, the gradient of $\mathbf{\theta}_{\chi}\bbx$ is expressed as
	\begin{equation} \label{eq_grad_temp}
		{\bm{\nabla} \theta}_{\chi}\bbx\equiv\mathbf{B}\bbx\hat{\pmb{\theta}}_\chi
	\end{equation}
	where $\mathbf{B}\bbx$ denotes the gradient matrix. Then, introducing expressions (\ref{eq_shape_temp}) and (\ref{eq_grad_temp}) into the Fourier's law, the heat flux, $\bf{q}_{\chi}\bbx$, can be written as
	\begin{equation} \label{eq_fourier_fem}
		\bm{q}_{\chi}\bbx\equiv-{\Kappalarge}_{\chi}\bbx \ \mathbf{B}\bbx\hat{\pmb{\theta}}_\chi \mdot
	\end{equation}
		
	Finally, the state equation (\ref{eq_weak_problem2}), once the previous expressions are replaced, yields to
	\begin{equation} \label{eq_equilibrium}
		{\mathbb K}_{\chi}\hat{\pmb{\theta}}_{\chi}=\mathbf{f} 
	\end{equation}
	with
	\begin{equation} \label{eq_equilibrium_stiff_force}
		\left\{
		\begin{split} 
			&\begin{split}
				{\mathbb K}_{\chi}=&\int_{\Omega}\mathbf{B}^T\bbx\ {\Kappalarge}_{\chi}\bbx\ \mathbf{B}\bbx\,d{\Omega}-\\
							&-\int_{\partial_h\Omega}\mathbf{N_{\theta}}^T\bbx h\mathbf{N_{\theta}}\bbx\dgamma
			\end{split} \\
			&\begin{split}
				\mathbf{f}=&\int_{\Omega} \mathbf{N_{\theta}}^T\bbx \hS_{\chi}\bbx\domega -\\
								&-\int_{\partial_{q}\Omega}\mathbf{N_{\theta}}^T\bbx{\overline{q}}\bbx\dgamma -\\
								&-\int_{\partial_h\Omega}\mathbf{N_{\theta}}^T\bbx h{\theta_{amb}\bbx}\dgamma
			\end{split}
		\end{split} \right. \mcolon
	\end{equation}
	where $\mathbb{K}_{\chi}$ and $\mathbf{f}$ stand for the stiffness matrix and the external forces vector, respectively.\footnote{From now on, the sub-index $\theta$ of ${\bf N}_{\theta}$ shall be omitted.}
	
	A Laplacian smoothing is used to smooth the topology, control the filament size and avoid checkerboard patterns. The smooth discrimination function, $\psi_\tau$, corresponds to the solution of
	\begin{equation} \label{eq_regularized_laplacean_smoothing}
		\left\{
		\begin{split}
			&\psi_{\tau}\bbx-\epsilon^2\Delta_{\bf x}\psi_{\tau}\bbx=\psi\bbx& &\quad in \; \Omega\\
			&\nabla_{\bf x}\psi_{\tau}\bbx\cdot\mathbf{n}={0}& &\quad on \; \partial\Omega
		\end{split}
		\right. \mcolon
	\end{equation}
	where, $\Delta_{\bf x}({\bf x},\cdot)$ and $\nabla_{\bf x}({\bf x},\cdot)$ stand for the Laplacian and gradient operators, respectively, and $\mathbf{n}$ is the outwards normal to the boundary of the analysis domain, $\partial\Omega$. The FE discretization of equation (\ref{eq_regularized_laplacean_smoothing}), considering $\psi_\tau\bbx=\mathbf{N}\bbx{\bf\hat{\mathbf \psi}_\tau}$, leads to the following system
	\begin{equation} \label{eq_discrete_laplacean_smoothing}
		\hat{\mathbf \psi}_\tau={\tilde{\mathbb G}}^{-1}{\mathbf f}(\psi)	
	\end{equation}
	with
	\begin{equation}
		\left\{
		\begin{split}
			&\tilde{\mathbb G}=\tilde{\mathbb M}+{\epsilon^2}\tilde{\mathbb K} \quad \rightarrow \\
			&\quad \rightarrow \quad
			\left\{
				\begin{split}
					&\tilde{\mathbb M}=\int_{\Omega}{\mathbf N}^T\bbx\mathbf{N}\bbx\domega ;\;\\
					&\tilde{\mathbb K}=\int_{\Omega}\nabla {\mathbf N}^T\bbx{\nabla {\mathbf N}\bbx \domega} ;\;\\
				\end{split}
				\right. &(a) \\
			&{\mathbf f}(\psi)=\int_\Omega{\mathbf{N}^T\bbx\psi\bbx\domega} &(b)
		\end{split} \right.
	\end{equation}
	where $\mathbf{N}(\bf{x})$ stands for the standard interpolation matrix and $\hat{\mathbf \psi}_\tau$ is the vector of nodal values of the field $\psi_\tau\bbx$. 
	
	\section{Thermal compliance minimization: cost function derivative} \label{App_thermal_complaince}
		The topological sensitivity of the thermal compliance optimization problem (equation (\ref{eq_discrete_form})) is computed in detail in this section via the \emph{adjoint method} and the Relaxed Topological Derivative (RTD). Let first rephrase the objective function, ${\cal J}^{(h_e)}(\chi)$, to incorporate the state equation (\ref{eq_equilibrium})
		\begin{equation} \label{eq_rephrased0}
			\overline{\cal J}^{(h_e)}(\chi)=\frac{1}{2} \mathbf{f}^{T}\hat{\pmb{\theta}}_{\chi}-
							\hat{\mathbf{w}}^T \underbrace{\left( {\mathbb K}_{\chi}\hat{\pmb{\theta}}_{\chi}-\mathbf{f}\right)}_{\textstyle{=\mathbf{0}}} \mcolon
		\end{equation}
		where $\hat{\mathbf{w}}$ corresponds to the solution of the \emph{adjoint state problem}, as aforementioned. Computing the RTD of equation (\ref{eq_rephrased0}) and reordering terms, one arrives to
		\begin{equation}\label{eq_diferentiation0}
			\begin{split}
				&\dfrac{\delta\overline{\cal J}^{(h_e)}(\chi)}{\delta\chi}\bbxhat=		\left(\frac{1}{2}\mathbf{f}^{T}-\hat{\mathbf{w}}^T{\mathbb K}_{\chi}\right)\dfrac{\delta\hat{\pmb{\theta}}_{\chi}}{\delta{\chi}}\bbxhat+ \\
				&\hspace{0.25cm}+ \left(\frac{1}{2}\dfrac{\delta\mathbf{f}^T _{\chi}}{\delta{\chi}}\bbxhat\hat{\pmb{\theta}}_{\chi}-\hat{\mathbf{w}}^T\dfrac{\delta {\mathbb K}_{\chi}}{\delta\chi}\bbxhat\hat{\pmb{\theta}}_{\chi} + \hat{\mathbf{w}}^T\dfrac{\delta\mathbf{f}_{\chi}}{\delta{\chi}}\bbxhat\right) \mdot
			\end{split}
		\end{equation}
		
		Substituting $\hat{\mathbf{w}}\equiv\dfrac{1}{2}\hat{\pmb{\theta}}_{\chi}$ in equation (\ref{eq_diferentiation0}), and considering the state equation (\ref{eq_equilibrium}), the expression can be simplified to
		\begin{equation} \label{eq_diferentiation01}
			\begin{split}
				\dfrac{\delta\overline{\cal J}^{(h_e)}(\chi)}{\delta\chi}\bbxhat=&\frac{1}{2}\underbrace{(\mathbf{f}^{T}-\hat{\pmb{\theta}}_{\chi}^T{\mathbb K}_{\chi})}_{\textstyle{=\mathbf{0}}}\dfrac{\delta\hat{\pmb{\theta}}_{\chi}}{\delta{\chi}}\bbxhat+\\
				&+\left(\dfrac{\delta\mathbf{f}^T _{\chi}}{\delta{\chi}}\bbxhat\hat{\pmb{\theta}}_{\chi}-\hat{\pmb{\theta}}_{\chi}^T\dfrac{\delta {\mathbb K}_{\chi}}{\delta\chi}\bbxhat\hat{\pmb{\theta}}_{\chi}\right)=\\
				=&\left[\dfrac{\delta\mathbf{f}^T _{\chi}}{\delta{\chi}}\bbxhat\hat{\pmb{\theta}}_{\chi}- \hat{\pmb{\theta}}_{\chi}^T\dfrac{\delta {\mathbb K}_{\chi}}{\delta\chi}\bbx\hat{\pmb{\theta}}_{\chi}\right]_{\mathbf{x}=\hat{\mathbf{x}}} \mdot
			\end{split}
		\end{equation}
		
		Then, considering equations (\ref{eq_conductivity_interp})-(\ref{eq_chi_heat_source}) and replacing the corresponding terms into equation (\ref{eq_diferentiation01}), the Relaxed Topological Derivative of equation (\ref{eq_rephrased0}) can be expressed as 
		\begin{equation}\label{eq_final_derivative0}
			\begin{split}
				&\dfrac{\delta\overline{\cal J}^{(h_e)}(\chi)}{\delta\chi}\bbxhat
				=\dfrac{\partial {\hS_{\chi}}}{\partial\chi}\bbxhat\mathbf{N}\bbxhat{\hat{\pmb{\theta}}}_{\chi} \Delta\chi_{_{\shS}}\bbxhat- \\
				 &\hspace{1cm}-{\hat{\pmb{\theta}}}^T_{\chi}\mathbf{B}^T\bbxhat\dfrac{\partial {\Kappalarge_{\chi}}}{\partial\chi}\bbxhat\mathbf{B}\bbxhat{\hat{\pmb{\theta}}}_{\chi}\Delta\chi_{\kappa}\bbxhat=\\
				&\hspace{1cm}=\left[\dfrac{\partial {\hS_{\chi}}}{\partial\chi}\mathbf{N}\bbx{\hat{\pmb{\theta}}}_{\chi}  \right]_{\mathbf{x}=\hat{\mathbf{x}}}\Delta\chi_{_{\shS}}\bbxhat -\\
				&\hspace{1cm}-\left[ \bm{\nabla}\theta^T_{\chi}\bbx\dfrac{\partial \Kappalarge_{\chi}}{\partial{\chi}}\bm{\nabla}\theta_{\chi}\bbx\right]_{\mathbf{x}=\hat{\mathbf{x}}}\Delta\chi_{\kappa}\bbxhat=\\
				&\hspace{1cm}=\left[m_{\shS}{\chi}^{m_{\shS}-1}\bbx \hS\bbx\mathbf{N}\bbx{\hat{\pmb{\theta}}}_{\chi}  \right]_{\mathbf{x}=\hat{\mathbf{x}}}\Delta\chi_{_{\shS}}\bbxhat-\\
				&\hspace{1cm}-\left[ m_{\kappa}{\chi}^{m_{\kappa}-1}\bbx\bm{\nabla}\theta^T_{\chi}\bbx \Kappalarge\bbx\bm{\nabla}\theta_{\chi}\bbx\right]_{\mathbf{x}=\hat{\mathbf{x}}}{{\Delta\chi_{\kappa}}({\hat{\bf x}}) }  \mcolon
			\end{split} 
		\end{equation}
		which is then written in terms of \emph{energy densities}, to recover equation (\ref{eq_compliance_topological_derivative}), as
		\begin{equation} 
			\begin{split}
				\dfrac{\delta{\overline{\cal J}^{(h_e)}(\theta_{\chi})}}{\delta\chi}\bbxhat 
							=&m_{\shS}\left({\chi}_{\shS}\bbxhat\right) ^{m_{\shS}-1}{\overline{\cal U}_{\shS}}\bbxhat\Delta\chi_{_{\shS}}\bbxhat-\\
							&-2m_{\kappa} \left(\chi_{\kappa}\bbxhat\right) ^{m_{\kappa}-1}{\overline{\cal U}}\bbxhat{{\Delta\chi}_{\kappa}({\hat{\bf x}}) }
							\mcolon
			\end{split}
		\end{equation}
		where $\overline{\cal U}\bbxhat$ is \textit{\Uconductivity} and ${\overline{\cal U}_{\shS}}\bbxhat$ is \textit{\Usource}, as described in equation (\ref{eq_compliance_energies}).

	\section{Thermal cloaking via heat flux manipulation: cost function derivative} \label{App_flux_cloaking}
		This section describes step-by-step the topological sensitivity computation of the thermal cloaking optimization problem (\ref{eq_cost_cloaking}), mimicking the procedure explained in Appendix \ref{App_thermal_complaince}. Let us then define the extended cost function, $\overline{{\cal J}}^{(h_e)}(\chi)$, i.e. 
		\begin{equation} \label{eq_rephrased_1_app}
			\begin{split}
				\overline{{\cal J}}^{(h_e)}(\chi)=&\Bigg(\underbrace{\int_{\Omega} {1_{\Omega_c}\bbx}\left|{\bf q}_{\chi}\left({\bf x},{\theta}_\chi^{(1)}\right)-{\overline{\bf q}}\bbx\right|^2 \domega}_{E\left(\chi,{\theta}_\chi^{(1)}\right)}\Bigg)^\frac{1}{2}- \\
				&-\hat{\bf w}^T\underbrace{\left( {\mathbb K}_{\chi}\hat{\pmb{\theta}}_{\chi}^{(1)}-\mathbf{f}^{(1)}\right)}_{\textstyle{=\mathbf{0}}} \mcolon
			\end{split}
		\end{equation}
		which is subsequently derived through the RTD, yielding to
		\begin{equation}\label{eq_diferentiation_1_app}
			\begin{split}
				\dfrac{\delta\overline{\cal J}^{(h_e)}(\chi)}{\delta\chi}\bbxhat=
				&\frac{1}{2}\dfrac{1}{{\cal J}^{(h_e)}(\chi)}\dfrac{\delta E(\chi)}{\delta\chi}\bbxhat - \\
				&-\hat{\mathbf{w}}^T\dfrac{\delta {\mathbb K}_{\chi}}{\delta\chi}\bbxhat\hat{\pmb{\theta}}_{\chi}^{(1)} - \\
				&-\hat{\mathbf{w}}^T{\mathbb K}_{\chi}\dfrac{\delta\hat{\pmb{\theta}}_{\chi}^{(1)}}{\delta{\chi}}\bbxhat+
				\hat{\mathbf{w}}^T\dfrac{\delta\mathbf{f}_{\chi}^{(1)}}{\delta{\chi}}\bbxhat 
			\end{split}
		\end{equation}
		where
		\begin{equation}\label{eq_diferentiation_12}
			\left\{
			\begin{split}
				& \dfrac{\delta E(\chi)}{\delta\chi}\bbxhat = 
				\left[ 2 \ \resizebox{0.6\hsize}{!}{%
								${1_{\Omega_c}\bbx}\left({\bf q}_{\chi}({\bf x},{\theta}_\chi^{(1)})-{\overline{\bf q}}\bbx\right)\dfrac{\delta {\bf q}_\chi(\chi)}{\delta\chi}\bbx$} \right]_{\bx=\bxhat} \mcolon\\
				& \dfrac{\delta {\bf q}_\chi(\chi)}{\delta\chi}\bbxhat = 
				-\dfrac{\delta {\Kappalarge}_{\chi}(\chi)}{\delta\chi}\bbxhat{\bm{\nabla}\theta}_{\chi}^{(1)}\bbxhat-{\Kappalarge}_{\chi}\nabla\dfrac{\delta {\pmb{\theta}}_{\chi} ^{(1)}}{\delta\chi}\bbxhat \mdot
			\end{split} \right.
		\end{equation}
		Introducing expressions (\ref{eq_diferentiation_12}) into equation (\ref{eq_diferentiation_1_app}), and manipulating the terms, we obtain
		\begin{equation} \label{eq_diferentiation_1_app2}
			\begin{split}	
				\dfrac{\delta\overline{\cal J}^{(h_e)}(\chi)}{\delta\chi}\bbxhat=&\resizebox{0.55\hsize}{!}{%
												$\underbrace{\left(-\hat{\mathbf{w}}^T{\mathbb K}_{\chi}-{\bf{C_1}}\left(\chi,\bxhat,{\theta}_\chi^{(1)}\right){\Kappalarge}_{\chi}\nabla 
				\right)}_{\displaystyle {\bf =0}}$} \dfrac{\delta{\hat{\pmb{\theta}}}_{\chi}^{(1)}}{\delta{\chi}}\bbxhat-\\
				&-{\bf{C_1}}\left(\chi,\bxhat,{\theta}_\chi^{(1)}\right)
				\dfrac{\delta {\Kappalarge}_{\chi}(\chi)}{\delta\chi}\bbxhat{\bm{\nabla}\theta}^{(1)}_{\chi}\bbxhat- \\
				&-\hat{\mathbf{w}}^T\dfrac{\delta {\mathbb K}_{\chi}}{\delta\chi}\bbxhat\hat{\pmb{\theta}}_{\chi}^{(1)}
				+\hat{\mathbf{w}}^T\dfrac{\delta\mathbf{f}_{\chi}^{(1)}}{\delta{\chi}}\bbxhat \mcolon
			\end{split}
		\end{equation}
		with
		\begin{equation}
			{\bf{C_1}}\left(\chi,\bxhat,{\theta}_\chi^{(1)}\right) = \dfrac{{1_{\Omega_c}\bbxhat} \left({\bf q}_{\chi}\left(\bxhat,{\theta}_\chi^{(1)}\right)-{\overline{\bf q}}\bbxhat\right)}{{\cal J}^{(h_e)}(\chi)} \mdot
		\end{equation}
		
		Now, the adjoint problem of equation (\ref{eq_diferentiation_1_app2}) is solved for $\hat{\bf{w}}\equiv\hat{\pmb{\theta}}_{\chi}^{(2)}$, leading to
		\begin{equation}\label{eq_diferentiation_2_app}
			\begin{split}
				\dfrac{\delta\overline{\cal J}^{(h_e)}(\chi)}{\delta\chi}&\bbxhat=
					-{\bf{C_1}}\left(\chi,\bxhat,{\theta}_\chi^{(1)}\right)\dfrac{\delta {\Kappalarge}_{\chi}(\chi)}{\delta\chi}\bbxhat{\bm{\nabla}\theta}_{\chi}^{(1)}\bbxhat- \\
					&-\left(\hat{\pmb{\theta}}_{\chi}^{(2)}\right)^T\dfrac{\delta {\mathbb K}_{\chi}}{\delta\chi}\bbxhat\hat{\pmb{\theta}}_{\chi}^{(1)}
					+\left(\hat{\pmb{\theta}}_{\chi}^{(2)}\right)^T\dfrac{\delta\mathbf{f}_{\chi}^{(1)}}{\delta{\chi}}\bbxhat \mdot
			\end{split}
		\end{equation}	
		
		After applying the RTD to the corresponding terms, equation (\ref{eq_diferentiation_2_app}) reads as
		\begin{align}\label{eq_final_derivative_1_app}
			&\dfrac{\delta\overline{\cal J}^{(h_e)}(\chi)}{\delta\chi}\bbxhat
				=\left[\left(\hat{\pmb{\theta}}_{\chi}^{(2)}\right)^T\mathbf{N}^T\bbx\dfrac{\partial {\hS_{\chi}}}{\partial\chi}\bbx\right]_{\mathbf{x}=\hat{\mathbf{x}}} \hspace{-0.6cm}\Delta\chi_{_{\shS}}\bbxhat- \nonumber\\
				&\hspace{0.5cm}-\left[\left(\hat{\pmb{\theta}}_{\chi}^{(1)}\right)^T\mathbf{B}^T\bbx\dfrac{\partial {\Kappalarge_{\chi}}}{\partial\chi}\bbx\mathbf{B}\bbx\hat{\pmb{\theta}}_{\chi}^{(2)}\right]_{\mathbf{x}=\hat{\mathbf{x}}}\hspace{-0.6cm}\Delta\chi_{\kappa}\bbxhat-\nonumber\\
				&\hspace{0.5cm}-\left[{\bf{C_1}}\left(\chi,{\bf x},{\theta}_\chi^{(1)}\right)\dfrac{\partial {\Kappalarge}_{\chi}}{\partial\chi}\bbx\mathbf{B}\bbx\hat{\pmb{\theta}}_{\chi}^{(1)}\right]_{\mathbf{x}=\hat{\mathbf{x}}}\hspace{-0.6cm}\Delta\chi_{\kappa}\bbxhat \mdot		
		\end{align}
		Subsequently, relations (\ref{eq_conductivity_interp}) and (\ref{eq_heat_sources_interp}) are considered in equation (\ref{eq_final_derivative_1_app}), which yields to
		\begin{align}\label{eq_final_derivative_1_app2}
			&\dfrac{\delta\overline{\cal J}^{(h_e)}(\chi)}{\delta\chi}\bbxhat
				=\left[m_{\shS}{\chi}^{m_{\shS}-1}\left(\hat{\pmb{\theta}}_{\chi}^{(2)}\right)^T\mathbf{N}^T\bbx \hS\bbx  \right]_{\mathbf{x}=\hat{\mathbf{x}}}\hspace{-0.6cm}\Delta\chi_{_{\shS}}\bbxhat- \nonumber\\
				&\hspace{0.5cm}-\left[ m_{\kappa}{\chi}^{m_{\kappa}-1}\left(\bm{\nabla}\theta^{(1)}_{\chi}\right)^T\bbx \Kappalarge\bbx\bm{\nabla}\theta_{\chi}^{(2)}\bbx\right]_{\mathbf{x}=\hat{\mathbf{x}}}{{\hspace{-0.6cm}\Delta\chi_{\kappa}}({\hat{\bf x}}) }- \nonumber\\
				&\hspace{0.5cm}-\left[m_{\kappa}{\chi}^{m_{\kappa}-1}{\bf{C_1}}\left(\chi,{\bf x},{\theta}_\chi^{(1)}\right){\Kappalarge}\bbx\mathbf{B}\bbx\hat{\pmb{\theta}}_{\chi}^{(1)}\right]_{\mathbf{x}=\hat{\mathbf{x}}}\hspace{-0.6cm}\Delta\chi_{\kappa}\bbxhat \mdot
		\end{align}
		
		Finally, equation (\ref{eq_final_derivative_1_app2}) can be reformulated, in terms of \emph{pseudo-energies}, as
		\begin{equation}
			\begin{split}
				\dfrac{\delta\overline{\cal J}^{(h_e)}(\chi)}{\delta\chi}\bbxhat&=
									m_{\shS}\left({\chi}_{\shS}\bbxhat\right) ^{m_{\shS}-1}{\overline{\cal U}_{\shS}}\bbxhat\Delta\chi_{_{\shS}}\bbxhat- \\
									&-2m_{\kappa} \left(\chi_{\kappa}\bbxhat\right) ^{m_{\kappa}-1}{\overline{\cal U}}_{1-2}\bbxhat{{\Delta\chi}_{\kappa}({\hat{\bf x}}) }- \\
									&-m_{\kappa} \left(\chi_{\kappa}\bbxhat\right) ^{m_{\kappa}-1}{\overline{\cal U}}_{\bf q}\bbxhat{{\Delta\chi}_{\kappa}({\hat{\bf x}}) } \mcolon
			\end{split}
		\end{equation}
		where $\overline{\cal U}_{1-2}\bbxhat$ is \textit{\Uconductivity}, ${\overline{\cal U}_{\shS}}\bbxhat$ is \textit{\Usource} and ${\overline{\cal U}}_{\bf q}\bbxhat$ corresponds to \textit{\Uflux}, as defined in equation (\ref{eq_energies_cloaking}).

	\section{Average temperature minimization: cost function derivative} \label{App_temp_min}
		Let us now proceed with the computation of the topological sensitivity of the average temperature minimization problem (\ref{eq_temperature_min}). As before, let $\overline{\cal J}^{(h_e)}_{\text{av}}(\chi)$ be the extended cost function, considering the \emph{state equation} through the Lagrange multiplier vector, $\hat{\mathbf{w}}$, defined as
		\begin{equation} \label{eq_rephrased_3_app}
			\overline{\cal J}^{(h_e)}_{\text{av}}(\chi)=C_2 {\bf 1}_{\partial_c\Omega}^T\hat{\pmb{\theta}}_\chi^{(1)}-\hat{\bf w}^T\underbrace{\left( {\mathbb K}_{\chi}\hat{\pmb{\theta}}_{\chi}^{(1)}-\mathbf{f}^{(1)}\right)}_{\textstyle{=\mathbf{0}}} \mcolon
		\end{equation}
		where $C_2=\left(\int_{\partial_c\Omega} \dgamma\right)^{-1}$.
		
		Applying the RTD to equation (\ref{eq_rephrased_3_app}) and reordering its terms, one obtains
		\begin{equation}\label{eq_diferentiation_5_app}
			\begin{split}
				\dfrac{\delta\overline{\cal J}^{(h_e)}_{\text{av}}(\chi)}{\delta\chi}\bbxhat=&
							\left(-\hat{\mathbf{w}}^T{\mathbb K}_{\chi}+C_2 {\bf 1}_{\partial_c\Omega}^T \right)\dfrac{\delta{\hat{\pmb{\theta}}}_{\chi}^{(1)}}{\delta{\chi}}\bbxhat- \\
							-&\hat{\mathbf{w}}^T\dfrac{\delta {\mathbb K}_{\chi}}{\delta\chi}\bbxhat\hat{\pmb{\theta}}_{\chi}^{(1)}
							+\hat{\mathbf{w}}^T\dfrac{\delta\mathbf{f}_{\chi} ^{(1)}}{\delta{\chi}}\bbxhat \mcolon
			\end{split}
		\end{equation}
		which is then simplified by choosing $\hat{\bf{w}}\equiv-C_2 \hat{\pmb{\theta}}_{\chi}^{(2)}$, yielding to
		\begin{equation}\label{eq_diferentiation_6_app}
			\begin{split}
				\dfrac{\delta\overline{\cal J}^{(h_e)}_{\text{av}}(\chi)}{\delta\chi}\bbxhat=&
				C_2 \underbrace{\left(\left(\hat{\pmb{\theta}}_{\chi}^{(2)}\right)^T{\mathbb K}_{\chi}+{\bf 1}_{\partial_c\Omega}^T \right)}_{\displaystyle {\bf =0}}\dfrac{\delta{\hat{\pmb{\theta}}}_{\chi}^{(1)}}{\delta{\chi}}\bbxhat+ \\
				&+C_2 \left(\hat{\pmb{\theta}}_{\chi}^{(2)}\right)^T\dfrac{\delta {\mathbb K}_{\chi}}{\delta\chi}\bbxhat\hat{\pmb{\theta}}_{\chi}^{(1)} -\\
				&-C_2 \left(\hat{\pmb{\theta}}_{\chi}^{(2)}\right)^T\dfrac{\delta\mathbf{f}_{\chi}^{(1)}}{\delta{\chi}}\bbxhat=\\
				=&C_2 \Bigg(\left(\hat{\pmb{\theta}}_{\chi}^{(2)}\right)^T\dfrac{\delta {\mathbb K}_{\chi}}{\delta\chi}\bbxhat\hat{\pmb{\theta}}_{\chi}^{(1)} -\\
				&\hspace{1cm}- \left(\hat{\pmb{\theta}}_{\chi}^{(2)}\right)^T\dfrac{\delta\mathbf{f}_{\chi}^{(1)}}{\delta{\chi}}\bbxhat \Bigg) \mdot
			\end{split}
		\end{equation}
		
		Equation (\ref{eq_diferentiation_6_app}) is finally discretized using the expressions in Section \ref{sec_finite_element_discretization}, which then reads as
		\begin{align}\label{eq_final_derivative_3_app}
			&\dfrac{\delta\overline{\cal J}^{(h_e)}_{\text{av}}(\chi)}{\delta\chi}\bbxhat
				=\resizebox{0.7\hsize}{!}{%
						$C_2 \left(\hat{\pmb{\theta}}_{\chi}^{(2)}\right)^T\mathbf{B}^T\bbxhat\dfrac{\partial {\Kappalarge_{\chi}}}{\partial\chi}\bbxhat\mathbf{B}\bbxhat\hat{\pmb{\theta}}_{\chi}^{(1)}\Delta\chi_{\kappa}\bbxhat$}- \nonumber\\
			&-C_2 \left(\hat{\pmb{\theta}}_{\chi}^{(2)}\right)^T\mathbf{N}^T\bbxhat\dfrac{\partial {\hS_{\chi}}}{\partial\chi}\bbxhat \Delta\chi_{_{\shS}}\bbxhat=\nonumber\\
			&=\resizebox{0.9\hsize}{!}{%
										$C_2 \left[ m_{\kappa}{\chi}^{m_{\kappa}-1}\bbx\left(\hat{\pmb{\theta}}_{\chi}^{(2)}\right)^T\mathbf{B}^T\bbx \Kappalarge\bbx\mathbf{B}\bbx\hat{\pmb{\theta}}_{\chi}^{(1)}\right]_{\mathbf{x}=\hat{\mathbf{x}}}{{\hspace{-0.6cm}\Delta\chi_{\kappa}}({\hat{\bf x}}) }$}- \nonumber\\
			&-C_2 \left[m_{\shS}{\chi}^{m_{\shS}-1}\bbx\left(\hat{\pmb{\theta}}_{\chi}^{(2)}\right)^T\mathbf{N}^T\bbx \hS\bbx  \right]_{\mathbf{x}=\hat{\mathbf{x}}}\hspace{-0.6cm}\Delta\chi_{_{\shS}}\bbxhat \mdot	
		\end{align}
		
		The Relaxed Topological Derivative of the cost function (\ref{eq_temperature_min}) can be finally expressed in terms of \emph{energy densities} as
		\begin{equation}
			\begin{split}
				\dfrac{\delta\overline{\cal J}^{(h_e)}_{\text{av}}(\chi)}{\delta\chi}\bbxhat 
								=&2 C_2 m_{\kappa} \left(\chi_{\kappa}\bbxhat\right) ^{m_{\kappa}-1}{\overline{\cal U}}_{1-2}\bbxhat{{\Delta\chi}_{\kappa}({\hat{\bf x}}) } -\\
								&-C_2 m_{\shS}\left({\chi}_{\shS}\bbxhat\right) ^{m_{\shS}-1}{\overline{\cal U}_{\shS-2}}\bbxhat\Delta\chi_{_{\shS}}\bbxhat \mcolon
			\end{split}
		\end{equation}
		where $\overline{\cal U}_{1-2}\bbxhat$ and ${\overline{\cal U}_{\shS-2}}\bbxhat$ are, respectively, \textit{\Uconductivity} and \textit{\Usource}, both defined in equation (\ref{eq_energy_average_temp}).

	\section{Temperature variance minimization: cost function derivation} \label{App_temp_var_min}
		Let us now address the corresponding RTD computation of the cost function for the minimization of the temperature variance (equation (\ref{eq_temperature_var_min2})), starting by defining the extended cost function as 
		\begin{equation} \label{eq_rephrased_4_app}
			\begin{split}
				\overline{\cal J}^{(h_e)}_{\text{vr}}(\chi)=&C_3 \left({\mathcal T}_\chi\left(\theta_\chi^{(1)}\right)\right)^T \mathbb{M}_{\partial_c\Omega}  {\mathcal T}_\chi\left(\theta_\chi^{(1)}\right)- \\
				&-\hat{\bf w}^T\underbrace{\left( {\mathbb K}_{\chi}\hat{\pmb{\theta}}_{\chi}^{(1)}-\mathbf{f}^{(1)}\right)}_{\textstyle{=\mathbf{0}}} \mcolon
			\end{split}
		\end{equation}
		where ${\mathcal T}_\chi\left(\theta_\chi^{(1)}\right)$ and $\mathbb{M}_{\partial_c\Omega}$ are respectively defined as
		\begin{align*}
			& {\mathcal T}_\chi\left(\theta_\chi^{(1)}\right) = \hat{\pmb{\theta}}_\chi^{(1)}-{\mathbb{I}}\,{\cal J}^{(h_e)}_{\text{av}}\left(\theta_\chi^{(1)}\right) \mcolon\\
			& \mathbb{M}_{\partial_c\Omega} = \int_{\partial\Omega} {\bf N}^T\bbx {1}_{\partial_c\Omega} \bbx {\bf N}\bbx \dgamma \mdot
		\end{align*}
		
		Applying the RTD to equation (\ref{eq_rephrased_4_app}) and rearranging the expression, one arrives to
		\begin{equation}\label{eq_diferentiation_7_app}
			\begin{split}
				\dfrac{\delta\overline{\cal J}^{(h_e)}_{\text{vr}}(\chi)}{\delta\chi}\bbxhat
				=&\resizebox{0.58\hsize}{!}{%
					$\overbrace{\left(-\hat{\mathbf{w}}^T{\mathbb K}_{\chi}+2 C_3 \left({\mathcal T}_\chi\left(\theta_\chi^{(1)}\right)\right)^T \mathbb{M}_{\partial_c\Omega} \right)}^{\displaystyle {\bf =0}}$}\dfrac{\delta{\hat{\pmb{\theta}}}_{\chi}^{(1)}}{\delta{\chi}}\bbxhat- \\
				&-2 C_3 \left({\mathcal T}_\chi\left(\theta_\chi^{(1)}\right)\right)^T \mathbb{M}_{\partial_c\Omega} {\mathbb{I}} \dfrac{\delta{\cal J}^{(h_e)}_{\text{av}}(\chi)}{\delta\chi}\bbxhat- \\
				&-\hat{\mathbf{w}}^T\dfrac{\delta {\mathbb K}_{\chi}}{\delta\chi}\bbxhat\hat{\pmb{\theta}}_{\chi}^{(1)}
				+\hat{\mathbf{w}}^T\dfrac{\delta\mathbf{f}_{\chi}^{(1)}}{\delta{\chi}}\bbxhat \mdot
			\end{split}
		\end{equation}
		
		Then, the \emph{adjoint state equation} can be readily identified from equation (\ref{eq_diferentiation_7_app}) and solved for $\hat{\bf{w}}\equiv-C_3 \hat{\pmb{\theta}}_{\chi}^{(3)}$, resulting in
		\begin{equation}\label{eq_diferentiation_8_app}
			\begin{split}
				\dfrac{\delta\overline{\cal J}^{(h_e)}_{\text{vr}}(\chi)}{\delta\chi}\bbxhat=
							&-2 C_3 \left({\mathcal T}_\chi\left(\theta_\chi^{(1)}\right)\right)^T \mathbb{M}_{\partial_c\Omega} {\mathbb{I}} \dfrac{\delta{\cal J}^{(h_e)}_{\text{av}}(\chi)}{\delta\chi}\bbxhat+ \\
							&+C_3 \left(\hat{\pmb{\theta}}_{\chi}^{(3)}\right)^T\dfrac{\delta {\mathbb K}_{\chi}}{\delta\chi}\bbxhat\hat{\pmb{\theta}}_{\chi}^{(1)} \\
							&-C_3 \left(\hat{\pmb{\theta}}_{\chi}^{(3)}\right)^T\dfrac{\delta\mathbf{f}_{\chi}^{(1)}}{\delta{\chi}}\bbxhat \mcolon
			\end{split}
		\end{equation}
		which can be, after inserting the RTD of ${\cal J}^{(h_e)}_{\text{av}}(\chi)$ (\ref{eq_diferentiation_6_app}), expressed as	 
		\begin{equation} \label{eq_diferentiation_8_app2}
			\begin{split}
				&\dfrac{\delta\overline{\cal J}^{(h_e)}_{\text{vr}}(\chi)}{\delta\chi}\bbxhat=-2 C_3 \left({\mathcal T}_\chi\left(\theta_\chi^{(1)}\right)\right)^T \mathbb{M}_{\partial_c\Omega} {\mathbb{I}}\Bigg( \\
				&\hspace{1cm}C_2 \left(\hat{\pmb{\theta}}_{\chi}^{(2)}\right)^T\dfrac{\delta {\mathbb K}_{\chi}}{\delta\chi}\bbxhat\hat{\pmb{\theta}}_{\chi}^{(1)} 
				-C_2 \left(\hat{\pmb{\theta}}_{\chi}^{(2)}\right)^T\dfrac{\delta\mathbf{f}_{\chi}^{(1)}}{\delta{\chi}}\bbxhat\Bigg)+ \\
				&\hspace{1cm}+C_3 \left(\hat{\pmb{\theta}}_{\chi}^{(3)}\right)^T\dfrac{\delta {\mathbb K}_{\chi}}{\delta\chi}\bbxhat\hat{\pmb{\theta}}_{\chi}^{(1)}
				-C_3 \left(\hat{\pmb{\theta}}_{\chi}^{(3)}\right)^T\dfrac{\delta\mathbf{f}_{\chi}^{(1)}}{\delta{\chi}}\bbxhat \mdot
			\end{split}
		\end{equation}
		
		Replacing the RTD of the stiffness matrix and the force vector into equation (\ref{eq_diferentiation_8_app2}), one arrives to
		\begin{equation}\label{eq_final_derivative_4_app}
			\begin{split}
				&\dfrac{\delta\overline{\cal J}^{(h_e)}_{\text{vr}}(\chi)}{\delta\chi}\bbxhat
				=-2 C_3 {\mathcal A}\left(\theta_\chi^{(1)}\right) \Bigg(\\
				&C_2 \left(\hat{\pmb{\theta}}_{\chi}^{(2)}\right)^T\mathbf{B}^T\bbxhat\dfrac{\partial {\Kappalarge_{\chi}}}{\partial\chi}\bbxhat\mathbf{B}\bbxhat\hat{\pmb{\theta}}_{\chi}^{(1)}\Delta\chi_{\kappa}\bbxhat- \\
				&\hspace{0.75cm}-C_2 \left(\hat{\pmb{\theta}}_{\chi}^{(2)}\right)^T\mathbf{N}^T\bbxhat\dfrac{\partial {\hS_{\chi}}}{\partial\chi}\bbxhat \Delta\chi_{_{\shS}}\bbxhat\Bigg)+\\
				+&C_3 \left(\hat{\pmb{\theta}}_{\chi}^{(3)}\right)^T\mathbf{B}^T\bbxhat\dfrac{\partial {\Kappalarge_{\chi}}}{\partial\chi}\bbxhat\mathbf{B}\bbxhat\hat{\pmb{\theta}}_{\chi}^{(1)}\Delta\chi_{\kappa}\bbxhat- \\
				&\hspace{0.75cm}-C_3 \left(\hat{\pmb{\theta}}_{\chi}^{(3)}\right)^T\mathbf{N}^T\bbxhat\dfrac{\partial {\shS_{\chi}}}{\partial\chi}\bbxhat \Delta\chi_{_{\shS}}\bbxhat \mcolon
			\end{split} 
		\end{equation}
		where ${\mathcal A}\left(\theta_\chi^{(1)}\right)$ is equal to $\left({\mathcal T}_\chi\left(\theta_\chi^{(1)}\right)\right)^T \mathbb{M}_{\partial_c\Omega} {\mathbb{I}}$. Now we introduce the definition of the conductivity and the heat source with respect to the topology (equations (\ref{eq_conductivity_interp}) and (\ref{eq_heat_sources_interp})) into expression (\ref{eq_final_derivative_4_app}), yielding to
		\begin{equation}
			\begin{split}
				&\dfrac{\delta\overline{\cal J}^{(h_e)}_{\text{vr}}(\chi)}{\delta\chi}\bbxhat
				=-2 C_3 {\mathcal A}\left(\theta_\chi^{(1)}\right) \Bigg( \\
				&\hspace{0.5cm}C_2 \left[ m_{\kappa}{\chi}^{m_{\kappa}-1}\left(\hat{\pmb{\theta}}_{\chi}^{(2)}\right)^T\mathbf{B}^T\bbx \Kappalarge\bbx\mathbf{B}\bbx\hat{\pmb{\theta}}_{\chi}^{(1)}\right]_{\mathbf{x}=\hat{\mathbf{x}}}{{\hspace{-0.6cm}\Delta\chi_{\kappa}}({\hat{\bf x}}) }-\\
				&\hspace{0.5cm}-C_2 \left[m_{\shS}{\chi}^{m_{\shS}-1}\left(\hat{\pmb{\theta}}_{\chi}^{(2)}\right)^T\mathbf{N}^T\bbx \hS\bbx  \right]_{\mathbf{x}=\hat{\mathbf{x}}}\hspace{-0.6cm}\Delta\chi_{_{\shS}}\bbxhat\Bigg)+\\
				&\hspace{0.5cm}+C_3 \left[ m_{\kappa}{\chi}^{m_{\kappa}-1}\left(\hat{\pmb{\theta}}_{\chi}^{(3)}\right)^T\mathbf{B}^T\bbx \Kappalarge\bbx\mathbf{B}\bbx\hat{\pmb{\theta}}_{\chi}^{(1)}\right]_{\mathbf{x}=\hat{\mathbf{x}}}{{\hspace{-0.6cm}\Delta\chi_{\kappa}}({\hat{\bf x}}) }- \\
				&\hspace{0.5cm}-C_3 \left[m_{\shS}{\chi}^{m_{\shS}-1}\left(\hat{\pmb{\theta}}_{\chi}^{(3)}\right)^T\mathbf{N}^T\bbx \hS\bbx  \right]_{\mathbf{x}=\hat{\mathbf{x}}}\hspace{-0.6cm}\Delta\chi_{_{\shS}}\bbxhat \mdot
			\end{split}
		\end{equation}

		Finally, the sensitivity $\dfrac{\delta\overline{\cal J}^{(h_e)}_{\text{vr}}(\chi)}{\delta\chi}$ at point $\hat{\bf x}$ can be written as a sum of \emph{actual energies}, which yields to
		\begin{equation} \label{eq_sensitivity_cloaking_var_temp_app}
			\begin{split}
				\dfrac{\delta\overline{\cal J}^{(h_e)}_{\text{vr}}(\chi)}{\delta\chi}\bbxhat =& 
				-4C_3C_2m_{\kappa} \left(\chi_{\kappa}\bbxhat\right) ^{m_{\kappa}-1}{\overline{\cal U}}_{1-2}\bbxhat{{\Delta\chi}_{\kappa}({\hat{\bf x}}) }+ \\
				&+C_3C_2m_{\shS}\left({\chi}_{\shS}\bbxhat\right) ^{m_{\shS}-1}{\overline{\cal U}_{\shS-2}}\bbxhat\Delta\chi_{_{\shS}}\bbxhat+\\
				&+2C_3m_{\kappa} \left(\chi_{\kappa}\bbxhat\right) ^{m_{\kappa}-1}{\overline{\cal U}}_{1-3}\bbxhat{{\Delta\chi}_{\kappa}({\hat{\bf x}}) }-\\
				&-C_3m_{\shS}\left({\chi}_{\shS}\bbxhat\right) ^{m_{\shS}-1}{\overline{\cal U}_{\shS-3}}\bbxhat\Delta\chi_{_{\shS}}\bbxhat \mcolon
			\end{split}
		\end{equation}
		where $\overline{\cal U}_{i-j}\bbxhat$ is \textit{\Uconductivity} for i-th and j-th temperature fields ($i,j=\{1,2,3\}$) and ${\overline{\cal U}_{\shS-k}}\bbxhat$ corresponds to \textit{\Usource} for the k-th temperature field ($k=\{1,2,3\}$).
	
\end{appendices}

\bibliographystyle{abbrvnat}
\bibliography{thermal_papers}   

\end{document}